\documentclass{ieeeaccess}
\usepackage{cite}
\usepackage{amsmath,amssymb,amsfonts}
\usepackage{graphicx}
\usepackage{textcomp}
\def\BibTeX{{\rm B\kern-.05em{\sc i\kern-.025em b}\kern-.08em
    T\kern-.1667em\lower.7ex\hbox{E}\kern-.125emX}}

\usepackage{cite}
\usepackage{amsmath,amssymb,amsfonts}

\usepackage{algorithm}
\usepackage[noend]{algpseudocode}
\algnewcommand\algorithmicinput{\textbf{Input:}}
\algnewcommand\algorithmicoutput{\textbf{Output:}}
\algnewcommand\Input{\item[\algorithmicinput]}%
\algnewcommand\Output{\item[\algorithmicoutput]}%
\algnewcommand{\LineComment}[1]{\State \(\triangleright\) #1}

\newlength{\subcolumnwidth}

\newcommand{\nextsubcolumn}[1][]{%
  \cr\noalign{\hfill}
  \if\relax\detokenize{#1}\relax\else\hsize=#1\setlength{\subcolumnwidth}{\hsize}\fi
}

\usepackage{graphicx}
\usepackage{textcomp}

\ifCLASSOPTIONcompsoc
  \usepackage[caption=false,font=normalsize,labelfont=sf,textfont=sf]{subfig}
\else
  \usepackage[caption=false,font=footnotesize]{subfig}
\fi
\usepackage{placeins}
         
\usepackage[font={sf,scriptsize},
            labelfont={bf},
            justification= raggedright, 
            singlelinecheck=false]
            {caption}

\usepackage{array}
\newcolumntype{C}{>{\centering\arraybackslash}p{0.04\textwidth}}
\newcolumntype{L}{>{\raggedright\arraybackslash}p{0.02\textwidth}}

\usepackage[table]{xcolor}
\definecolor{Grey}{gray}{0.93}
\definecolor{blue}{RGB}{0, 47, 167}

\usepackage{bm}
\usepackage{enumitem}
\usepackage{mathtools}

\begin{document}
\history{Date of publication xxxx 00, 0000, date of current version xxxx 00, 0000.}
\doi{10.1109/ACCESS.2022.DOI}

\title{Deep Learning for Predictive Analytics in Reversible Steganography}
\author{
\uppercase{Ching-Chun~Chang}\authorrefmark{1},
\uppercase{Xu~Wang}\authorrefmark{2,3},
\uppercase{Sisheng~Chen}\authorrefmark{2,4},
\uppercase{Isao~Echizen}\authorrefmark{1},
\uppercase{Victor~Sanchez}\authorrefmark{5}, 
\uppercase{and Chang-Tsun~Li}\authorrefmark{6}
}
\address[1]{National Institute of Informatics, Tokyo, Japan (e-mail: iechizen@nii.ac.jp)}
\address[2]{Department of Information Engineering and Computer Science, Feng Chia University, Taichung, Taiwan}
\address[3]{School of Information Science and Engineering, University of Jinan, Jinan, China (e-mail: xu.wang.phd@gmail.com)}
\address[4]{School of Big Data and Artificial Intelligence, Fujian Polytechnic Normal University, Fuzhou, China (e-mail: sisheng.chen.phd@gmail.com)}
\address[5]{Department of Computer Science, University of Warwick, Coventry, UK (e-mail: v.f.sanchez-silva@warwick.ac.uk)}
\address[6]{School of Information Technology, Deakin University, Geelong, Australia (e-mail: changtsun.li@deakin.edu.au)}
\tfootnote{This work was supported in part by the Japan Society for the Promotion of Science (JSPS) under KAKENHI Grants (JP16H06302, JP18H04120, JP20K23355, JP21H04907 and JP21K18023) and the Japan Science and Technology Agency (JST) under CREST Grants (JPMJCR18A6 and JPMJCR20D3).}

\markboth
{C.-C. Chang \headeretal: Deep Learning for Predictive Analytics in Reversible Steganography}
{C.-C. Chang \headeretal: Deep Learning for Predictive Analytics in Reversible Steganography}

\corresp{Corresponding author: C.-C. Chang (e-mail: ccchang@nii.ac.jp).}

\begin{abstract}
Deep learning is regarded as a promising solution for reversible steganography. There is an accelerating trend of representing a reversible steo-system by monolithic neural networks, which bypass intermediate operations in traditional pipelines of reversible steganography. This end-to-end paradigm, however, suffers from imperfect reversibility. By contrast, the modular paradigm that incorporates neural networks into modules of traditional pipelines can stably guarantee reversibility with mathematical explainability. Prediction-error modulation is a well-established reversible steganography pipeline for digital images. It consists of a predictive analytics module and a reversible coding module. Given that reversibility is governed independently by the coding module, we narrow our focus to the incorporation of neural networks into the analytics module, which serves the purpose of predicting pixel intensities and a pivotal role in determining capacity and imperceptibility. The objective of this study is to evaluate the impacts of different training configurations upon predictive accuracy of neural networks and provide practical insights. In particular, we investigate how different initialisation strategies for input images may affect the learning process and how different training strategies for dual-layer prediction respond to the problem of distributional shift. Furthermore, we compare steganographic performance of various model architectures with different loss functions.

\end{abstract}


\titlepgskip=-15pt

\maketitle

\section{Introduction}\label{sec:intro}
\PARstart{S}{teganography} is recognised as an important research field in multimedia forensics and security. It is the practice of concealing a message within a \emph{cover} object to produce a \emph{stego} object~\cite{668971}. Steganographic technology has a wide range of applications, including but not limited to covert communications~\cite{2005_1511007}, ownership identification~\cite{1997_650120} and tamper proofing~\cite{1999_771070}. Reversible steganography is used primarily for data authentication in fidelity-sensitive applications such as creative production, forensic science, legal proceedings, medical diagnosis and military reconnaissance~\cite{2001_Fridrich_Invertible, 2003_1196739, 2013_6329433, 2016_RDH_Survey}. It lightens the burden of metadata management by invisibly inserting auxiliary authentication information (e.g. a tamper-proof fingerprint, a hash digest or a digital signature) into the data. Message embedding is carried out in a reversible manner; that is, steganographic distortion can be removed and the original data reconstructed in a time-reversed fashion. The ability to reverse a steganographic process, namely \emph{reversibility}, is a desirable feature in this era of data-centric artificial intelligence. Many real-world intelligent systems require colossal amounts of data to be collected and used in the training and validation processes. However, hostile attacks may occur during data communications. It has been reported that learning machinery is susceptible to data-driven adversarial attacks~\cite{2015_Perturb_Goodfellow, 2016_DeepFool, Poisoning17}. For instance, adversarial perturbation is an invisible noise crafted to fool a deep-learning system. Another example is adversarial contamination that imperceptibly injects poisonous samples into the training pool, thereby compromising the accuracy of the system. To combat such cybersecurity threats, steganography-based authentication schemes can be applied to verify the integrity of training and validation data, thereby ensuring secure data exchange. If an irreversible steganographic algorithm is incorporated into a data exchange protocol for authentication purposes, steganographic distortion might present uncontrollable risks to the reliability of data-centric autonomous machines~\cite{9753668}. Therefore, the ability to remove steganographic distortion and restore data integrity is of paramount importance.

The core of reversible steganography, in common with lossless compression, is predictive coding~\cite{1948_6773024}. Prediction-error modulation is one of the main pillars of contemporary reversible steganography due to its optimum rate-distortion performance~\cite{2011_5762603, 2014_6746082, Hwang:2016aa}. A prediction-error modulation scheme has an analytics module to serve the purpose of context-aware pixel intensity prediction, and a coding module to perform encoding/decoding in the residual domain. Deep learning is currently at the very heart of multimedia analytics and it offers appealing solutions for pixel intensity prediction. In this paper, we investigate the cause-effect relationships between the variables of interest regarding the use of deep learning models for pixel intensity prediction. In particular, we investigate the impacts of different initialisation and training strategies upon predictive accuracy. Our contributions are summarised as follows:

\begin{itemize}
\item We compare zero initialisation and local-mean initialisation to gain an insight into whether pre-processing of input images with a local-mean filter can help the models perform better or has a counterproductive effect. 

\item We explore universal training, independent training, and causal training to gain an insight into the problem of distributional shift in dual-layer prediction caused by steganographic distortion.

\item We carry out an ablation study to analyse the impacts of different loss functions upon steganographic performance with the aim of providing insights regarding the applicability of neural network models originally proposed for low-level computer vision, and demonstrate the state-of-the-art steganographic performance with the residual dense network (RDN) which is an advanced model for image super-resolution~\cite{2018_8578360} and image restoration~\cite{8964437}.
\end{itemize}

The remainder of this paper is organised as follows. Section~\ref{sec:review} provides a literature review on the recent development of steganography with deep learning. Section~\ref{sec:pre} delineates modules of a reversible steganographic scheme. Section~\ref{sec:method} introduces the methodology concerning intensity initialisation strategies, dual-layer training strategies, and neural network models. Section~\ref{sec:exp} evaluates the impact of different variables of interest upon predictive accuracy and steganographic performance. Limitations of this study are discussed in Section~\ref{sec:limit}. Conclusions are drawn in Section~\ref{sec:con}.


\section{Literature Review}\label{sec:review}
Artificial intelligence has been evolving over the years, being at the forefront of transformations of the world we live in~\cite{10.1093/mind/LIX.236.433}. Deep learning, as a new class of multi-purpose intelligent algorithms, learns how to solve complicated tasks through the observation of a large amount of data, and has brought about groundbreaking advances in many branches of science~\cite{LeCun:2015aa}. The foundations of deep learning are neural networks, or connectionist systems, which are capable of discovering intricate structures in high-dimensional data via multiple layers of artificial neurons. The recent development of steganographic methods is centred on the use of deep-learning models to enhance performance. Significant breakthroughs have been achieved in some basic properties such as \emph{capacity} (the allowable size of the embedded message) and \emph{imperceptibility} (the perceptual quality of the stego object), as well as some application-oriented properties such as \emph{secrecy} (the degree to which the stego object can elude detection), and \emph{robustness} (the degree to which the embedded message can survive distortions of various forms)~\cite{Zhu:2018aa}. In this section, we briefly review some seminal studies of secure and robust steganography with deep learning and then discuss the use of neural networks in reversible steganography with a contrast drawn between end-to-end and modular frameworks. Note that our study focuses on the basic properties (i.e. capacity and imperceptibility) and reversibility since application-oriented requirements can be contradictory and difficult to achieve concurrently.

\subsection{Secrecy}
Secrecy is at the heart of covert communications. The ability to pass secret information under surveillance is essential to espionage and military operations. It has been reported that deep learning can be used to identify locations in a cover image at which message embedding would not arouse suspicion~\cite{2017_8017430, 2019_8603808, 2020_8735922}. More technically, a deep learning algorithm assigns a cost to every pixel to quantify the effect of making modifications. In this way, a high degree of perceptual and statistical undetectability can be reached by minimising the cost. Other approaches include using generative models to convert a given message into a cover image which is less vulnerable to steganalysis~\cite{Volkhonskiy:2019aa}, to transform a cover image along with a secret message into a stego image~\cite{NIPS2017_6791}, and to encode a message directly into a realistic stego image in the absence of an explicit cover image~\cite{2018_8403208}.

\subsection{Robustness}
Robustness is a prioritised requirement for copyright protection. In commercial applications, entrepreneurs can prevent copyright infringement by embedding a registered watermark into digital assets. Deep learning has been used to embed invisible watermarks into images and videos in a durable way in order to identify copyright ownership and deter illegitimate copying~\cite{Luo:2020aa}. Unauthorised screen recording is a new scenario in which an adversary attempts to capture an electronically displayed still photograph or video footage with a digital camera, causing optical interference in watermark extraction. It has been shown that neural networks can be trained to simulate optical interference and then learn to encode and decode watermarks in a robust manner~\cite{Wengrowski_2019_CVPR}. Decoding accuracy can also be enhanced by using neural networks to remove light artefacts such as moir\'{e} fringes, a type of textile with a rippled or wavy appearance, prior to watermark extraction~\cite{Fang:2021aa}. To facilitate augmented reality, deep learning has been used to decode hyperlinks embedded in physical photographs rather than digital media, subject to real-world variations in print quality, illumination, occlusion and viewing distance~\cite{Tancik:2020aa}.

\subsection{Reversibility\textemdash End-to-End Framework}
Reversibility is a desirable characteristic for applications in which accuracy and consistency of data are important. It is a requisite for preventing an accumulation of steganographic distortion over each data transmission. A reversible steganographic method often relies on intricate logical operations to regulate imperceptibility and guarantee reversibility. Such operations are hard to achieve using existing neural network models. One of the earliest approaches uses a U-shaped network (U-Net) to automatically encode a cover image and a secret message into a stego image and a separate neural network model for decoding~\cite{Duan:2019aa}. Another approach encodes a message bitstream into a cover image via a generative model, trains a cycle-consistent generative adversarial network (CycleGAN) to learn the back-and-forth transformation between the cover and stego images, and extracts the message bitstream from the stego image by using another neural network model~\cite{Jung:2019aa}. Furthermore, an intriguing study shows that both encoding (forward mapping) and decoding (inverse mapping) can be performed by using a single invertible neural network (INN)~\cite{Lu_2021_CVPR}. While a monolithic end-to-end system generally offers a large steganographic capacity, a common limitation within this type of framework is imperfect reversibility due to the presence of an information bottleneck (i.e. a form of lossy compression) in neural networks. The lack of transparency, interpretability and explainability in neural networks adds an extra dimension to the problem.

\begin{figure*}[t]
    \centering
    \includegraphics[width=0.9\textwidth]{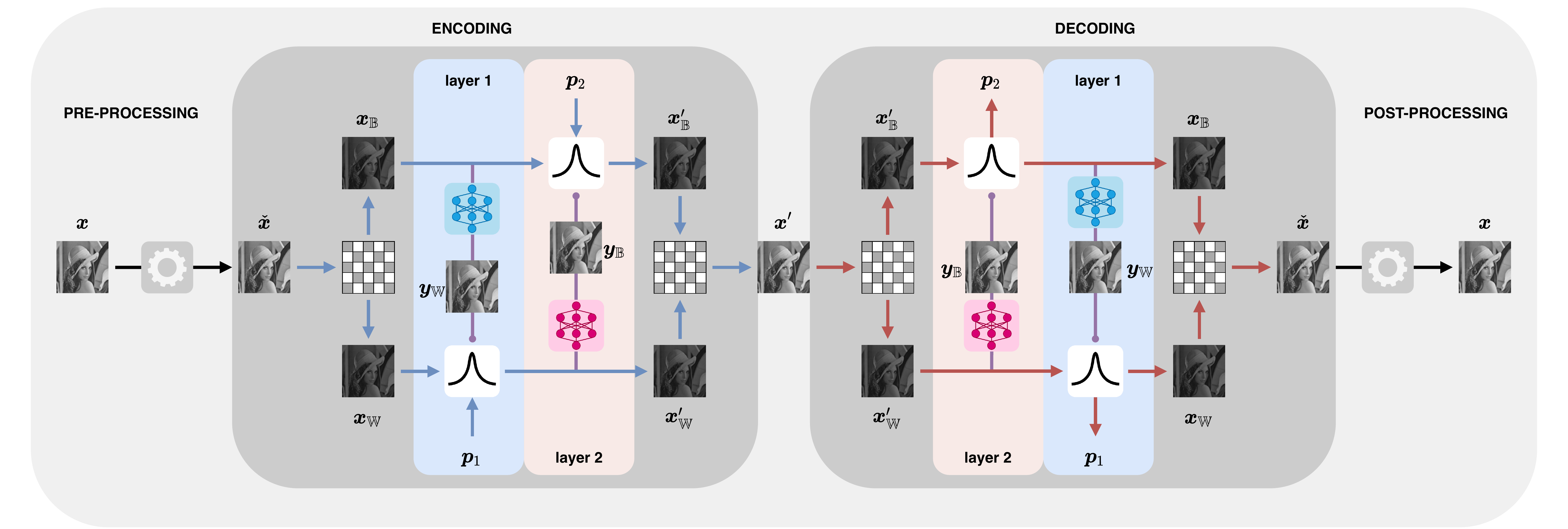}
    \caption{Schematic workflow of prediction-error modulation. From left to right: pre-processing, encoding, decoding and post-processing stages.}
    \label{fig:scheme}
\end{figure*}

\subsection{Reversibility\textemdash Modular Framework}
Although reversibility is not an all-or-nothing proposition, the unreliability of end-to-end learning may pose uncontrollable risks in certain circumstances. Problem decomposition or modularisation is essential to addressing complex problems. Existing reversible steganographic schemes often consist of \emph{coding} and \emph{analytics} modules. In general, the coding module handles the encoding and decoding mechanisms, which are designed to achieve perfect reversibility subject to an imperceptibility constraint. The analytics module models the data distribution and exploits data redundancy in order to optimise steganographic rate-distortion performance. For instance, the regular{\textendash}singular (RS) scheme introduced by Fridrich \textit{et al.} constructs a discrimination function to categorise image blocks into regular, singular and unusable groups on the basis of the smoothness prior, and uses invertible flipping of the least significant bits to achieve invisible and erasable embedding~\cite{Fridrich:2002aa}. We can define the former part as the analytics module and the latter part as the coding module. The problem of limited reversibility in previous deep-learning approaches lies in the difficulty of lossless coding in an end-to-end learning fashion. A recent study achieves perfect reversibility with a deep-learning-based RS scheme following this established \emph{modular} framework~\cite{2020_9245471}. A conditional generative adversarial network (GAN), referred to as pix2pix~\cite{2017_pix2pix_8100115}, is applied to improve the performance of the discrimination function. It has been shown that by partitioning a scheme into coding and analytics modules and deploying neural networks in the latter, reversibility can be reliably guaranteed.

\subsection{Analysis of Prior Art}
A combination of prediction-error modulation with deep learning has been described in two studies~\cite{Hu:2021aa} and~\cite{Chang:2021aa}. Neural networks are incorporated into the schemes based upon \emph{modularity}. The most notable feature in common is that both studies divide pixels into \emph{query} and \emph{context} sets by using a chequered pattern and develop neural network models for predicting the query from the context. The main difference is neural network architectures. The former constructs a multi-scale convolutional neural network (MS-CNN) to address the issue of restricted receptive fields in traditional predictors. The latter applies a persistent memory network (MemNet) originally developed for image denoising and image super-resolution~\cite{8237748}. From a certain perspective, context-aware pixel intensity prediction is closely related to \emph{low-level computer vision} tasks such as image denoising and image super-resolution because they all rely largely on low-level (or pixel-level) features such as edges, contours and textures, rather than high-level semantics~\cite{8723565}. Therefore, it is expected that advanced neural networks from the low-level computer vision domain can be applied directly with minor modifications. Another difference is the setting of the initial intensities of the query pixels. The former simply sets the query pixels to zero and learns to predict from scratch. By contrast, the latter initialises them to the mean of neighbouring pixels and performs prediction in a coarse-to-fine manner. Consequently, a problem of pixel-intensity prediction is, in a certain sense, transformed into a problem of image denoising or image super-resolution. Last but not least, the chequered pattern for context/query splitting features a dual-layer prediction: pixels assigned as the query in the first round, after partial embedding, can be assigned as the context in the second round. Such a practice gives rise to the problem of \emph{distributional shift}\textemdash the data distribution in the target (online) domain for deployment shifts from that in the training (offline) domain. In consequence of the distortion introduced in the preceding steganographic process, assigning those distorted pixels as the context could introduce discrepancy and bias into the contextual information, thereby undermining predictive performance. The expected performance relies on the extrapolation (generalisation) capability of neural networks. To summarise, this work aims to seek answers to the following questions:
\begin{itemize}
\item Can different intensity initialisation strategies affect predictive accuracy of neural networks substantially?
\item Can the problem of distributional shift be mitigated, thereby enabling efficient dual-layer prediction?
\item Can off-the-shelf neural networks from the low-level computer vision domain be transferred to perform pixel intensity prediction?
\end{itemize}



\begin{table}[t]
\caption{List of variables and operations.} 
\centering 
\begin{tabular}{l l} 
\hline\hline 
\textbf{var. \& op.} & \textbf{definition}\\ [0.5ex] 
\hline 
$\bm{x}$ 								& cover image \\
$\bm{x}^{\prime}$ 						& stego image \\
$\check{\bm{x}}$ 				& processed image \\
$\bm{x}_{\mathcal{Q}}$ 					& initialised query pixels \\
$\bm{y}_{\mathcal{Q}}$ 					& predicted query pixels \\
$\bm{\varepsilon}$ 						& prediction errors\\
$\bm{\varepsilon}^{\prime}$ 				& modulated prediction errors\\
$\bm{v}$ 								& overflow-status register \\
$\bm{m}$ 								& message \\
$\bm{p}$ 							& payload \\
$t$ 							& index for $\bm{v}$ and $\bm{p}$ \\
$\vartheta$ 							& stego-channel parameter\\
$\mathbb{B}$ 							& black set of pixel coordinates\\
$\mathbb{W}$ 							& white set of pixel coordinates\\
$\operatorname{net}_{1}$ 			& neural network for first-layer prediction\\
$\operatorname{net}_{2}$ 			& neural network for second-layer prediction\\
$\operatorname{scale}_{\Downarrow}$	& downscaling of possibly overflowing pixels\\
$\operatorname{scale}_{\Uparrow}$ 	& upscaling of possibly overflowing pixels\\
$\operatorname{split}$ 				& context-query splitting \\
$\operatorname{merge}$ 				& context-query merger\\
$\operatorname{encode}$ 				& prediction-error modulation (payload embedding)\\
$\operatorname{decode}$ 				& prediction-error de-modulation (payload extraction)\\
$\operatorname{concat}$ 				& concatenation of bit-streams\\
$\operatorname{deconcat}$ 			& de-concatenation of bit-streams\\
[1ex] 
\hline 
\end{tabular}
\label{tab:def} 
\end{table}

\section{Reversible Steganography}\label{sec:pre}
In this section, we delineate a reversible steganographic scheme based on prediction-error modulation. We begin with an overview of the scheme and then dissect each scheme component.

\subsection{Scheme Overview}
A schematic workflow of the encoding and decoding procedures is provided in Figure~\ref{fig:scheme} with the definitions of variables and operations listed in Table~\ref{tab:def}. The scheme based on prediction-error modulation can be broken down into an analytics module and a coding module. The former models the distribution of natural images and predicts the intensity of query pixels on the basis of the available context pixels. The latter encodes/decodes messages into/from prediction errors. Consider the following scenario: A sender (encoder) embeds a message $\bm{m}$ into a cover image $\bm{x}$ to produce a stego image $\bm{x^{\prime}}$ and then transmits it to a receiver (decoder) who extracts the message and restores the cover image on receipt of the stego image.

\begin{figure*}[t]
\begin{minipage}[t]{0.48\textwidth}
\centering
\begin{algorithm}[H]
\centering
\caption{Pre-processing}\label{alg:downscl}
\begin{algorithmic}

\Input $\bm{x}$, $\vartheta$
\Output $\check{\bm{x}}, \bm{v}$

\LineComment{down-scaling}
\State $\bm{v} = \varnothing$
\For {$i \gets 1$ to $\operatorname{height}(\bm{x})$}
\For {$j \gets 1$ to $\operatorname{width}(\bm{x})$}
\State $\check{x}_{i,j} \gets x_{i,j}$

\If{$x_{i,j} \geq 255 - \vartheta + 1$}
\Comment{overflow}
	\State $\check{x}_{i,j} \gets x_{i,j} - \vartheta$
	\State $\bm{v} = \operatorname{concat}(\bm{v}, 1)$\;

\ElsIf{$255 - \vartheta \geq x_{i,j} \geq 255 - 2\vartheta + 1 $}
	\State $\bm{v} = \operatorname{concat}(\bm{v}, 0)$

\ElsIf{$x_{i,j} \leq 0 + \vartheta - 1$}
\Comment{overflow}
	\State $\check{x}_{i,j} \gets x_{i,j} + \vartheta$
	\State $\bm{v} = \operatorname{concat}(\bm{v}, 1)$

\ElsIf{$0 + \vartheta \leq x_{i,j} \leq 0 + 2\vartheta - 1$}
	\State $\bm{v} = \operatorname{concat}(\bm{v}, 0)$

\EndIf
\EndFor
\EndFor

\end{algorithmic}
\end{algorithm}
\end{minipage}
\hfill
\begin{minipage}[t]{0.48\textwidth}
\centering
\begin{algorithm}[H]
\centering
\caption{Post-processing}\label{alg:upscl}
\begin{algorithmic}

\Input $\check{\bm{x}}$, $\bm{v}$, $\vartheta$
\Output $\bm{x}$

\LineComment{up-scaling}
\State $t=1$
\For {$i \gets 1$ to $\operatorname{height}(\check{\bm{x}})$}
\For {$j \gets 1$ to $\operatorname{width}(\check{\bm{x}})$}
\State $x_{i,j} \gets \check{x}_{i,j}$

\If{$255 - \vartheta \geq x_{i,j} \geq 255 - 2\vartheta + 1 $}
	\If{$v_t = 1$}
	\Comment{overflow}
		\State $x_{i,j} \gets \check{x}_{i,j} + \vartheta$
	\EndIf
\State $t = t + 1$
\ElsIf{$0 + \vartheta \leq x_{i,j} \leq 0 + 2\vartheta - 1$}
	\If{$v_t = 1$}
	\Comment{overflow}
		\State $x_{i,j} \gets \check{x}_{i,j} - \vartheta$
	\EndIf
\State $t = t + 1$;
\EndIf

\EndFor
\EndFor

\end{algorithmic}
\end{algorithm}
\end{minipage}
\end{figure*}

At the encoder side, we pre-process the cover image to prevent pixel intensity overflow during message embedding at the cost of generating extra auxiliary information. We treat the auxiliary information as a part of the payload and concatenate it with the intended message. Then, a chequered pattern is applied to split the image into a black set and a white set of pixels, denoted respectively by $\bm{x}_\mathbb{B}$ and $\bm{x}_\mathbb{W}$. In the first-layer prediction, we designate the black set as the context set and the white set as the query set. The roles are reversed in the second-layer prediction. We make use of a neural network or a predictive algorithm to estimate the intensities of the query pixels based on the observed context pixels. The prediction errors or residuals, denoted by $\bm{\varepsilon}$, are computed by subtracting the predicted values from the actual values. A portion of the payload is embedded in the residual domain through arithmetic operations. The modulated prediction errors, denoted by $\bm{\varepsilon}^{\prime}$, are then added to the predicted values, resulting in a slight distortion to the query pixels. The process can be repeated once more to embed another part of the payload by swapping the roles between the context and the query.

At the decoder side, we carry out message extraction and image restoration in a \emph{last-in-first-out} manner. Similar to the encoding process, the decoding process begins by splitting pixels of the stego image into the context and the query, predicting the query from the context, and calculating the prediction errors. A partial payload is extracted and the prediction errors are de-modulated back to the original state through inverse operations. Steganographic distortion is removed by adding the restored prediction errors to the predicted values. The decoding process can likewise be repeated once more by swapping roles between the context and query. Finally, the restored image is post-processed with the extracted auxiliary information to undo the minute modifications caused by the overflow prevention measure.

To summarise, the algorithmic steps are enumerated as follows. For the encoding phase:
\begin{enumerate}
\item $\{\check{\bm{x}}, \bm{v}\} = \operatorname{scale}_{\Downarrow}(\bm{x})$;
\item $\bm{p} = \operatorname{concat}(\bm{v}, \bm{m})$;
\item $\{\bm{x}_{\mathbb{B}}, \bm{x}_{\mathbb{W}}\} = \operatorname{split}(\check{\bm{x}})$;
\item $\bm{y}_{\mathbb{W}} = \operatorname{net}_1(\bm{x}_{\mathbb{B}}, \bm{x}_{\mathcal{Q}}| \mathcal{Q} = \mathbb{W})$;
\item $\bm{x}^{\prime}_{\mathbb{W}} = \operatorname{encode}(\bm{x}_{\mathbb{W}}, \bm{y}_{\mathbb{W}}, \bm{p}_{1})$;
\item $\bm{y}_{\mathbb{B}} = \operatorname{net}_2(\bm{x}^{\prime}_{\mathbb{W}}, \bm{x}_{\mathcal{Q}}| \mathcal{Q} = \mathbb{B})$;
\item $\bm{x}^{\prime}_{\mathbb{B}} = \operatorname{encode}(\bm{x}_{\mathbb{B}}, \bm{y}_{\mathbb{B}}, \bm{p}_{2})$;
\item $\bm{x}^{\prime} = \operatorname{merge}(\bm{x}^{\prime}_{\mathbb{B}}, \bm{x}^{\prime}_{\mathbb{W}})$.
\end{enumerate}
For the decoding phase:
\begin{enumerate}
\item $(\bm{x}^{\prime}_{\mathbb{B}}, \bm{x}^{\prime}_{\mathbb{W}}) = \operatorname{split}(\bm{x}^{\prime})$;
\item $\bm{y}_{\mathbb{B}} = \operatorname{net}_2(\bm{x}^{\prime}_{\mathbb{W}}, \bm{x}_{\mathcal{Q}}| \mathcal{Q} = \mathbb{B})$;
\item $\{ \bm{x}_{\mathbb{B}}, \bm{p}_{2} \} = \operatorname{decode}(\bm{x}^{\prime}_{\mathbb{B}}, \bm{y}_{\mathbb{B}})$;
\item $\bm{y}_{\mathbb{W}} = \operatorname{net}_1(\bm{x}_{\mathbb{B}}, \bm{x}_{\mathcal{Q}}| \mathcal{Q} = \mathbb{W})$;
\item $\{ \bm{x}_{\mathbb{W}}, \bm{p}_{1} \} = \operatorname{decode}(\bm{x}^{\prime}_{\mathbb{W}}, \bm{y}_{\mathbb{W}})$;
\item $\check{\bm{x}} = \operatorname{merge}(\bm{x}_{\mathbb{B}}, \bm{x}_{\mathbb{W}})$;
\item $\{\bm{v}, \bm{m}\} = \operatorname{deconcat}(\bm{p})$;
\item $\bm{x} = \operatorname{scale}_{\Uparrow}(\check{\bm{x}},\bm{v})$.
\end{enumerate}
All images are assumed to be 8-bit greyscale throughout this study. Let us denote by $\vartheta$ a threshold for the stego channel such that
\begin{equation}
\textrm{stego channel: } \{\varepsilon \mid \operatorname{abs}(\varepsilon) < \vartheta\} ,
\end{equation}
where $\operatorname{abs}$ denotes the absolute value. In other words, we use the parameter $\vartheta$ to determine for which prediction error values the payload embedding process can take place. The following presents the algorithmic details surrounding the notion of the stego channel.

\begin{table}[t!]
\caption{Code charts for overflow prevention w.r.t. different settings of $\vartheta$.} 

\subfloat[$\vartheta = 1$]{
\scalebox{0.7}{
\begin{tabular}{L | C | C } 
\hline\hline 
$x$ & $0$ & $255$\\ [0.5ex]
\hline
$\check{x}$ & $1$ & $254$\\ [0.5ex]
\hline 
\end{tabular}}}
\\

\subfloat[$\vartheta = 2$]{
\scalebox{0.7}{
\begin{tabular}{L | C | C | C | C } 
\hline\hline 
$x$ & $0$ & $1$ & $254$ & $255$\\ [0.5ex]
\hline
$\check{x}$ & $2$ & $3$ & $252$ & $253$\\ [0.5ex]
\hline 
\end{tabular}}}
\\

\subfloat[$\vartheta = 3$]{
\scalebox{0.7}{
\begin{tabular}{L | C | C | C | C | C | C } 
\hline\hline 
$x$ & $0$ & $1$ & $2$ & $253$ & $254$ & $255$ \\ [0.5ex]
\hline
$\check{x}$ & $3$ & $4$ & $5$ & $250$ & $251$ & $252$ \\ [0.5ex]
\hline 
\end{tabular}}}

\label{tab:Overflow_Coding} 
\end{table}

\begin{figure*}[t]
\begin{minipage}[t]{0.48\textwidth}
\centering
\begin{algorithm}[H]
\centering
\caption{Modulation}\label{alg:mod}
\begin{algorithmic}

\Input $\varepsilon$, $\bm{p}$, $t$, $\vartheta$
\Output $\varepsilon^{\prime}$, $t$

\LineComment{modulation}
\If {$\varepsilon = 0$}
	\If {$p_t = 0$}
	\Comment{embed 1 bit}
		\State $\varepsilon^{\prime} \gets 0$
		\State $t \gets t + 1$
	\ElsIf {$p_t = 1$ and $p_{t+1} = 0$}
	\Comment{embed 2 bits}
		\State $\varepsilon^{\prime} \gets -1$
		\State $t \gets t + 2$
	\ElsIf{$p_t = 1$ and $p_{t+1} = 1$}
	\Comment{embed 2 bits}
		\State $\varepsilon^{\prime} \gets +1$
		\State $t \gets t + 2$
	\EndIf
\ElsIf{$0 < \operatorname{abs}(\varepsilon) < \vartheta$}
\Comment{embed 1 bit}
	\If{$p_t = 0$}
		\State $\varepsilon^{\prime} \gets 2\varepsilon$
	\Else
		\State $\varepsilon^{\prime} \gets 2\varepsilon + \operatorname{sgn}(\varepsilon)\cdot 1$
	\EndIf
	\State $t \gets t + 1$
\Else
\Comment{non-embeddable}
	\State $\varepsilon^{\prime} \gets \varepsilon + \operatorname{sgn}(\varepsilon)\cdot \vartheta$
	\State $t \gets t$
\EndIf

\end{algorithmic}
\end{algorithm}
\end{minipage}
\hfill
\begin{minipage}[t]{0.48\textwidth}
\centering
\begin{algorithm}[H]
\centering
\caption{De-modulation}\label{alg:demod}
\begin{algorithmic}

\Input $\varepsilon^{\prime}$, $\bm{p}$, $t$, $\vartheta$
\Output $\varepsilon$, $\bm{p}$, $t$

\LineComment{de-modulation}
\If{$0 \leq \operatorname{abs}(\varepsilon^{\prime}) \leq 1 $}
	\State$\varepsilon \gets 0$\;
	\If{$\varepsilon^{\prime} = 0$}
	\Comment{extract 1 bit}
		\State $\bm{p} \gets \operatorname{concat}(\bm{p}, 0)$
		\State$t \gets t + 1$
	\ElsIf{$\varepsilon^{\prime} = -1$}
	\Comment{extract 2 bits}
		\State $\bm{p} \gets \operatorname{concat}(\bm{p}, 1, 0)$
		\State $t \gets t + 2$
	\ElsIf{$\varepsilon^{\prime} = +1$}
	\Comment{extract 2 bits}
		\State $\bm{p} \gets \operatorname{concat}(\bm{p}, 1, 1)$
		\State $t \gets t + 2$
	\EndIf
\ElsIf{$1 < \operatorname{abs}(\varepsilon^{\prime}) < 2\vartheta$}
\Comment{extract 1 bit}
	\State $\varepsilon \gets \operatorname{floor}(\operatorname{abs}(\varepsilon^{\prime}) / 2) \cdot \operatorname{sgn}(\varepsilon^{\prime})$
	\State $\bm{p} \gets \operatorname{concat}(\bm{p}, \operatorname{mod}(\operatorname{abs}(\varepsilon^{\prime}), 2))$
	\State $t \gets t + 1$
\Else
\Comment{non-extractable}
	\State $\varepsilon \gets \varepsilon^{\prime} - \operatorname{sgn}(\varepsilon^{\prime}) \cdot \vartheta$
	\State $\bm{p} \gets \bm{p}$
	\State $t \gets t$
\EndIf

\end{algorithmic}
\end{algorithm}
\end{minipage}
\end{figure*}

\begin{table*}[t]
\centering
\caption{Code charts for prediction-error modulation w.r.t. different settings of $\vartheta$.} 

\subfloat[$\vartheta = 1$]{
\scalebox{0.7}{
\begin{tabular}{L | C C | C | C C | C C | C C | C C C | C C | C C | C C | C | C C} 
\hline\hline 
$\varepsilon$ & {} & $-254$ & $\cdots$ & {} & $-3$ & {} & $-2$ & {} & $-1$ &

\cellcolor{Grey}{} & \cellcolor{Grey}$0$ & \cellcolor{Grey}{} &

$+1$ & {} & $+2$ & {} & $+3$ & {} & $\cdots$ & $+254$ & {} \\ [0.5ex] 
\hline
$p$ & {} & $\varnothing$ & $\cdots$ & {} & $\varnothing$ & {} & $\varnothing$ & {} & $\varnothing$ & 
\cellcolor{Grey}{$10_2$} & \cellcolor{Grey}$0_2$ & \cellcolor{Grey}{$11_2$} & 
$\varnothing$ & {} & $\varnothing$ & {}  & $\varnothing$ & {} & $\cdots$ & $\varnothing$ & {} \\ [0.7ex]

\hline
$\varepsilon^{\prime}$ & {} & $-255$ & $\cdots$ & {} & $-4$ & {} & $-3$ & {} & $-2$ & 
\cellcolor{Grey}$-1$ & \cellcolor{Grey}$0$ & \cellcolor{Grey}$+1$ & 
$+2$ & {} & $+3$ & {} & $+4$  & {} & $\cdots$ & $+255$ & {} \\ [0.7ex]

\hline 
\end{tabular}}}
\\

\subfloat[$\vartheta = 2$]{
\centering 
\scalebox{0.7}{
\begin{tabular}{L | C C | C | C C | C C | C C | C C C | C C | C C | C C | C | C C} 
\hline\hline 
$\varepsilon$ & {} & $-253$ & $\cdots$ & {} & $-3$ & {} & $-2$ & \cellcolor{Grey}{} & \cellcolor{Grey}$-1$ &

\cellcolor{Grey}{} & \cellcolor{Grey}$0$ & \cellcolor{Grey}{} &

\cellcolor{Grey}$+1$ & \cellcolor{Grey}{} & $+2$ & {} & $+3$ & {} & $\cdots$ & $+253$ & {} \\ [0.5ex] 
\hline
$p$ & {} & $\varnothing$ & $\cdots$ & {} & $\varnothing$ & {} & $\varnothing$ & \cellcolor{Grey}{$1_2$} & \cellcolor{Grey}$0_2$ & 
\cellcolor{Grey}{$10_2$} & \cellcolor{Grey}$0_2$ & \cellcolor{Grey}{$11_2$} & 
\cellcolor{Grey}$0_2$ & \cellcolor{Grey}{$1_2$} & $\varnothing$ & {}  & $\varnothing$ & {} & $\cdots$ & $\varnothing$ & {} \\ [0.7ex]

\hline
$\varepsilon^{\prime}$ & {} & $-255$ & $\cdots$ & {} & $-5$ & {} & $-4$ & \cellcolor{Grey}$-3$ & \cellcolor{Grey}$-2$ & 
\cellcolor{Grey}$-1$ & \cellcolor{Grey}$0$ & \cellcolor{Grey}$+1$ & 
\cellcolor{Grey}$+2$ & \cellcolor{Grey}$+3$ & $+4$ & {} & $+5$  & {} & $\cdots$ & $+255$ & {} \\ [0.7ex]

\hline 
\end{tabular}}}
\\

\subfloat[$\vartheta = 3$]{
\centering 
\scalebox{0.7}{
\begin{tabular}{L | C C | C | C C | C C | C C | C C C | C C | C C | C C | C | C C} 
\hline\hline 
$\varepsilon$ & {} & $-252$ & $\cdots$ & {} & $-3$ & \cellcolor{Grey}{} & \cellcolor{Grey}$-2$ & \cellcolor{Grey}{} & \cellcolor{Grey}$-1$ &

\cellcolor{Grey}{} & \cellcolor{Grey}$0$ & \cellcolor{Grey}{} &

\cellcolor{Grey}$+1$ & \cellcolor{Grey}{} & \cellcolor{Grey}$+2$ & \cellcolor{Grey}{} & $+3$ & {} & $\cdots$ & $+252$ & {} \\ [0.5ex] 
\hline
$p$ & {} & $\varnothing$ & $\cdots$ & {} & $\varnothing$ & \cellcolor{Grey}{$1_2$} & \cellcolor{Grey}$0_2$ & \cellcolor{Grey}{$1_2$} & \cellcolor{Grey}$0_2$ & 
\cellcolor{Grey}{$10_2$} & \cellcolor{Grey}$0_2$ & \cellcolor{Grey}{$11_2$} & 
\cellcolor{Grey}$0_2$ & \cellcolor{Grey}{$1_2$} & \cellcolor{Grey}$0_2$ & \cellcolor{Grey}{$1_2$}  & $\varnothing$ & {} & $\cdots$ & $\varnothing$ & {} \\ [0.7ex]

\hline
$\varepsilon^{\prime}$ & {} & $-255$ & $\cdots$ & {} & $-6$ & \cellcolor{Grey}$-5$ & \cellcolor{Grey}$-4$ & \cellcolor{Grey}$-3$ & \cellcolor{Grey}$-2$ & 
\cellcolor{Grey}$-1$ & \cellcolor{Grey}$0$ & \cellcolor{Grey}$+1$ & 
\cellcolor{Grey}$+2$ & \cellcolor{Grey}$+3$ & \cellcolor{Grey}$+4$ & \cellcolor{Grey}$+5$ & $+6$  & {} & $\cdots$ & $+255$ & {} \\ [0.7ex]

\hline 
\end{tabular}}}

\label{tab:PEM_Coding} 
\end{table*}

\subsection{Overflow Handling}
Both encoding and decoding are carried out in the residual domain. Addition and subtraction on residual values are equivalent to the same arithmetic operations on pixel values (i.e. a pro rata increment or decrement in pixel intensity). Given that computations are defined in a Galois finite field, such arithmetic operations may cause pixel intensity overflow: intensity values that are unexpectedly small or large wrap around the minimum or maximum after manipulations. We pre-process the image to prevent pixels from moving off boundary (or becoming saturated) and mark the locations where off-boundary pixels may occur. To distinguish between the processed pixels $\check{x}$ and unprocessed pixels $x$ with the same value, we flag $1$ (true) for the former and $0$ (false) for the latter in an overflow-status register, that is,
\begin{equation}
\check{x}_{i,j} = 
\begin{cases}
x_{i,j} + \vartheta &\textrm{if } x_{i,j} \in \mathbb{L}_{1},\\
x_{i,j} - \vartheta &\textrm{if } x_{i,j} \in \mathbb{U}_{1},\\
x_{i,j} &\textrm{otherwise},
\end{cases}
\end{equation}
and
\begin{equation}
\bm{v} = 
\begin{cases}
\operatorname{concat}(\bm{v}, 0) &\textrm{if } x_{i,j} \in \mathbb{L}_{0} \cup \mathbb{U}_{0},\\
\operatorname{concat}(\bm{v}, 1) &\textrm{if } x_{i,j} \in \mathbb{L}_{1} \cup \mathbb{U}_{1},\\
\end{cases}
\end{equation}
where
\begin{equation}
\begin{split}
\mathbb{L}_{0} &= [0+\vartheta, 0+2\vartheta-1] , \\
\mathbb{L}_{1} &= [0, 0+\vartheta-1] , \\
\mathbb{U}_{0} &= [255-2\vartheta+1, 255-\vartheta] , \\
\mathbb{U}_{1} &= [255-\vartheta+1,255] .
\end{split}	
\end{equation}
The parameter $\vartheta$ is associated with the modulation step in the message embedding process. A greater value of $\vartheta$ spans a wider range of pixel values that may be modified beyond the boundary and thus having a higher probability of increasing the size of the register. The overflow-status register is an overhead of reversibility. It has to be concatenated with the message and embedded as a part of the payload. As a consequence, its size has to be deducted from the overall payload when assessing steganographic capacity. See Algorithms~\ref{alg:downscl} and~\ref{alg:upscl} for the pseudo-codes. The overflow handling process is codified in Table~\ref{tab:Overflow_Coding}.

\subsection{Encoding/Decoding}
Let $y_{i,j}$ denote a prediction of the pixel at location $(i,j)$. For each query pixel, we compute its prediction error by
\begin{equation}
\varepsilon = \check{x}_{i,j} - y_{i,j} .
\end{equation}
For the prediction errors determined as the stego channel, we embed a bit of payload information by
\begin{equation}
\varepsilon^{\prime} = \varepsilon + \operatorname{sgn}(\varepsilon)\cdot p_t,
\end{equation}
where $\operatorname{sgn}$ extracts the sign of the prediction error (either positive or negative) and $p_t$ is the current payload bit. For the rest of the errors, we shift them by $\vartheta$ to prevent error values from overlapping (i.e. indistinguishable from the errors determined as stego channel); that is,
\begin{equation}
\varepsilon^{\prime} = \varepsilon + \operatorname{sgn}(\varepsilon)\cdot \vartheta.
\end{equation}
A modulated error is then added to the predicted value, resulting in a stego pixel
\begin{equation}
x^{\prime}_{i,j} = y_{i,j} + \varepsilon^{\prime}.
\end{equation}
Let us take the case in which $\vartheta = 2$ for example. When $\varepsilon = 0$, we can map it to one of the three states, $\varepsilon^{\prime} = 0$ or $\pm{1}$, to encode a ternary digit (a trit of information). When $\varepsilon = \pm{1}$, we shift it to $\pm{2}$ to avoid ambiguity and map it to one of the two states, $\varepsilon^{\prime} = \pm{2}$ or $\pm{3}$ (disregarding the sign), to represent one bit of information. For all other $\varepsilon$ such that $\varepsilon \notin (-\vartheta, +\vartheta)$, we shift each of them by $\vartheta$ in either a positive or negative direction, depending on the sign. However, converting the payload between binary (base $2$) and ternary (base $3$) numeral systems on the fly can be problematic. To circumvent this issue, we can map the errors with value $0$ to the digit $0$ if the observed payload bit $p_t$ is $0_2$; otherwise, we read the next payload bit $p_{t+1}$ and map the errors to the digit $-1$ if $[p_t, p_{t+1}] = [1_2,0_2]$ and to the digit $+1$ if $[p_t, p_{t+1}] = [1_2,1_2]$; that is,
\begin{equation}
\{\varepsilon^{\prime} \mid \varepsilon=0\} =
\begin{cases}
0 &\textrm{if $p_t = 0_2$} ,\\
-1 &\textrm{if $[p_t, p_{t+1}] = [1_2,0_2]$} ,\\
+1 &\textrm{if $[p_t, p_{t+1}] = [1_2,1_2]$} .
\end{cases}
\end{equation}
Theoretically, the mapping permits one trit or $\log_2 3 \approx 1.585$ bits of information to be embedded. In practice, the compromised solution embeds one or two bits with a probability of $0.5$ each, thereby being capable of embedding $1.5$ bits on average for $\varepsilon = 0$. Decoding is simply an inverse mapping. It begins by predicting the query pixel intensities and computing the prediction errors. Each error is de-modulated according to its magnitude. For errors between $\pm 1$, we map them back to $0$ and extract the corresponding payload bits. For errors whose magnitude is in $(1, 2\vartheta)$ regardless of the sign, we de-modulate them by the floor division of the magnitude by $2$ and extract the payload bits by the remainder of the division. For the rest errors, we shift them back by $\vartheta$. The original pixel intensity is recovered by adding the de-modulated error to the predicted value.

\begin{figure}[t!] 
\centering
\subfloat[Cover]{\includegraphics[width=0.25\columnwidth]{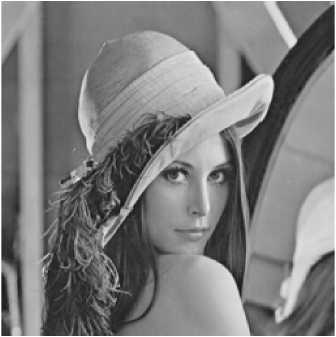}}
\hfil
\subfloat[Stego ($\vartheta = 1$)]{\includegraphics[width=0.25\columnwidth]{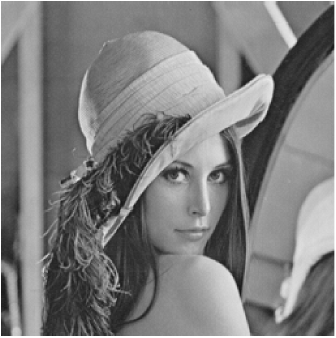}}
\hfil
\subfloat[Stego ($\vartheta = 2$)]{\includegraphics[width=0.25\columnwidth]{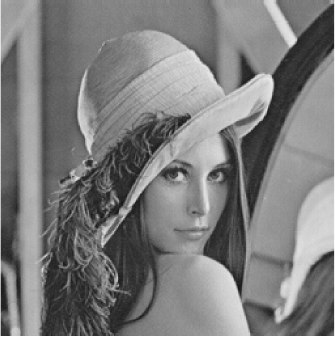}}
\hfil
\subfloat[Stego ($\vartheta = 3$)]{\includegraphics[width=0.25\columnwidth]{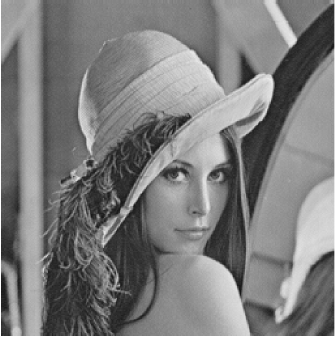}}
\\
\subfloat[NA]{\includegraphics[width=0.25\columnwidth]{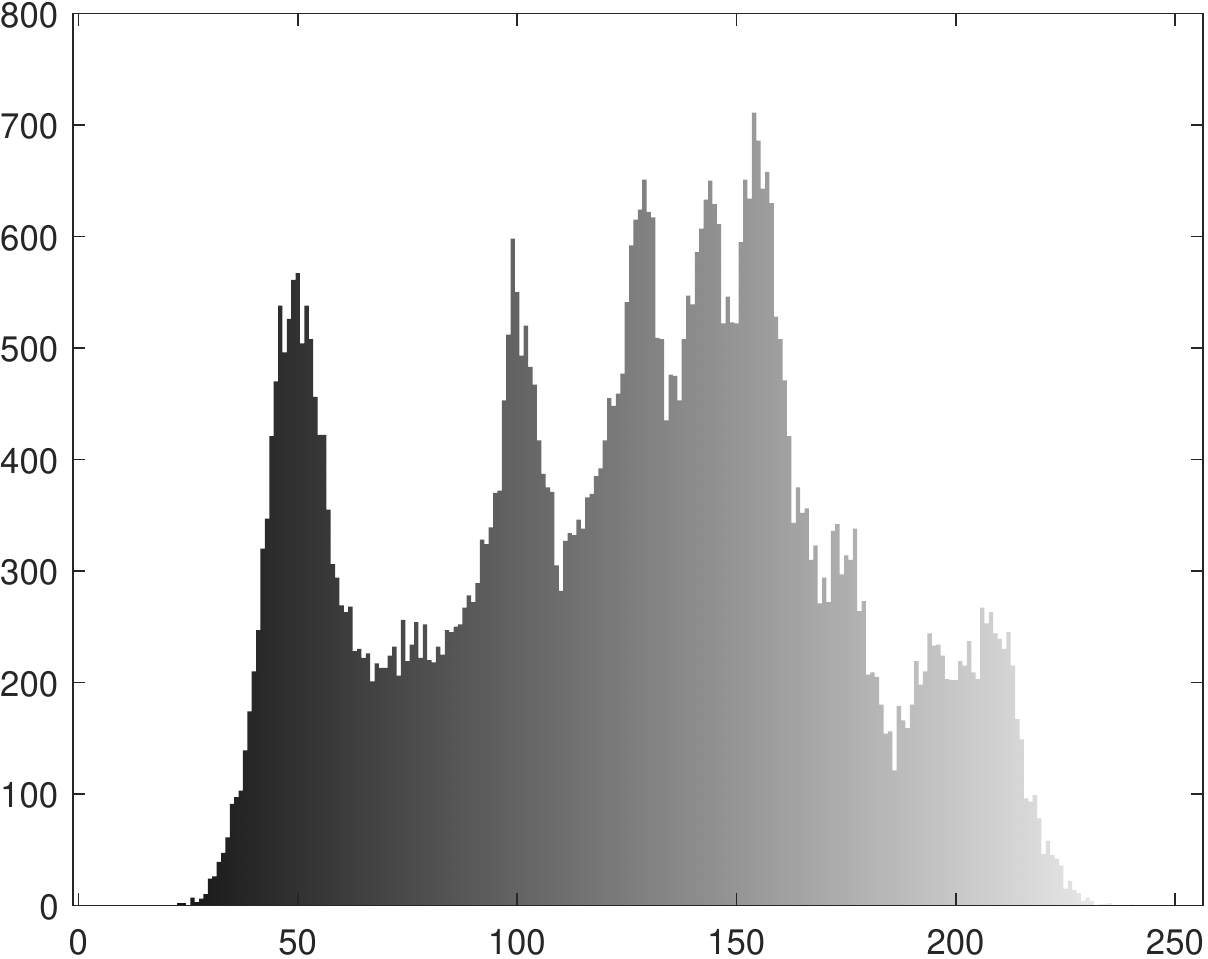}}
\hfil
\subfloat[48.3806 dB]{\includegraphics[width=0.25\columnwidth]{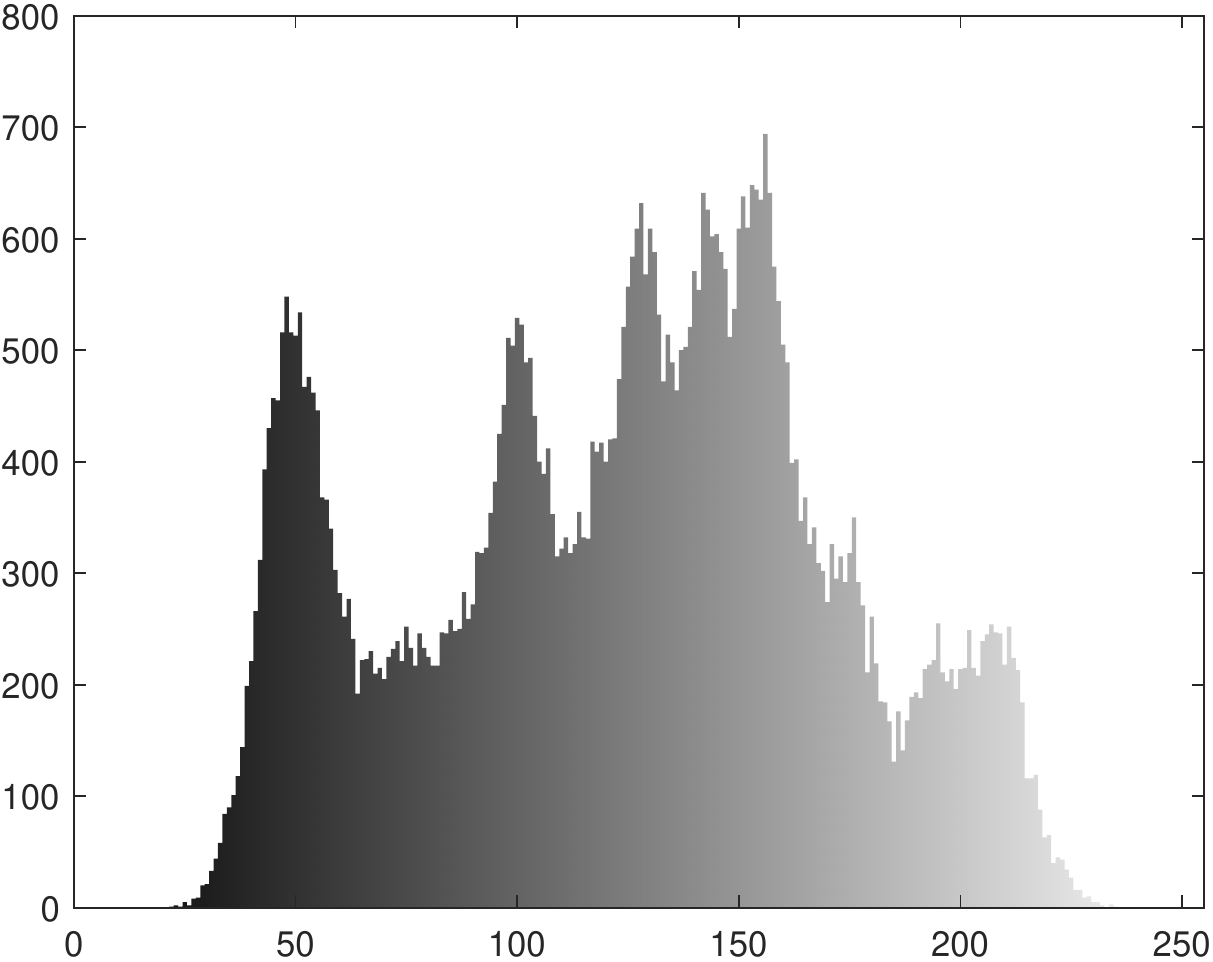}}
\hfil
\subfloat[43.6148 dB]{\includegraphics[width=0.25\columnwidth]{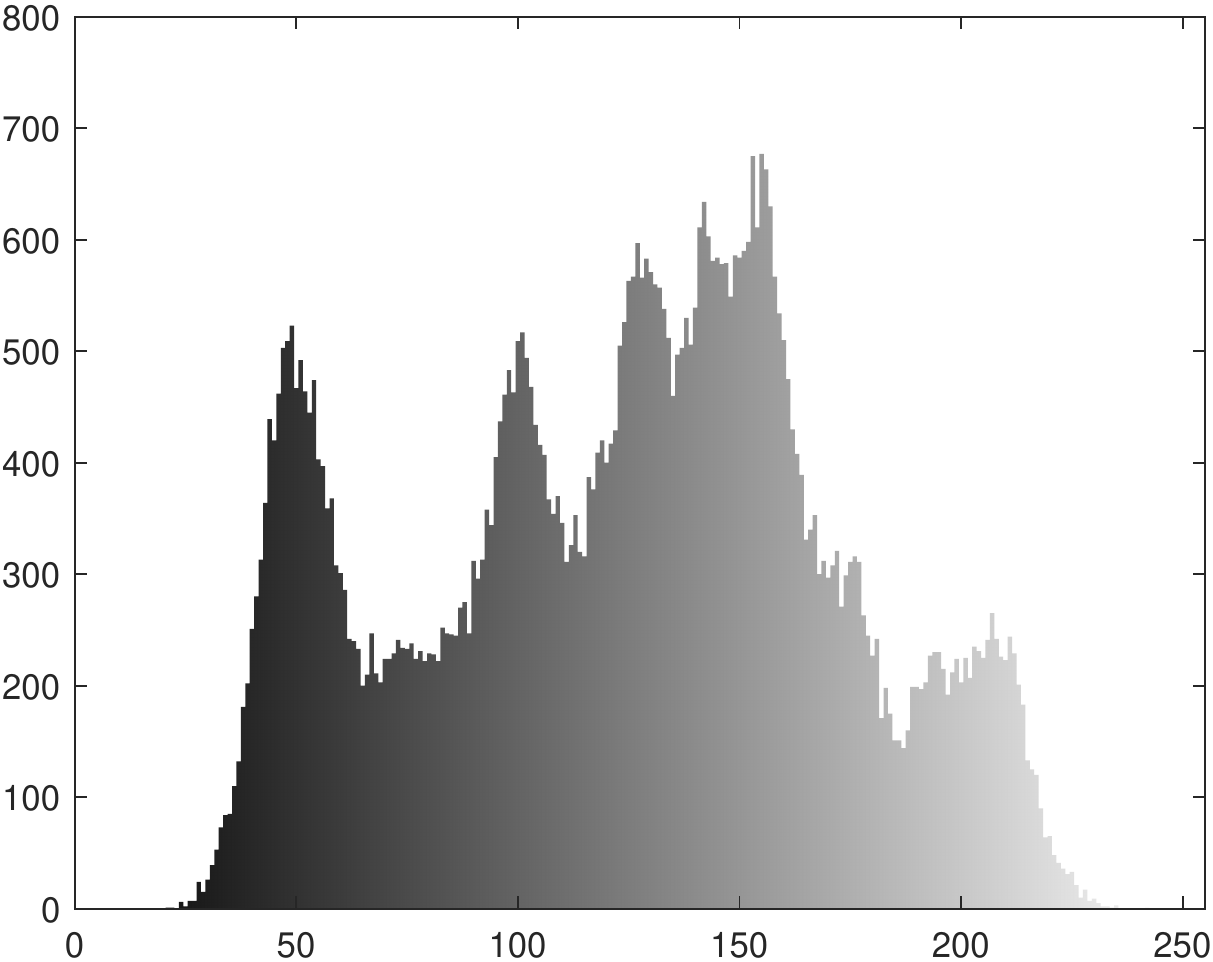}}
\hfil
\subfloat[41.4330 dB]{\includegraphics[width=0.25\columnwidth]{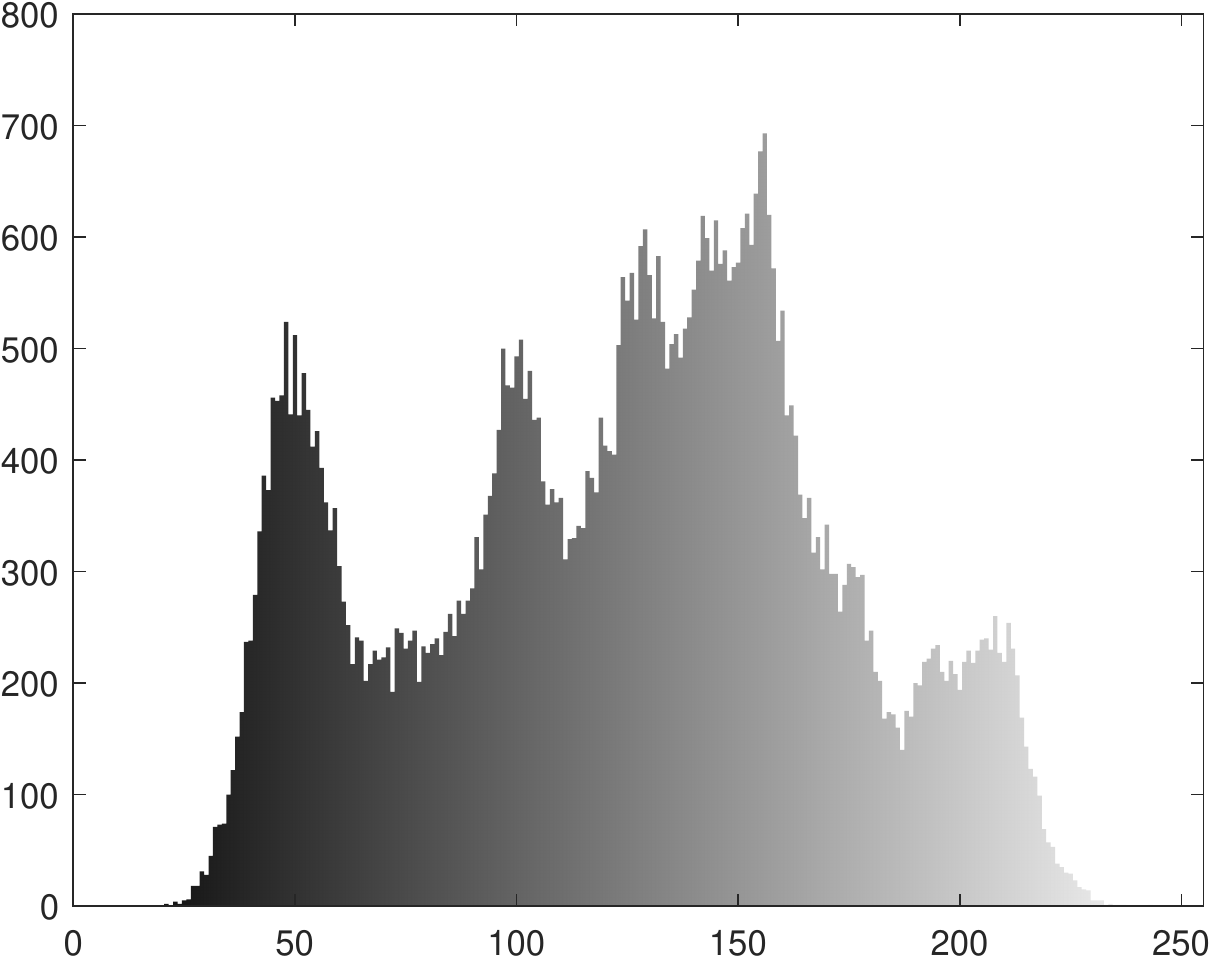}}
\\
\subfloat[NA]{\includegraphics[width=0.25\columnwidth]{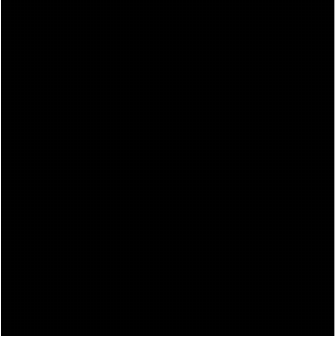}}
\hfil
\subfloat[0.3386 bpp]{\includegraphics[width=0.25\columnwidth]{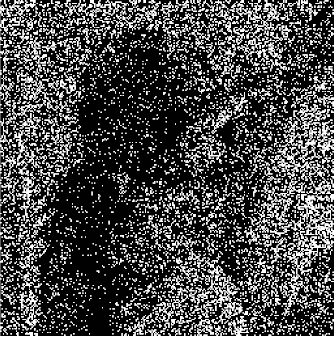}}
\hfil
\subfloat[0.6423 bpp]{\includegraphics[width=0.25\columnwidth]{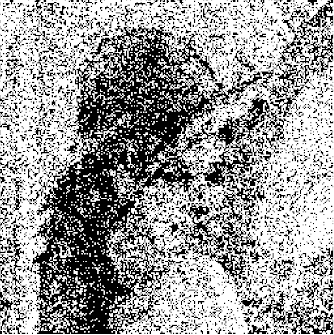}}
\hfil
\subfloat[0.8296 bpp]{\includegraphics[width=0.25\columnwidth]{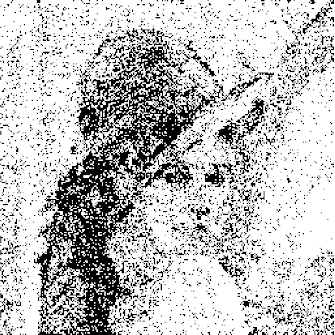}}

\caption{Stego images, their tonal distributions, and locations of carrier pixels (coloured in white). Numerical data expresses PSNR and embedding rate.}
\label{fig:stego}
\end{figure}

According to the \emph{Law of Error}~\cite{Wilson:1923aa}, we can hypothesise that the frequency of an error approximates an exponential function of its magnitude and therefore follows a zero-mean Laplacian distribution. This symmetrical modulation pattern is derived from the fact that prediction errors are expected to centre around $0$. Therefore, it is advisable to designate errors with a small absolute magnitude (of high frequency) as the stego channel. A rise in the width of the stego channel represents an increased steganographic capacity. See Algorithms~\ref{alg:mod} and~\ref{alg:demod} for the pseudo-codes. The prediction-error modulation process is codified in Table~\ref{tab:PEM_Coding}. Figure~\ref{fig:stego} displays stego images, their tonal distributions, and locations of carrier pixels w.r.t. different settings of stego-channel parameter $\vartheta$. Results are generated by spreading a pseudo-random binary message over the whole image at the maximal steganographic capacity. It can be observed that the quality degradation is virtually imperceptible, the disturbance to the tonal distribution is subtle, and the pixels selected for carrying the payload are mostly clustered in smooth areas. We would like to note that this coding method is simple, readily scalable and computationally efficient but not necessarily optimal. The main theme of this study is the analytics module, which is independent of the coding module. Any coding method that reflects more accurately the statistical distribution of residuals can be applied without conflicting with the findings of this study. For the subject of mathematical optimisation of reversible steganographic coding, the interested reader is referred to the research study~\cite{https://doi.org/10.48550/arxiv.2202.13133}.

\section{Predictive Analytics}\label{sec:method}
In this section, we take a closer look at the practice of training neural networks for predictive analytics in reversible steganography. We investigate different initialisation strategies for configuring the inputs of neural networks, different training strategies for fitting neural networks with dual-layer prediction, and different neural network architectures suitable for context-aware pixel intensity prediction.

\subsection{Context--Query Splitting}
\begin{figure}[t]
    \centering
    \includegraphics[width=0.23\textwidth]{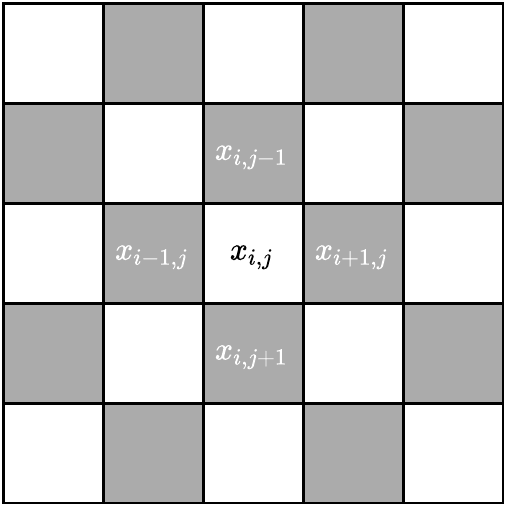}
    \caption{Illustration of chequered pattern for context--query splitting.}
    \label{fig:chequerboard}
\end{figure}

A convenient way to define pixel connectivity is to consider the von Neumann neighbourhood, that is, four adjacent pixels connected horizontally and vertically. If we sample the context and query pixels uniformly in such a way that each query pixel has $4$ connected context pixels, we end up forming a pattern analogous to a chequerboard, as illustrated in Figure~\ref{fig:chequerboard}. We can define the black set $\mathbb{B}$ as the context and the white set $\mathbb{W}$ as the query (or the other way round):
\begin{equation}
(i,j) \in 
\begin{cases}
\textrm{$\mathbb{B} \coloneqq \mathcal{C}$} &\textrm{if } i+j \textrm{ is odd} ,  \\
\textrm{$\mathbb{W} \coloneqq \mathcal{Q}$} &\textrm{if } i+j \textrm{ is even} .
\end{cases}
\end{equation}
A na{\"i}ve way to estimate the intensity of the central pixel is to calculate the mean of locally connected pixels. This heuristic predictive model, referred to as local-mean interpolation (LMI)~\cite{2009_4811982}, is based on the smoothness prior, a generic contextual assumption on real-world photographs. The recent studies have shown that neural network models (e.g. MS-CNN~\cite{Hu:2021aa} and MemNet~\cite{Chang:2021aa}) can be applied to improve predictive accuracy. Because predictive accuracy is instrumental in steganographic rate-distortion performance, the factors that could have the impacts upon predictive accuracy are primary concerns of this study.

\subsection{Initialisation Strategies}
The first step of the encoding phase is to split a cover image into $\bm{x}_{\mathbb{B}}$ and $\bm{x}_{\mathbb{W}}$. While the values of the query pixels are supposed to be predicted, the initial values of query pixels $\bm{x}_{\mathcal{Q}}$ have to be determined and cannot be set as null since the input to the applied neural networks must be a complete image. To this end, we consider two simple initialisation strategies: zero initialisation and local-mean initialisation. 

\subsubsection{Zero initialisation}
For zero initialisation, we set the initial value for each query pixel explicitly to $0$. This strategy is rather arbitrary but computationally efficient when compared with other initialisation strategies. Apart from this, it involves little subjective knowledge, preliminary analysis, or preconception about the intensities. In other words, it involves minimal human interference in the machine learning process. We may perceive prediction with zero initialisation as a special type of image super-resolution problem in which pixels are downsampled according to a special chequered pattern. The query pixels set to zero are viewed as missing data.

\subsubsection{Local-mean initialisation}
For local-mean initialisation, we assign the mean intensity of $4$ connected neighbours as the initial value of each query pixel. This strategy involves the computation of a local mean, but it is often the case that pre-estimation could help to accelerate the training process. We can test whether learning-based models tend to perform better and converge faster when such approximate values are pre-calculated. From another perspective, local-mean initialisation formulates the prediction problem as a image denoising problem by viewing the query pixels as noisy data.

\subsection{Training Strategies}
\begin{figure}[t]
    \centering
    \includegraphics[width=0.41\textwidth]{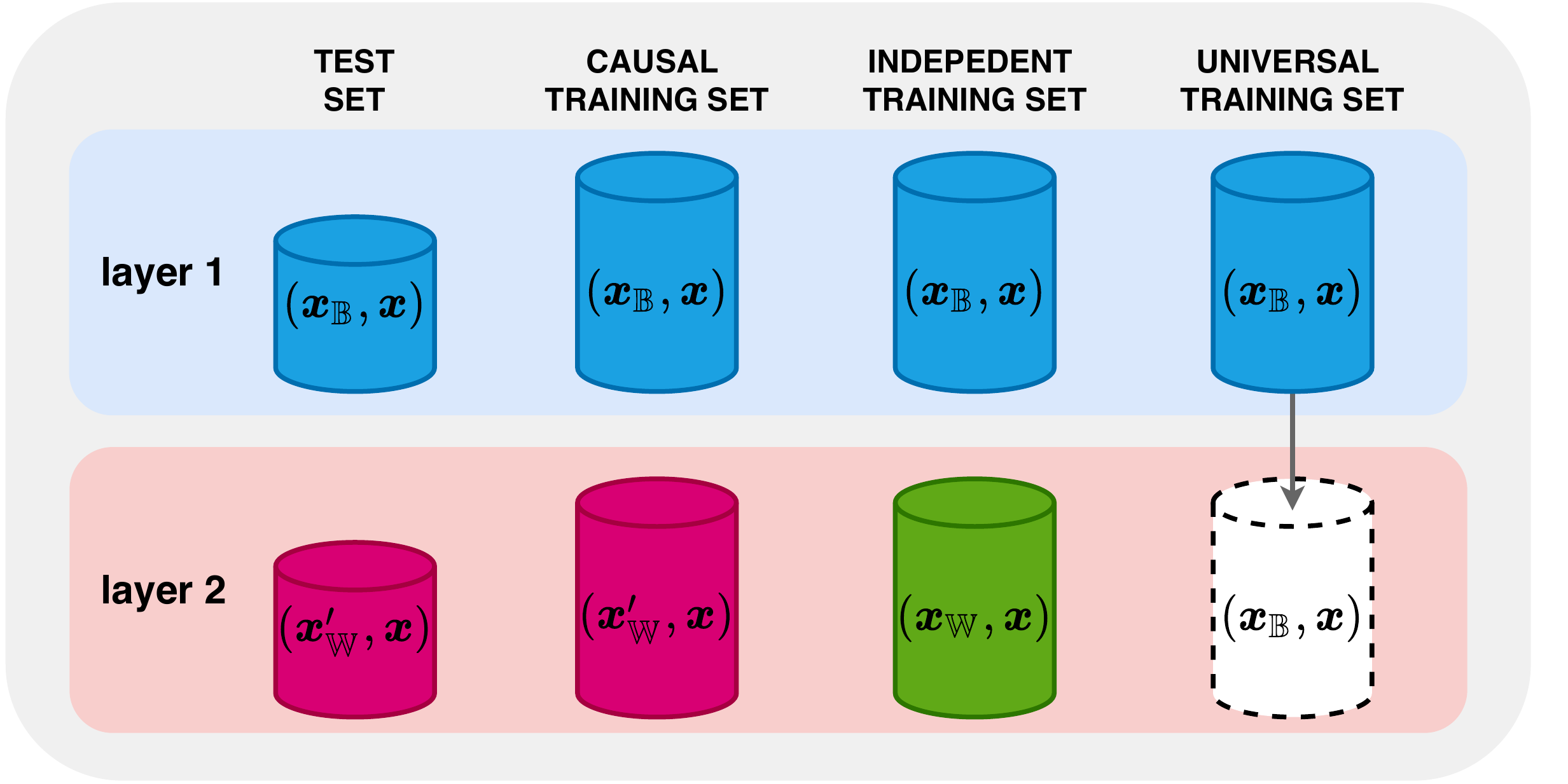}
    \caption{Training sets of causal, indepedent and universal strategies for dual-layer predictive analytics.}
    \label{fig:division}
\end{figure}

Dual-layer prediction entails the problem of distributional shift because steganographic distortion appears after the first-layer embedding. In the first layer, the intensities of query pixels are predicted by a neural network model and then modified with the prediction-error modulation algorithm. In the second layer, the roles of context and query are switched. As a result, the context set consists of distorted pixels. This can also be viewed as uncertainty propagation in the sense that errors in the previous layer propagate to the next layer, impairing the predictive performance. Formally, we aim to minimise the loss of two test sets:
\begin{equation}
\begin{split}
\textrm{test}_1 &: \mathcal{L}(\operatorname{net}(\bm{x}_{\mathbb{B}}, \bm{x}_{\mathcal{Q}}), \bm{x}_{\mathbb{W}}) ,\\
\textrm{test}_2 &: \mathcal{L}(\operatorname{net}(\bm{x}^{\prime}_{\mathbb{W}}, \bm{x}_{\mathcal{Q}}), \bm{x}_{\mathbb{B}}) .
\end{split}
\end{equation}
To this end, we explore three training strategies: universal training, independent training, and causal training. The training strategies differ by the configurations of the training set, as illustrated in Figure~\ref{fig:division}.

\subsubsection{Universal training}
The universal training is a cost-efficient way to manage dual-layer prediction by training a single model for performing prediction in both layers. The motivation is that the context/query switch between the first and second layers can be perceived as a simple geometric translation and convolutional neural networks (CNNs) are considered to be translation-invariant. For that reason, a single model may be sufficient and can generalise well for tackling dual-layer prediction. According to the chequered pattern, the context (or query) coordinates of the first layer, when shifted by one step (either horizontally or vertically), becomes the context (or query) coordinates of the second layer. Inspired by the translation-invariance property of classic CNN models, we conjecture that a model trained on the first set can be deployed directly to make inferences on the second test set. In other words, for both test sets, we train a universal model by
\begin{equation}
\operatorname{net}_{\textrm{uni}}: \underset{\operatorname{net}}{\operatorname{argmin}}\, \mathcal{L}(\operatorname{net}(\bm{x}_{\mathbb{B}}, \bm{x}_{\mathcal{Q}}), \bm{x}_{\mathbb{W}}) .
\end{equation}

\subsubsection{Independent training}
The independent training is to train two models in parallel for respective layers. For inference on the second test set, we structure the training set without taking account of steganographic distortion. This strategy can serve to check whether a single translation-invariant model is adequate. If the conjecture is valid, both would yield a similar result. Although the practice of training two models indicates an extra computational cost, this strategy permits parallel training because the unmodified images rather than stego images are used as the inputs for training the second-layer predictor, that is,
\begin{equation}
\begin{split}
\operatorname{net}_{\textrm{ind}}^1: \underset{\operatorname{net}}{\operatorname{argmin}}\, \mathcal{L}(\operatorname{net}(\bm{x}_{\mathbb{B}}, \bm{x}_{\mathcal{Q}}), \bm{x}_{\mathbb{W}}) ,\\
\operatorname{net}_{\textrm{ind}}^2: \underset{\operatorname{net}}{\operatorname{argmin}}\, \mathcal{L}(\operatorname{net}(\bm{x}_{\mathbb{W}}, \bm{x}_{\mathcal{Q}}), \bm{x}_{\mathbb{B}}) .
\end{split}
\end{equation}

\subsubsection{Causal training}
The causal training is to train the models in a consecutive manner such that the succeeding model is trained after the data has been modified with the preceding model. That is, a predictive model is trained and then deployed for embedding information into the training data for another model. The problem of distributional shift stems from the discrepancy between the training and deployment environments (i.e. distributions of the training and test sets). Reducing the deployment loss requires the distribution of the training set to be as close as possible to that of the test set. A potential way to remedy the problem is to inject steganographic distortion to the second-layer training set. At the first glance, it seems not surprising that the distributional shift can be mitigated by introducing steganographic distortion to the training of the second layer. However, steganographic distortion varies with the embedded message and is quite random (rather than being a fixed noise pattern). It implies that such distortion can not be simply filtered out by the models. Nevertheless, we hypothesis that neural networks can learn to predict from much more stable and reliable image representations if such distortion is presented during training. Implementing a steganographic algorithm incurs an extra computational cost and the succeeding model can only be trained once the training of the preceding model has been completed. Two models are causally connected as the prediction of the first model contributes to the prediction of the second model. Specifically, two models are trained by
\begin{equation}
\begin{split}
\operatorname{net}_{\textrm{cos}}^1: \underset{\operatorname{net}}{\operatorname{argmin}}\, \mathcal{L}(\operatorname{net}(\bm{x}_{\mathbb{B}}, \bm{x}_{\mathcal{Q}}), \bm{x}_{\mathbb{W}}) ,\\
\operatorname{net}_{\textrm{cos}}^2: \underset{\operatorname{net}}{\operatorname{argmin}}\, \mathcal{L}(\operatorname{net}(\bm{x}^{\prime}_{\mathbb{W}}, \bm{x}_{\mathcal{Q}}), \bm{x}_{\mathbb{B}}) .
\end{split}
\end{equation}

\subsection{Neural Networks}
\begin{figure}[t!] 
\centering
\subfloat[MS-CNN\label{fig:MSCNN}]{\includegraphics[width=0.81\columnwidth]{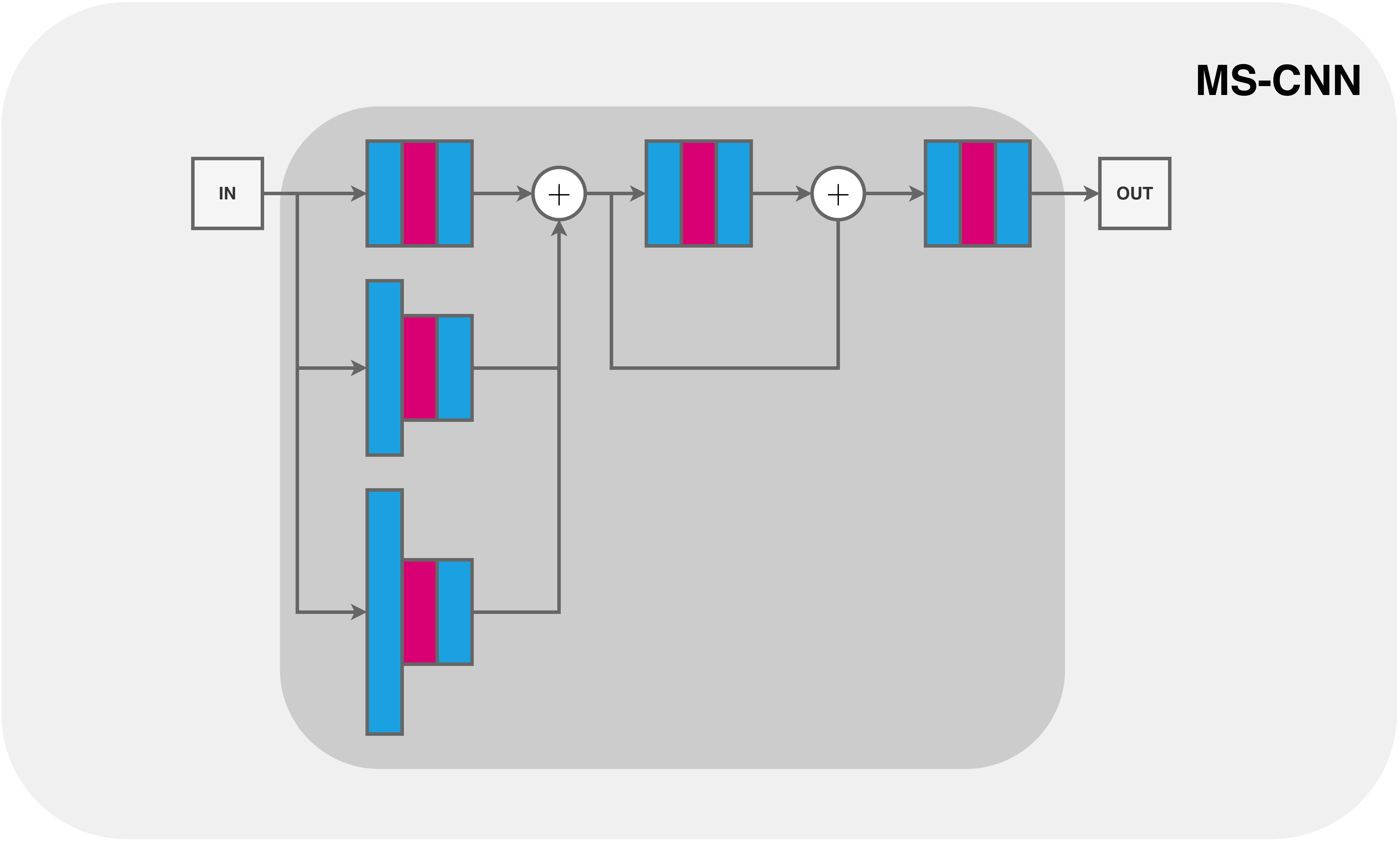}}
\\
\subfloat[MemNet\label{fig:MemNet}]{\includegraphics[width=0.81\columnwidth]{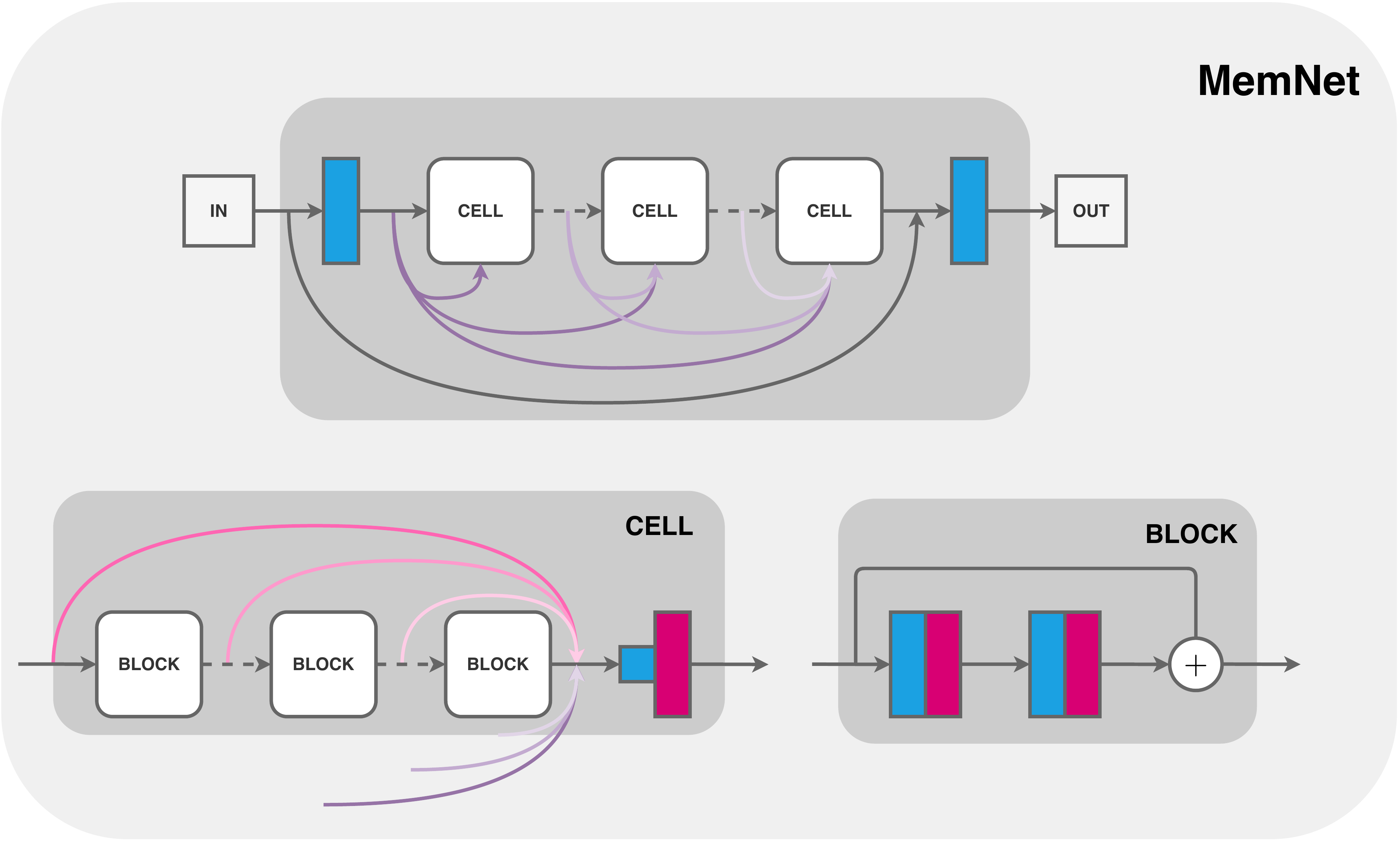}}
\\
\subfloat[RDN\label{fig:RDN}]{\includegraphics[width=0.81\columnwidth]{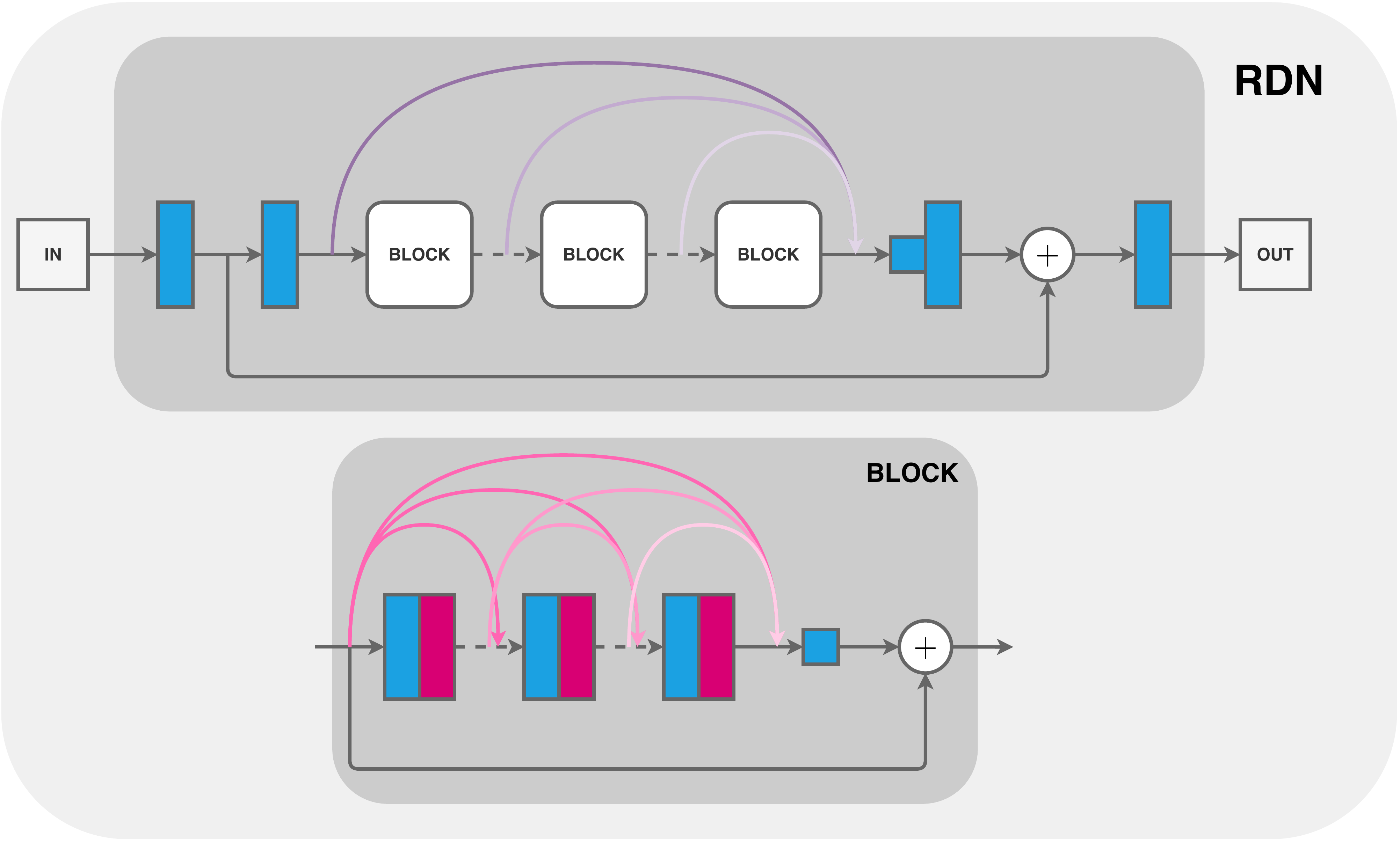}}

\caption{Architectural details of neural networks for context-aware pixel-intensity prediction. Blue boxs, red boxs and plus sign represent covlutional layers, activation functions and element-wise sum respectively.}
\label{fig:networks}
\end{figure}

An image whose query pixels are initialised to either zero or local mean can be viewed as a noisy and low-resolution version of the original image. The goal of context-aware pixel-intensity prediction can therefore be considered as refining the observed image into a clean and high-resolution one. Based on this perception, we can adopt advanced deep-learning models originally devised for image denoising and image super-resolution. A neural network is essentially a non-linear function that learns to minimise a pre-defined loss function by transforming the input into useful features or representations in a latent space. For the task of pixel-intensity prediction, we suggest training the models to minimise the $\ell_1$ norm or mean absolute error (MAE). The reason for choosing the $\ell_1$ norm over the $\ell_2$ norm is that the latter tends to produce overly smoothed outputs in image enhancement tasks~\cite{4775883, 7797130, 2017_8099502}. The $\ell_2$ norm, or mean squared error (MSE), encourages producing an average of plausible solutions and is prone to being affected by outliers. There are other preferable loss functions in the field of low-level computer vision. An example is the Euclidean distance between the high-level feature maps extracted by a $19$-layer VGG neural network~\cite{Simonyan15} pre-trained on the ImageNet (a database for large-scale visual recognition)~\cite{ILSVRC15}. Another example is the adversarial loss which uses a discriminator to estimate the likelihood that a generated instance is real~\cite{NIPS2014_5423}. Nevertheless, we do not anticipate these ad hoc loss functions improving steganographic performance because prediction-error modulation relies mainly on pixel-wise distance. Apart from the choice of loss functions, when applying off-the-shelf models, upsampling and transposed convolutional layers should be replaced with regular convolutional layers because the sizes of the input and output images are the same in pixel-intensity prediction. While the neural network architectures vary from one to another and it is difficult to foresee their performance in different tasks (due to the `black box' nature of neural networks), we attach relative importance to shortcut connection because it mitigates the performance degradation problem when increasing network depth~\cite{2016_7780459} and allows models to learn explicitly the minute differences between the input and output images~\cite{2017_7839189}. To ensure reproducibility, the architectural details of the MS-CNN, MemNet and RDN used for empirical evaluation are illustrated in Figure~\ref{fig:networks} and specified as follows.

\subsubsection{MS-CNN}
The MS-CNN extracts multi-scale image features with convolutional kernels of different sizes parallel to one another, to overcome the problem of restricted receptive fields in traditional linear predictors. The multi-scale features are aggregated in two successive convolutional layers to make a final prediction. In the implementation, the multi-scale kernel sizes are configured to $3 \times 3$, $5 \times 5$, and $7 \times 7$ with the number of channels set to $32$.

\begin{figure*}[t!] 
\centering
\subfloat[MS-CNN]{\includegraphics[width=0.6\columnwidth]{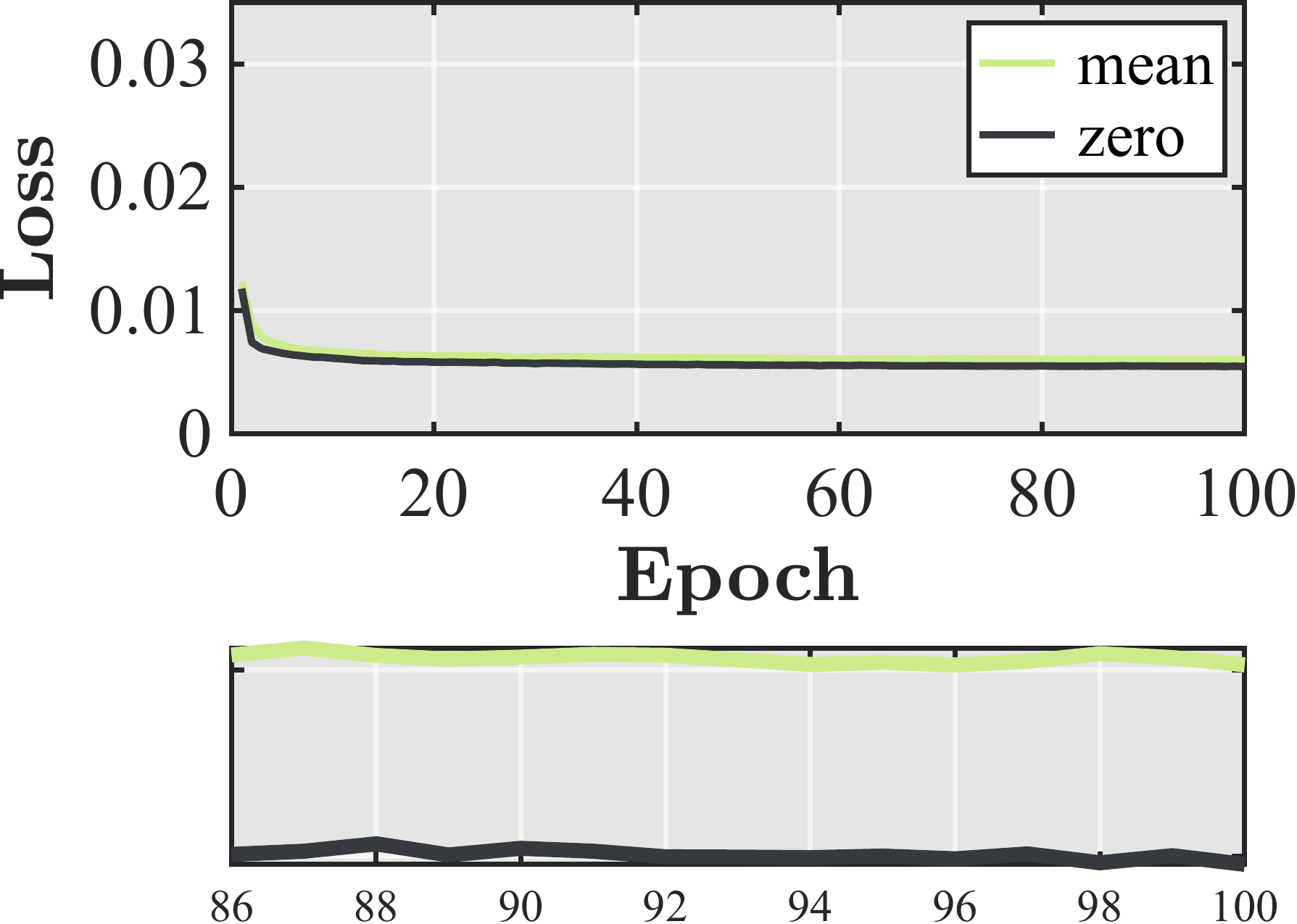}}
\hfil
\subfloat[MemNet]{\includegraphics[width=0.6\columnwidth]{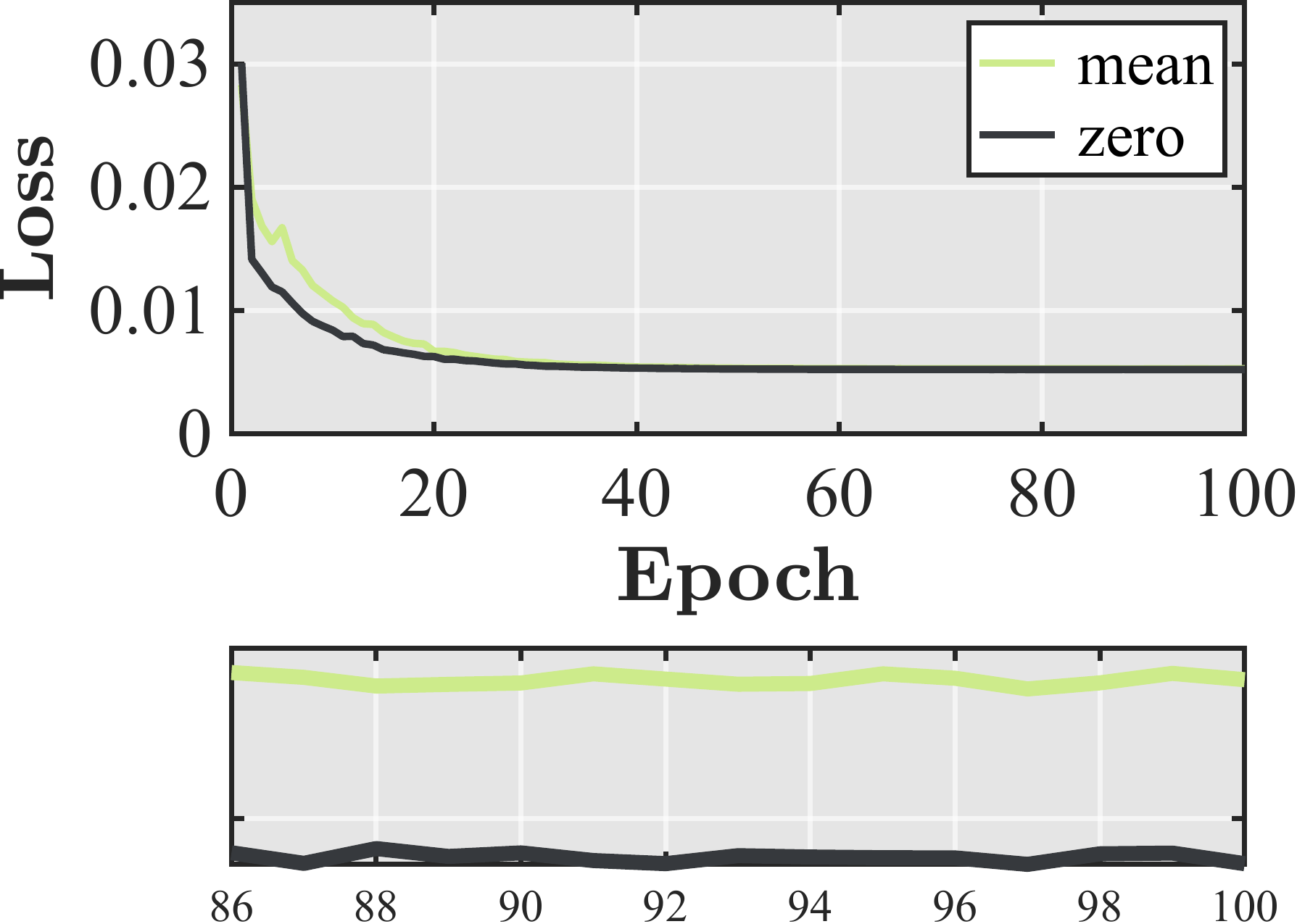}}
\hfil
\subfloat[RDN]{\includegraphics[width=0.6\columnwidth]{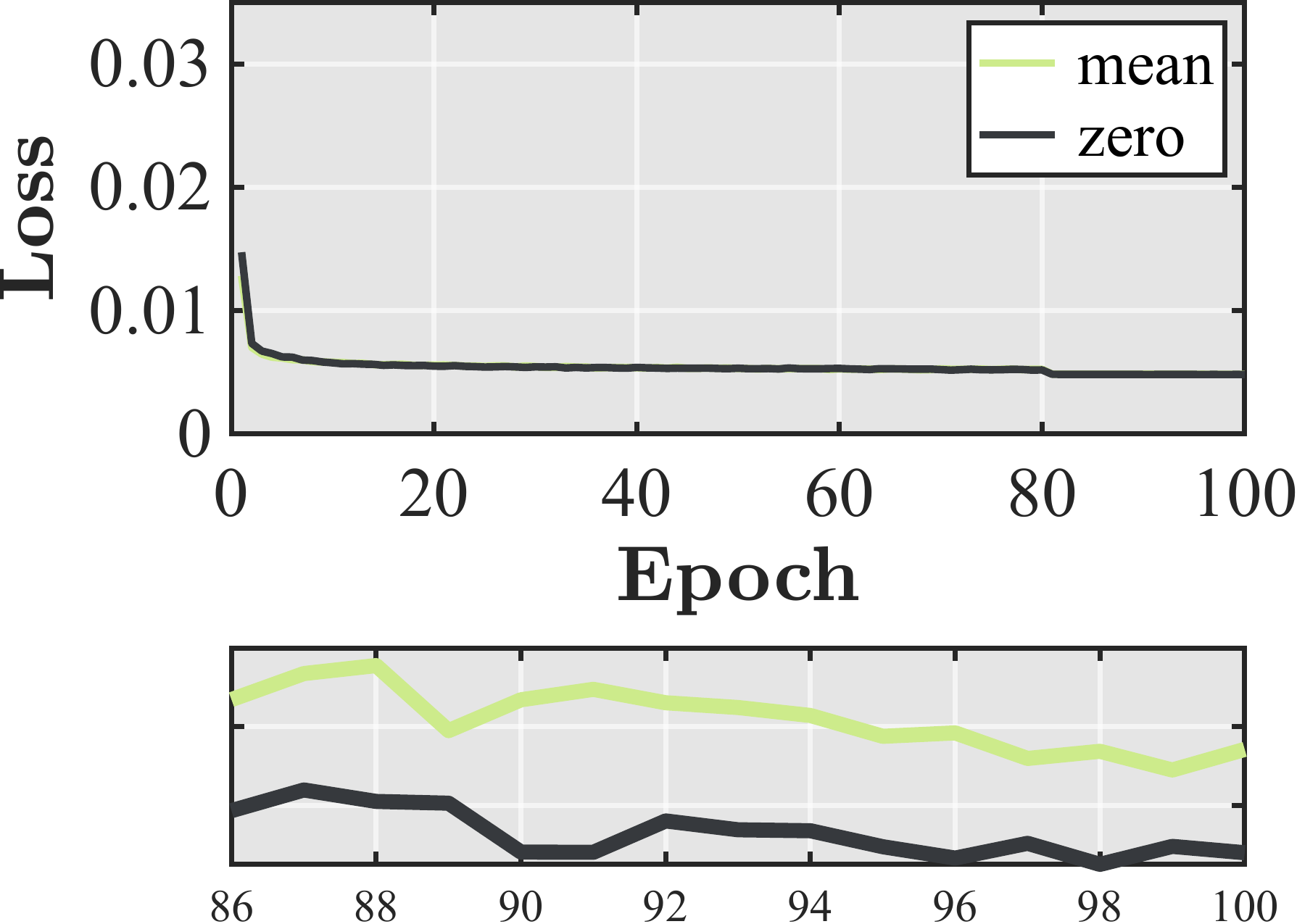}}

\caption{Learning curves for zero and local-mean initialisation strategies over 100 epochs. Top plot shows training loss over 100 epochs. Bottom plot shows zoomed-in section from 86th to 100th epoch.}
\label{fig:init}
\end{figure*}

\begin{figure}[t!] 
\centering
\subfloat[PSNR]{\includegraphics[width=0.48\columnwidth]{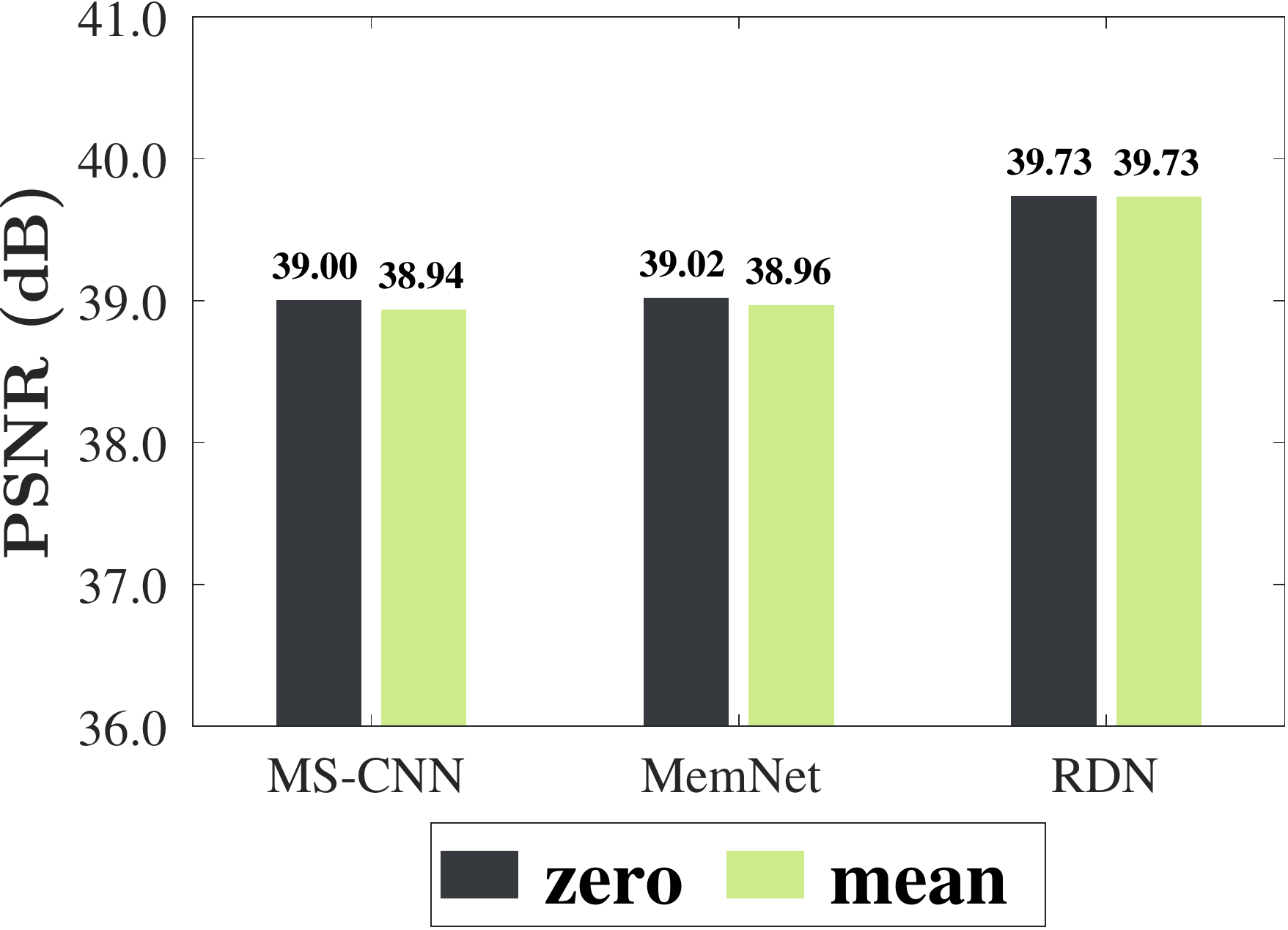}}
\hfil
\subfloat[SSIM]{\includegraphics[width=0.48\columnwidth]{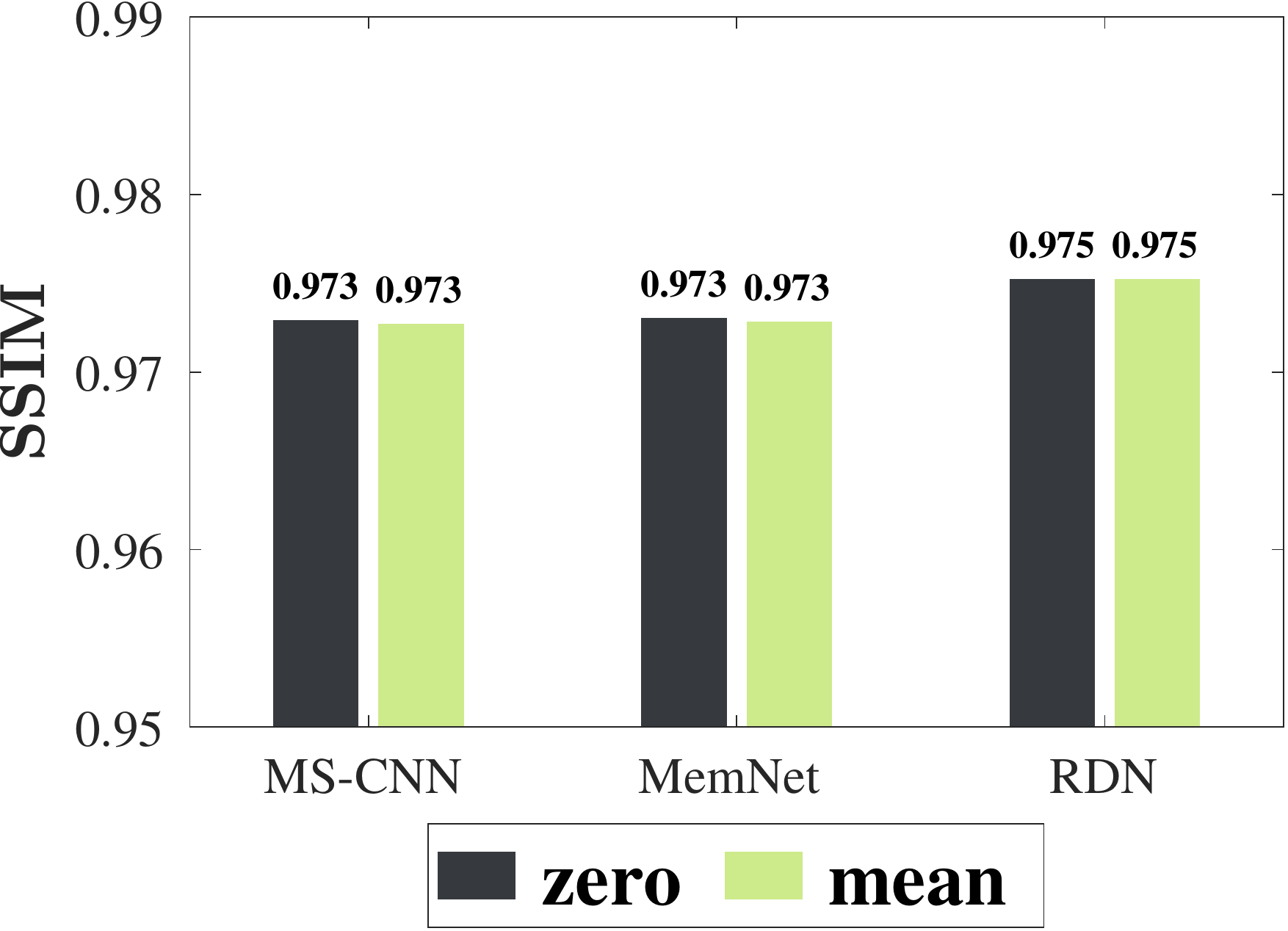}}

\caption{Visual quality assessment for first-layer prediction with zero and local-mean initialisation.}
\label{fig:init_vis}
\end{figure}

\subsubsection{MemNet}
The MemNet consists of interconnected memory cells, each comprising a recurrent unit and a gate unit. The recurrent connectivity substantially reduces the number of trainable parameters, enabling the formation of a lightweight model for storage. The gating mechanism regulates important latent states or persistent memories, thereby mitigating the vanishing gradient problem often encountered when training deep neural networks. The recurrent unit comprises a series of residual blocks with shared weights, whereas the gating mechanism is a convolutional layer with a kernel size of $1 \times 1$. For each memory cell, the outputs from the tightly structured residual blocks (short-term memories) along with the outputs from previous cells pass through a gate unit to attain a persistent state (long-term memory). In the implementation, the number of memory cells and the number of residual blocks in each cell are both configured to $5$ with the kernel size set to $3 \times 3$ and the number of channels set to $64$. 

\subsubsection{RDN}
The RDN combines residual connections and dense connections to learn hierarchical image representations. The model exploits hierarchical representations to the fullest by fusing features at both the global and local levels. At the local level, convolutional layers are densely connected, forming a residual dense block with a skip connection between the input and output. A convolutional layer with a kernel size of $1 \times 1$ is used to fuse the local feature maps. At the global level, the outputs of multiple blocks are, again, blended via a convolutional layer with kernel size $1$, followed by a skip connection from a shallow layer to a deep layer. In the implementation, the number of residual dense blocks is configured to $3$ and the number of convolutional layers in each block is configured to $5$ with the kernel size set to $3 \times 3$ and the number of channels set to $64$. 

\section{Experiments}\label{sec:exp}
In this section, we report and discuss experimental results w.r.t. different predictive models. We begin by discussing the choice of benchmarks and describing the general set-up of experiments. Then, we evaluate the impacts of different initialisation and training strategies on predictive accuracy. To further evaluate the performance of different models, we analyse the distribution of prediction errors. Finally, we verify a direct correlation between predictive accuracy and steganographic performance by examining the steganographic rate-distortion curves.

\begin{figure}[t!] 
\centering
\subfloat[MS-CNN (PSNR)]{\includegraphics[width=0.48\columnwidth]{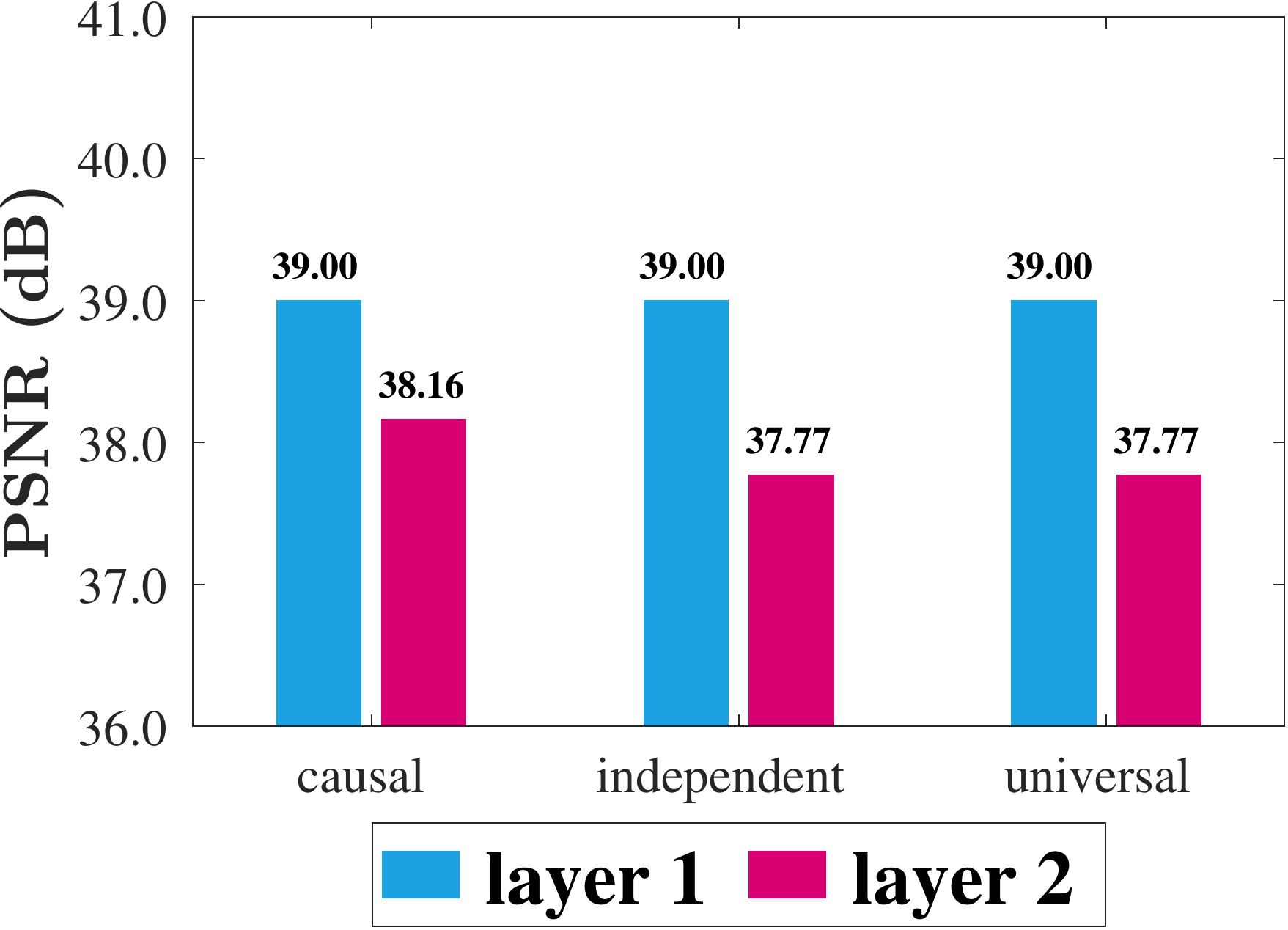}}
\hfil
\subfloat[MS-CNN (SSIM)]{\includegraphics[width=0.48\columnwidth]{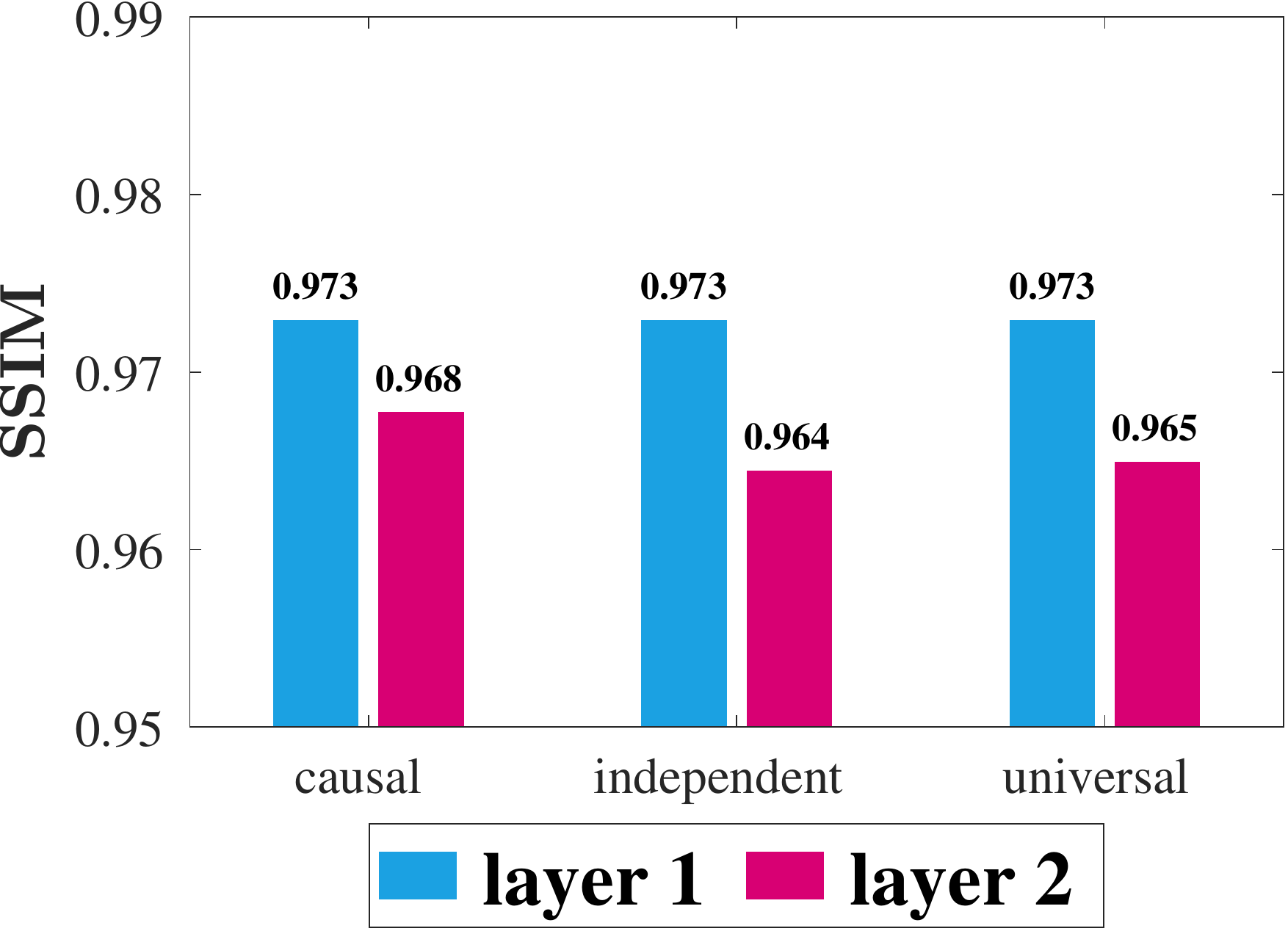}}
\\
\subfloat[MemNet (PSNR)]{\includegraphics[width=0.48\columnwidth]{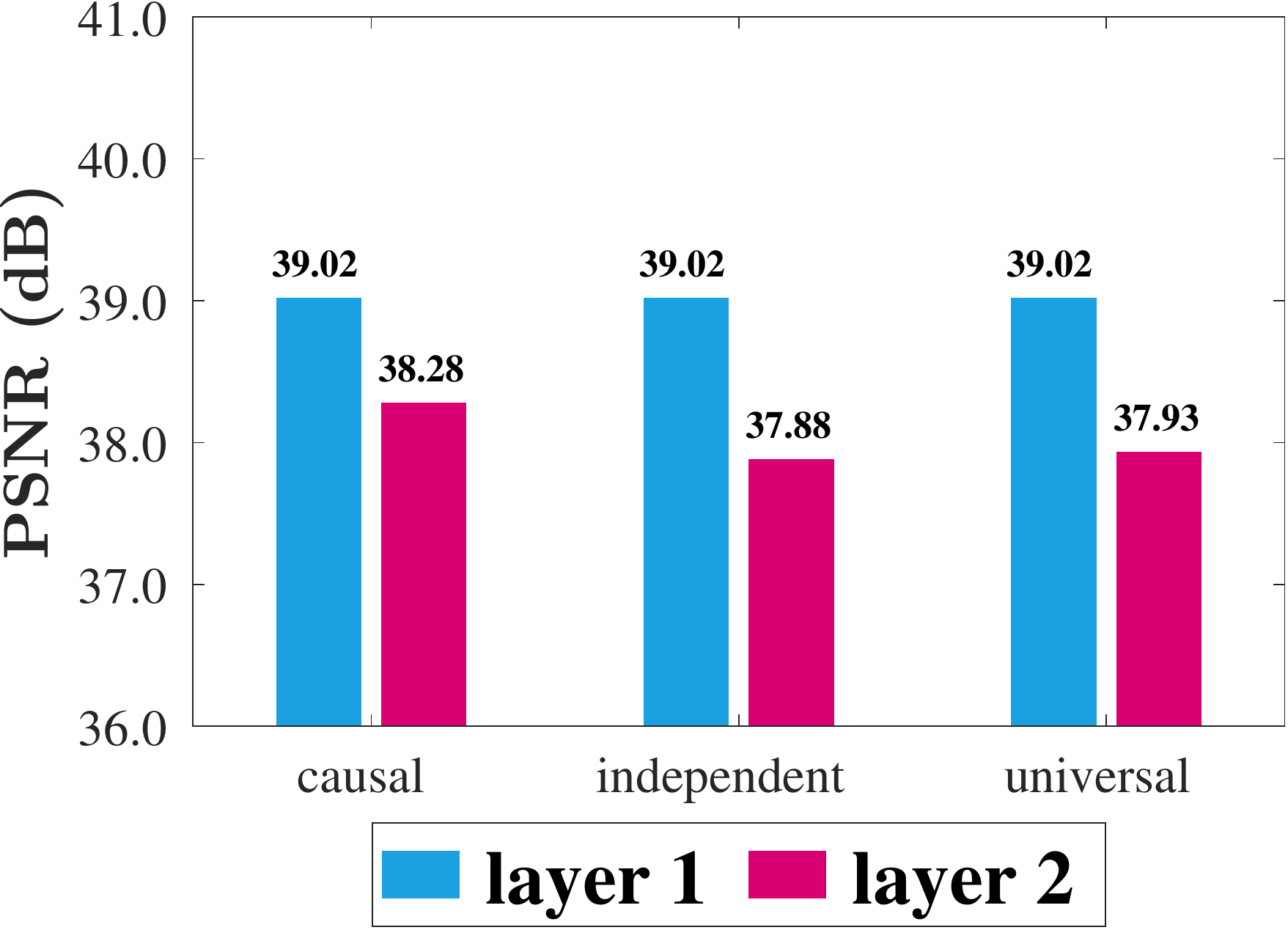}}
\hfil
\subfloat[MemNet (SSIM)]{\includegraphics[width=0.48\columnwidth]{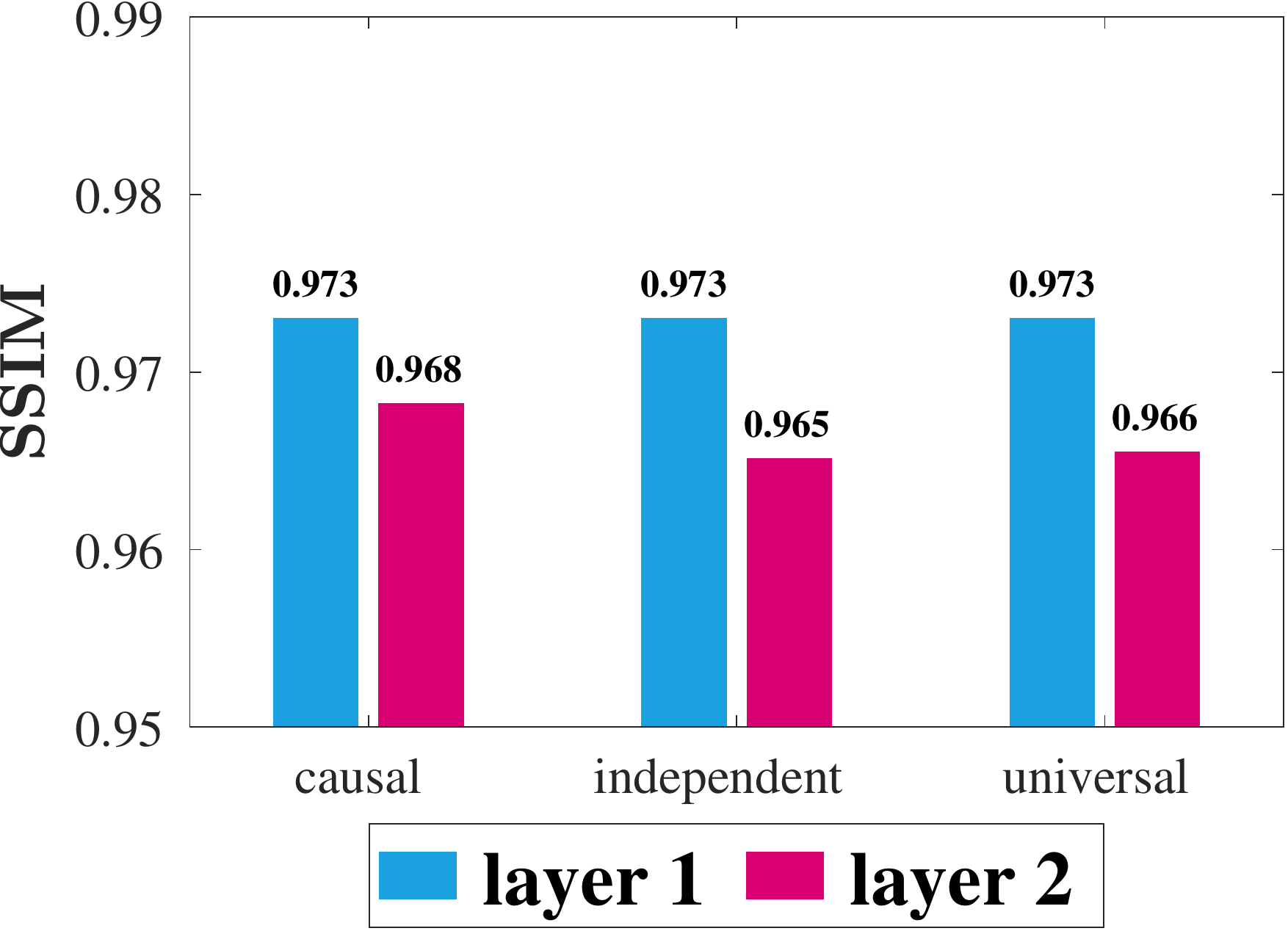}}
\\
\subfloat[RDN (PSNR)]{\includegraphics[width=0.48\columnwidth]{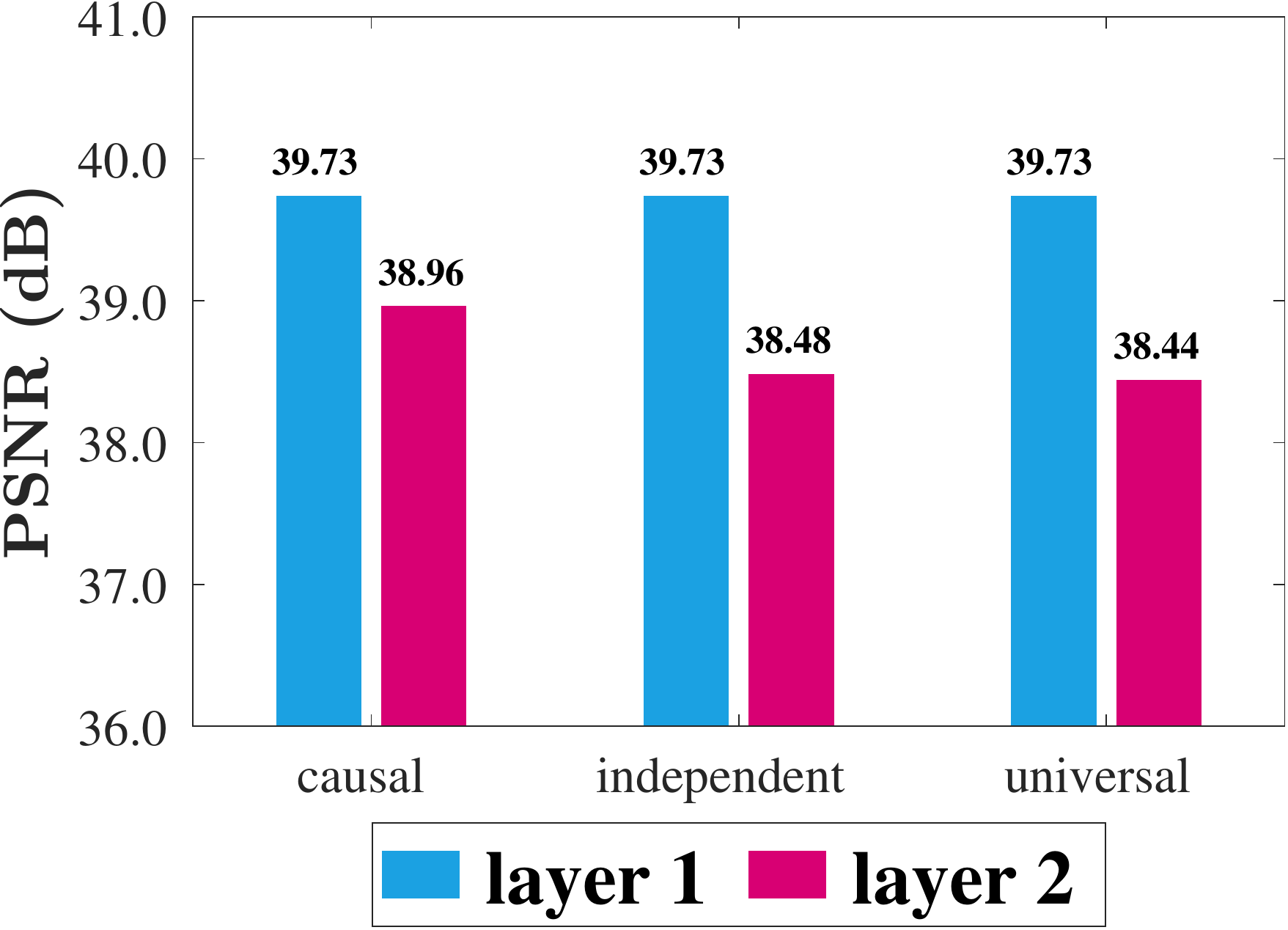}}
\hfil
\subfloat[RDN (SSIM)]{\includegraphics[width=0.48\columnwidth]{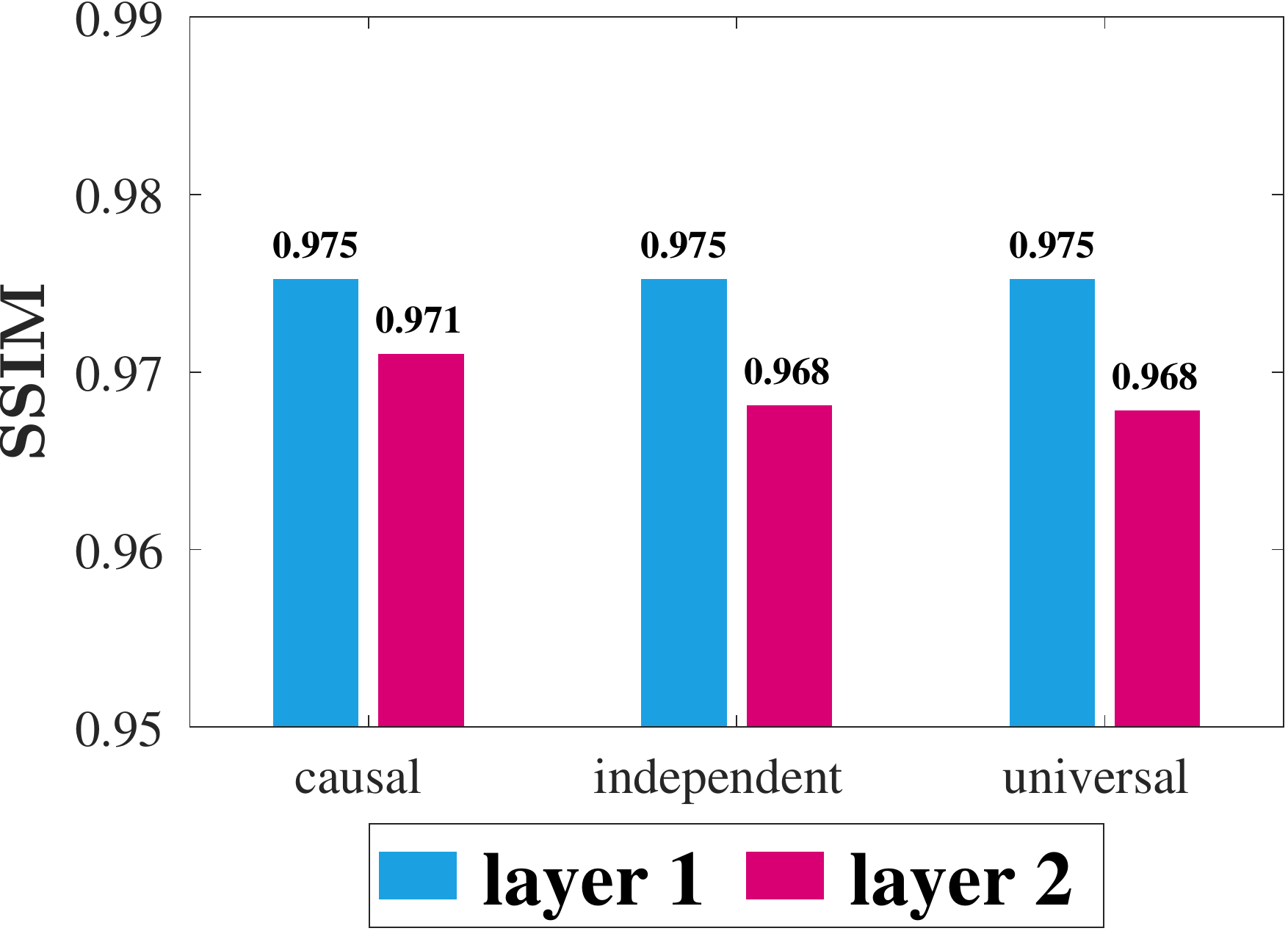}}

\caption{Visual quality assessment for dual-layer prediction with causal, independent and universal training.}
\label{fig:training}
\end{figure}

\subsection{Benchmarks}
The main theme of this study is predictive analytics. For a fair comparison, we evaluate the performance of different predictive models based on the same steganographic algorithm (i.e. prediction-error modulation). We do not compare steganographic performance with the end-to-end framework because it approaches the problem from a different standpoint. In general, the end-to-end framework cannot reliably satisfy the requirement for perfect reversibility given the limit of current deep-learning algorithms. For the predictive analytics module, we select LMI as a benchmark model~\cite{2009_4811982} based on heuristics and both MS-CNN~\cite{Hu:2021aa} and MemNet~\cite{Chang:2021aa} as state-of-the-art models based on deep learning. All the selected models are specifically designed and applied for the prediction task with the same context--query arrangement (i.e. the chequered pattern). The mentioned deep learning models have also been shown to outperform other traditional methods for either same or different context--query arrangement, including nearest-neighbour interpolation~\cite{2003_1227616}, median-edge detector~\cite{2007_4099409}, gradient-adjusted predictor~\cite{2008Fallahpour} and bilinear interpolation~\cite{5313862}. In summary, the coding algorithm and the context--query splitting pattern are held constant (as control variables) throughout the course of the investigation to prevent the interference in the experimental results. We select LMI, MS-CNN and MemNet as representative benchmarks and carry out a comparative study with the advanced RDN model. 

\subsection{Experimental Setup}
The neural network models are trained and tested on the BOSSbase dataset~\cite{2011_BOSSbase}, which originated from an academic competition for digital steganography. It comprises a collection of $10,000$ greyscale photographs covering a wide variety of subjects and scenes. To fit the models over a broad range of hardware options with reasonable training time, all the images are resized to a resolution of $256 \times 256$ pixels by using the Lanczos resampling algorithm~\cite{1979_Lanczos}. The training and test sets are randomly sampled at a ratio of $80$/$20$. All the neural network models are trained to minimise the $\ell_1$ loss unless specified otherwise. Further inferences and evaluations are made on a set of standard test images from the volume `miscellaneous' in the USC-SIPI dataset~\cite{2006_USC_SIPI}. To investigate predictive performance on challenging samples, we also take Brodatz texture images from the volume `textures' in the USC-SIPI dataset~\cite{BrodatzTextures}. We use peak signal-to-noise ratio (PSNR) (expressed in decibels; dB)~\cite{iet:/content/journals/10.1049/el_20080522} and structural similarity (SSIM)~\cite{1284395} for perceptual quality assessment. When evaluating the visual quality of predicted images, we take the whole image into account because the bias in the context pixels is almost negligible. The context pixels in each predicted image are almost always the same as those in the original image. We use embedding rate (expressed in bits per pixel; bpp) for steganographic capacity evaluation.

\begin{figure}[t!] 
\centering

\subfloat[Aeroplane]{\includegraphics[width=0.33\columnwidth]{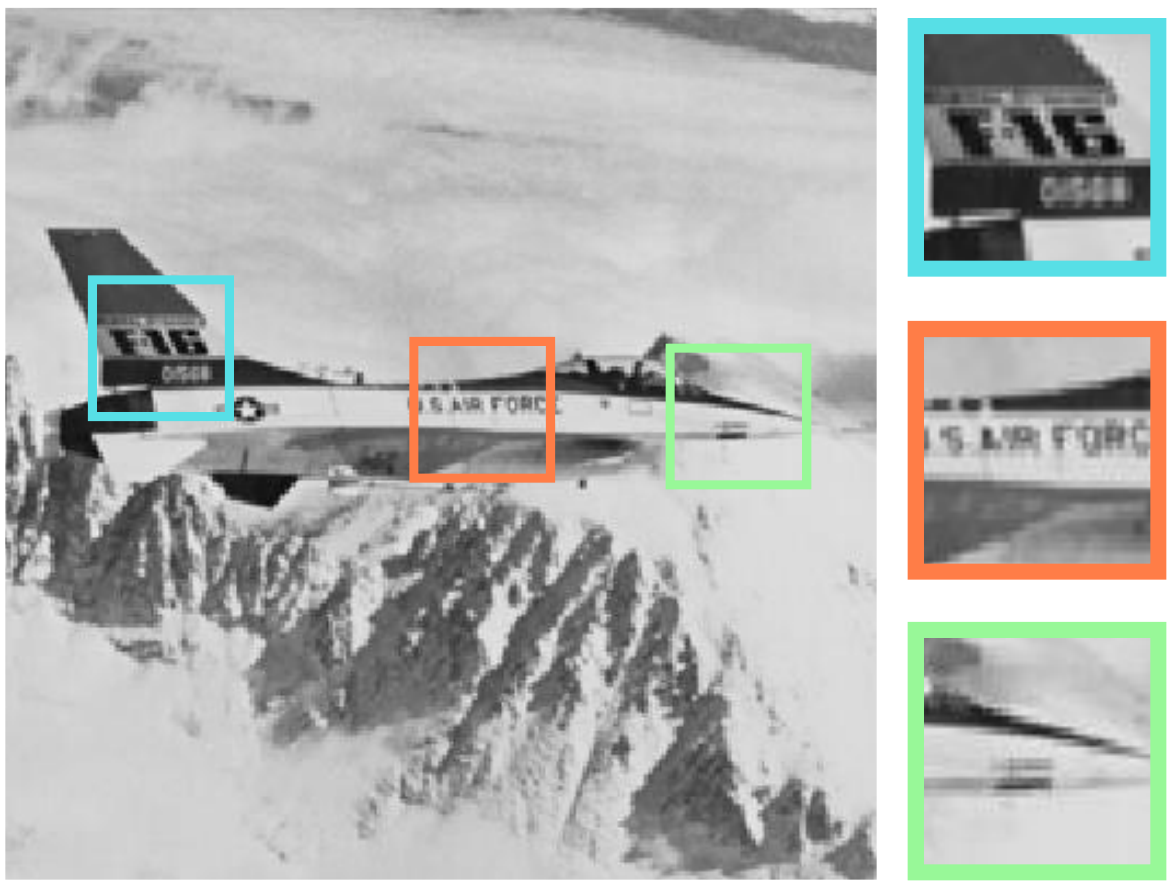}}
\hfil
\subfloat[RDN: 36.4811]{\includegraphics[width=0.33\columnwidth]{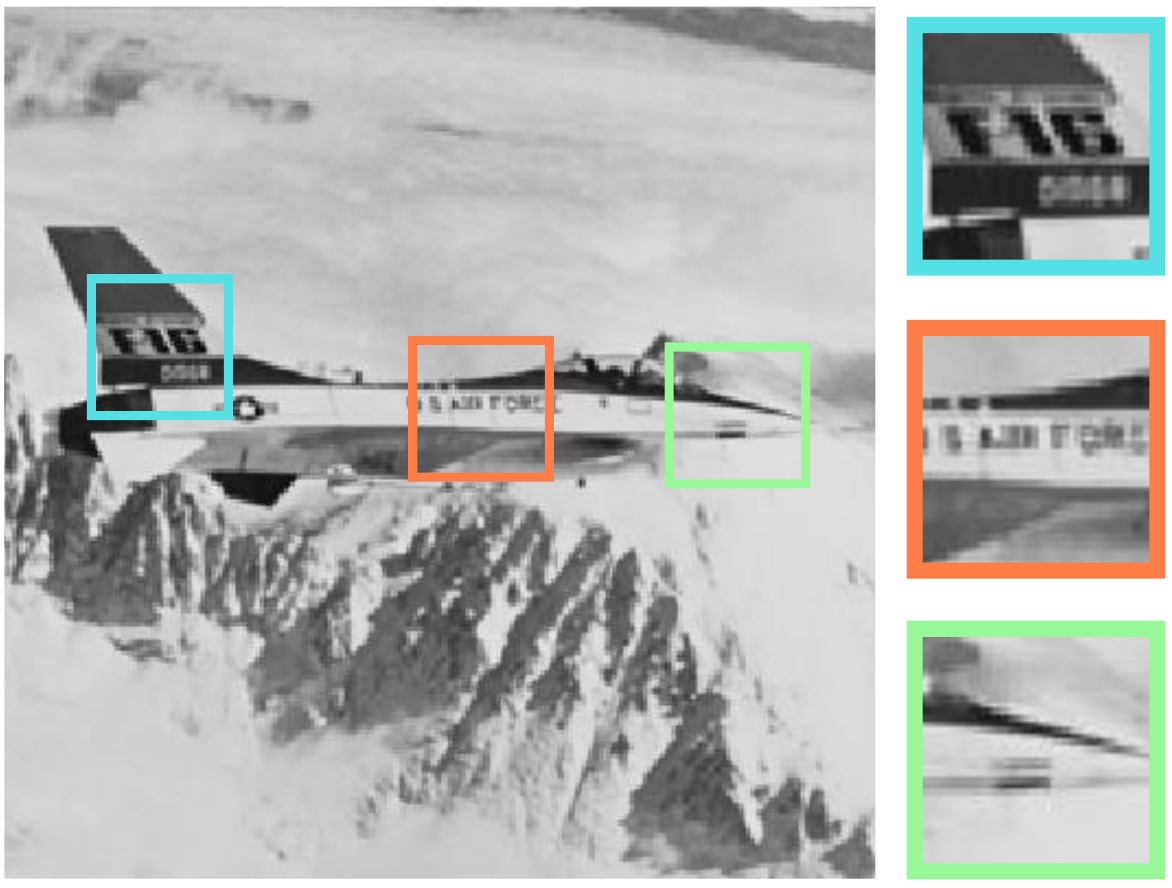}}
\hfil
\subfloat[LMI: 32.9728]{\includegraphics[width=0.33\columnwidth]{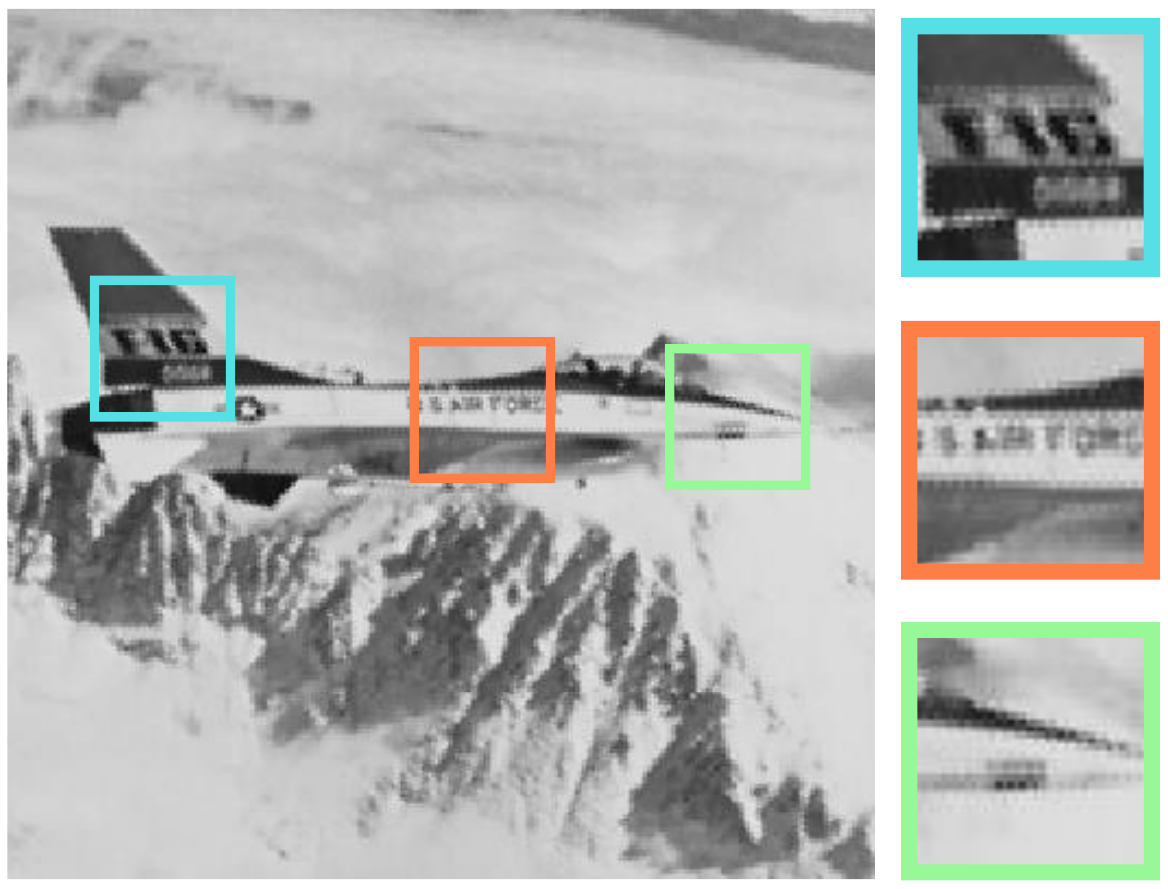}}
\\

\subfloat[Lena]{\includegraphics[width=0.33\columnwidth]{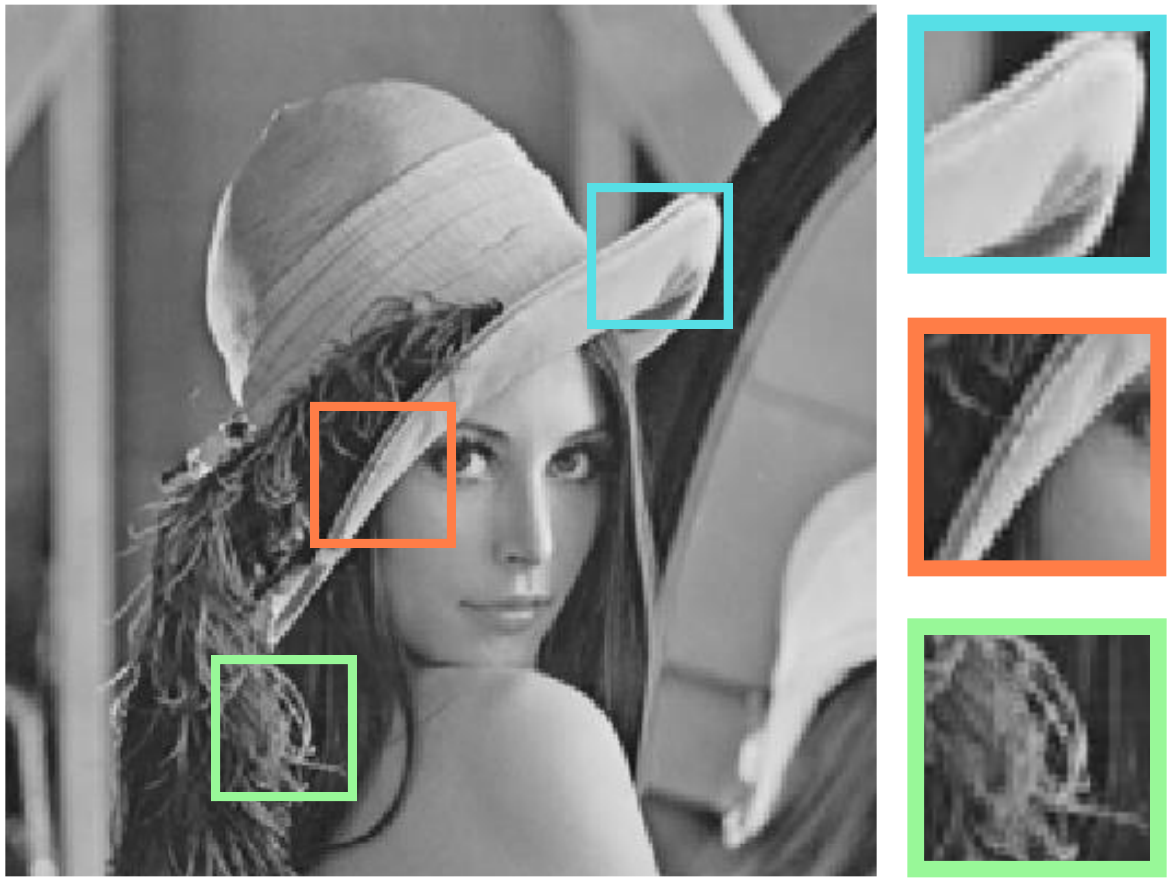}}
\hfil
\subfloat[RDN: 38.4287]{\includegraphics[width=0.33\columnwidth]{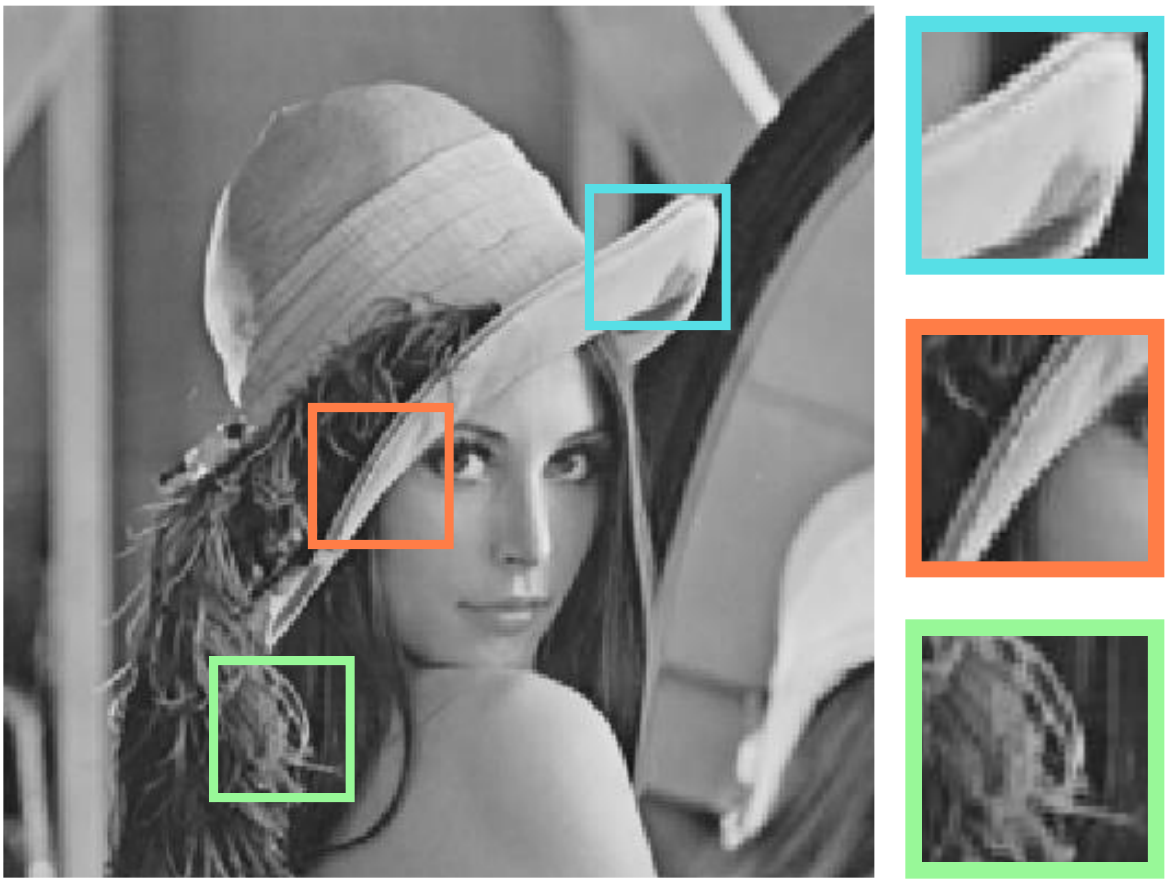}}
\hfil
\subfloat[LMI: 34.3331]{\includegraphics[width=0.33\columnwidth]{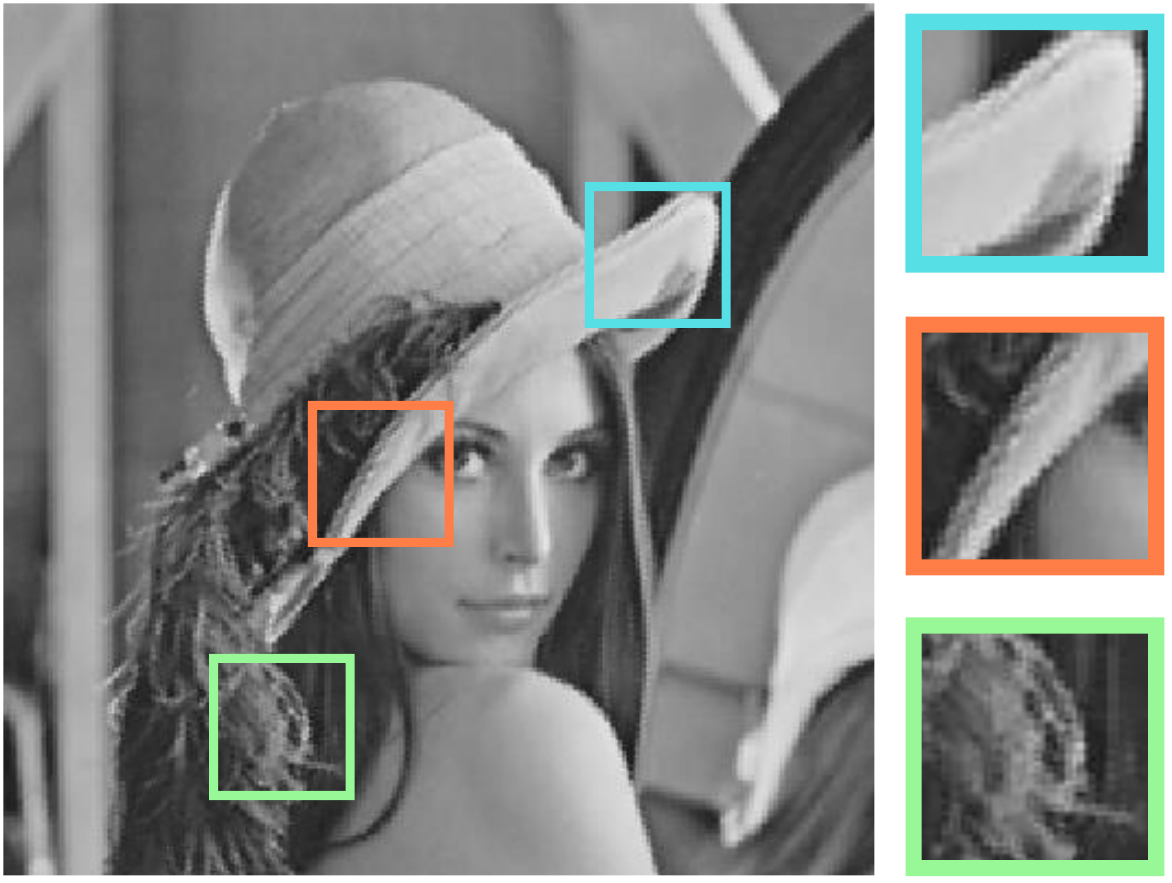}}
\\

\subfloat[Mandrill]{\includegraphics[width=0.33\columnwidth]{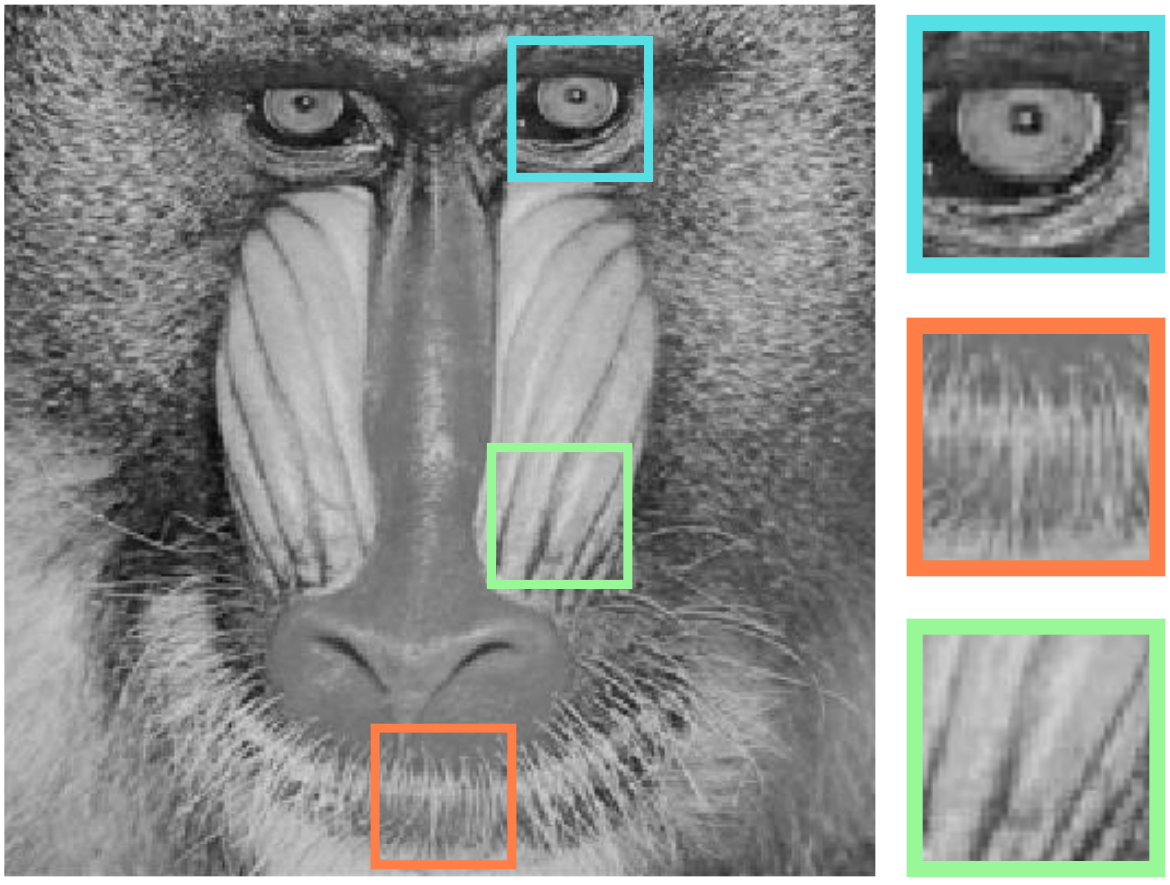}}
\hfil
\subfloat[RDN: 29.7238]{\includegraphics[width=0.33\columnwidth]{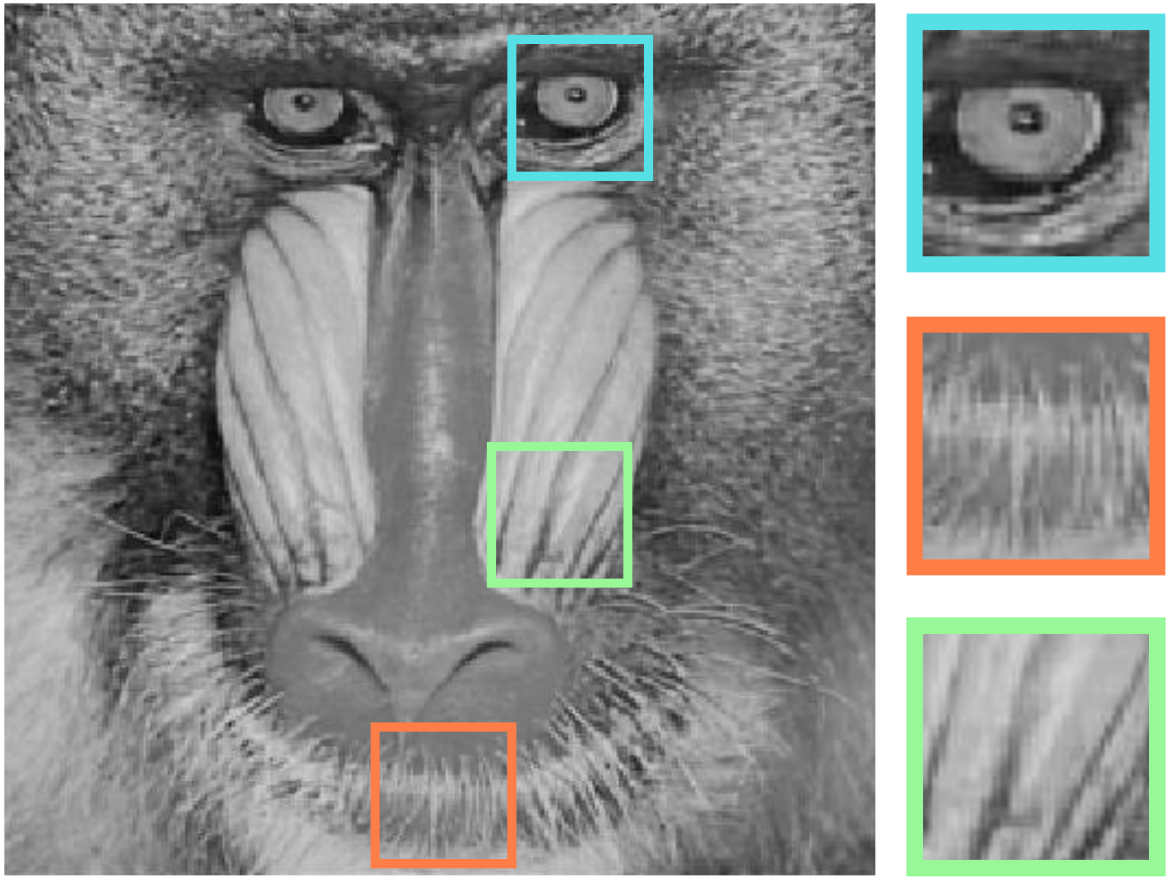}}
\hfil
\subfloat[LMI: 28.4906]{\includegraphics[width=0.33\columnwidth]{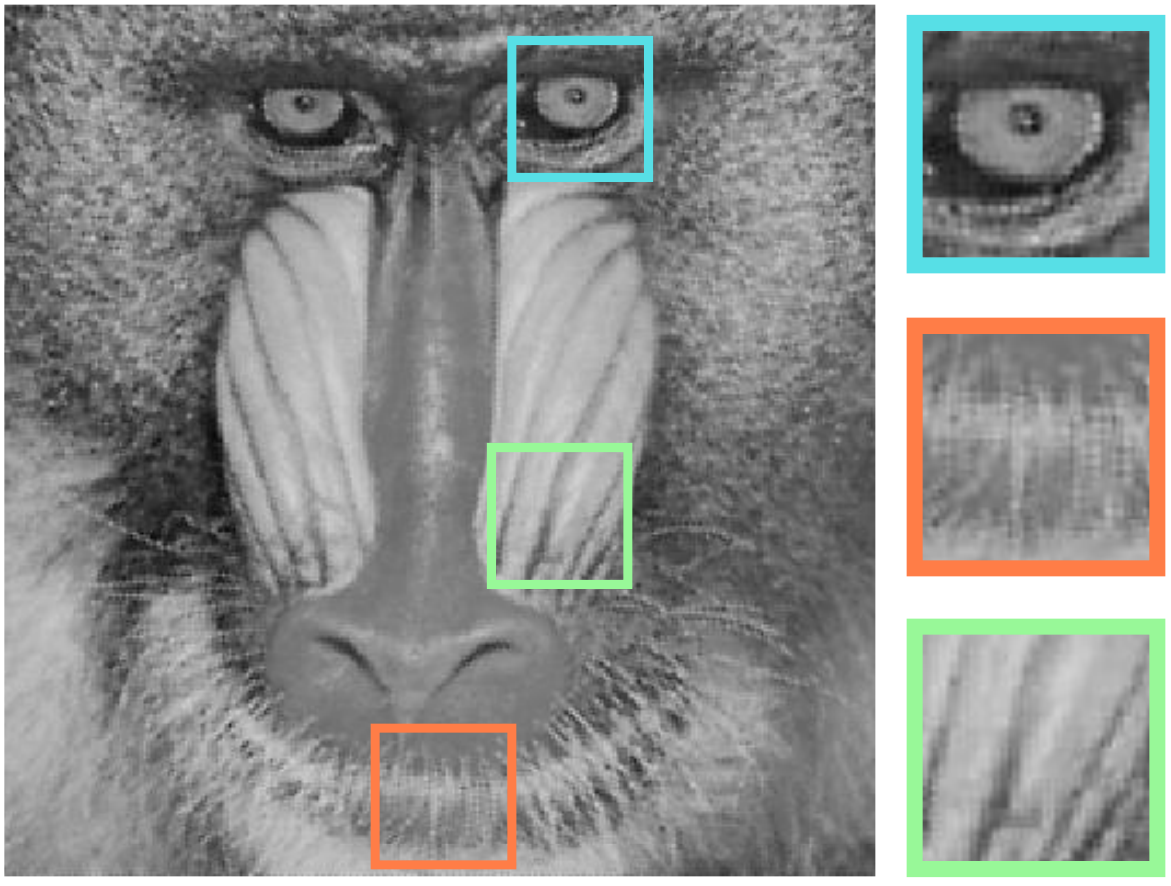}}

\caption{Visual comparisons for selected images from USC-SIPI `miscellaneous' volume. Numerical data expresses PSNR (in dB).}
\label{fig:vis_USCSIPI}
\end{figure}

\begin{figure}[t!] 
\centering

\subfloat[Textures 1.1.03]{\includegraphics[width=0.33\columnwidth]{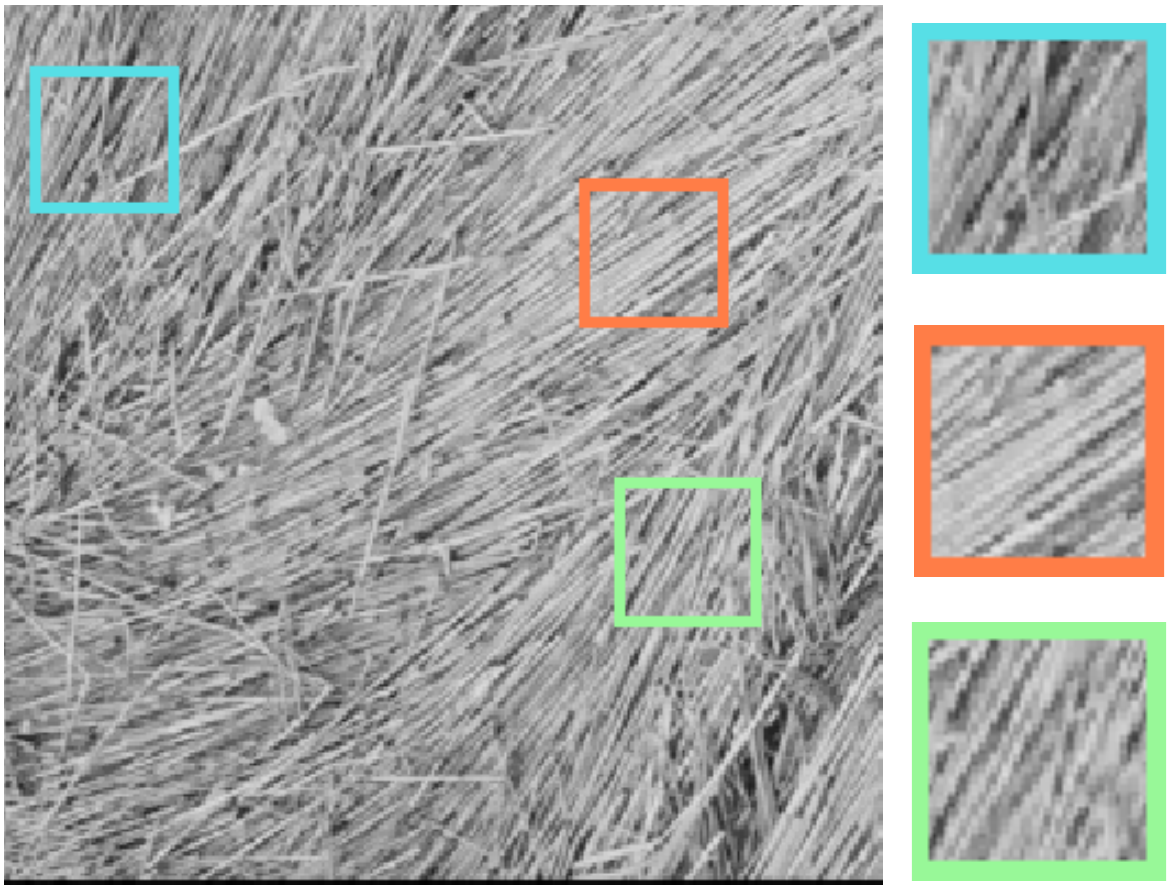}}
\hfil
\subfloat[RDN: 27.3730]{\includegraphics[width=0.33\columnwidth]{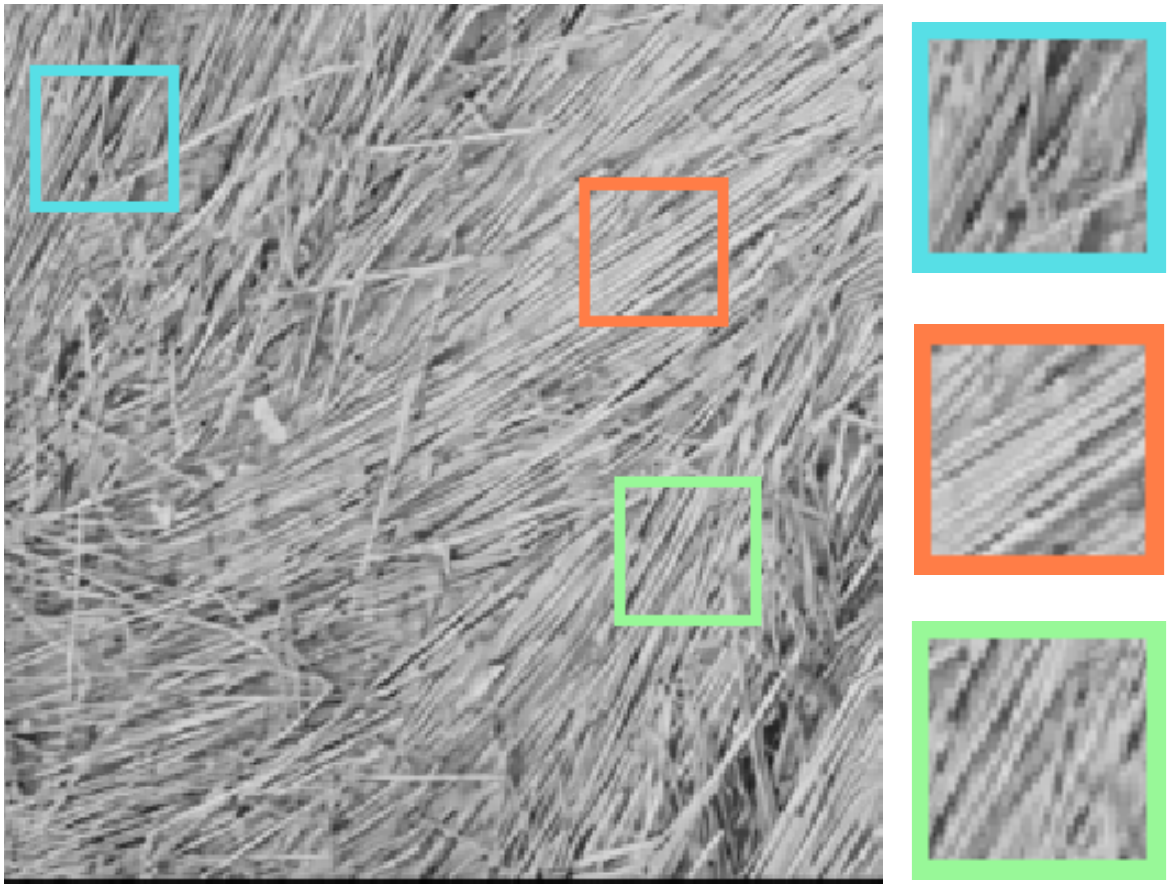}}
\hfil
\subfloat[LMI: 25.3123]{\includegraphics[width=0.33\columnwidth]{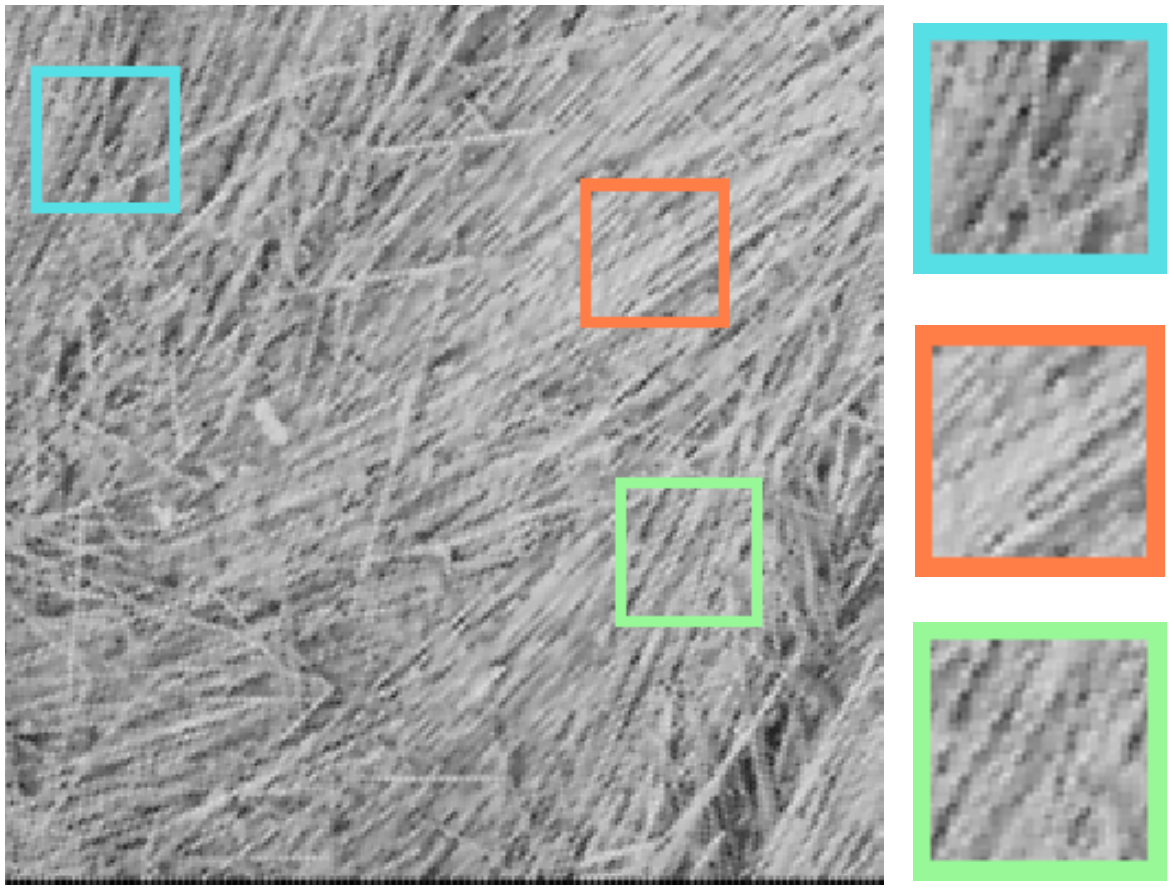}}
\\

\subfloat[Textures 1.2.08]{\includegraphics[width=0.33\columnwidth]{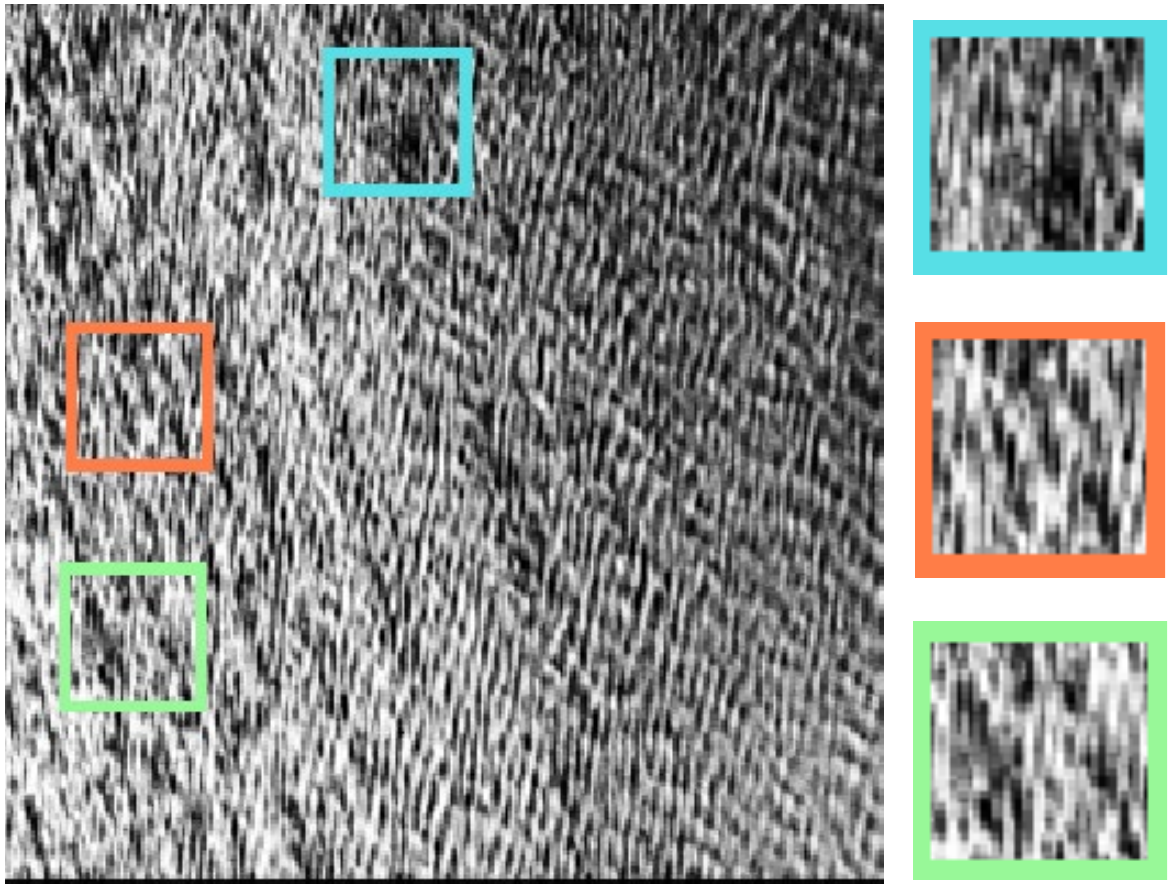}}
\hfil
\subfloat[RDN: 29.2578]{\includegraphics[width=0.33\columnwidth]{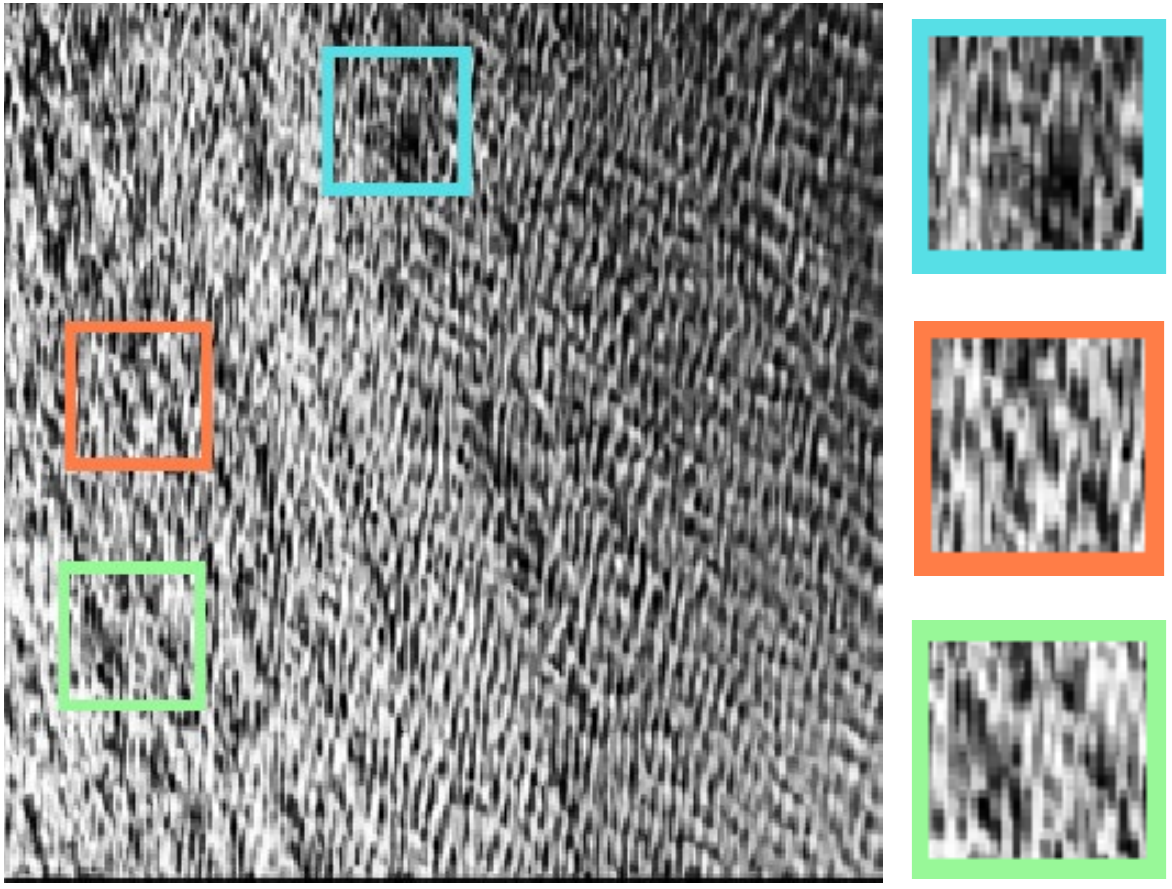}}
\hfil
\subfloat[LMI: 21.3223]{\includegraphics[width=0.33\columnwidth]{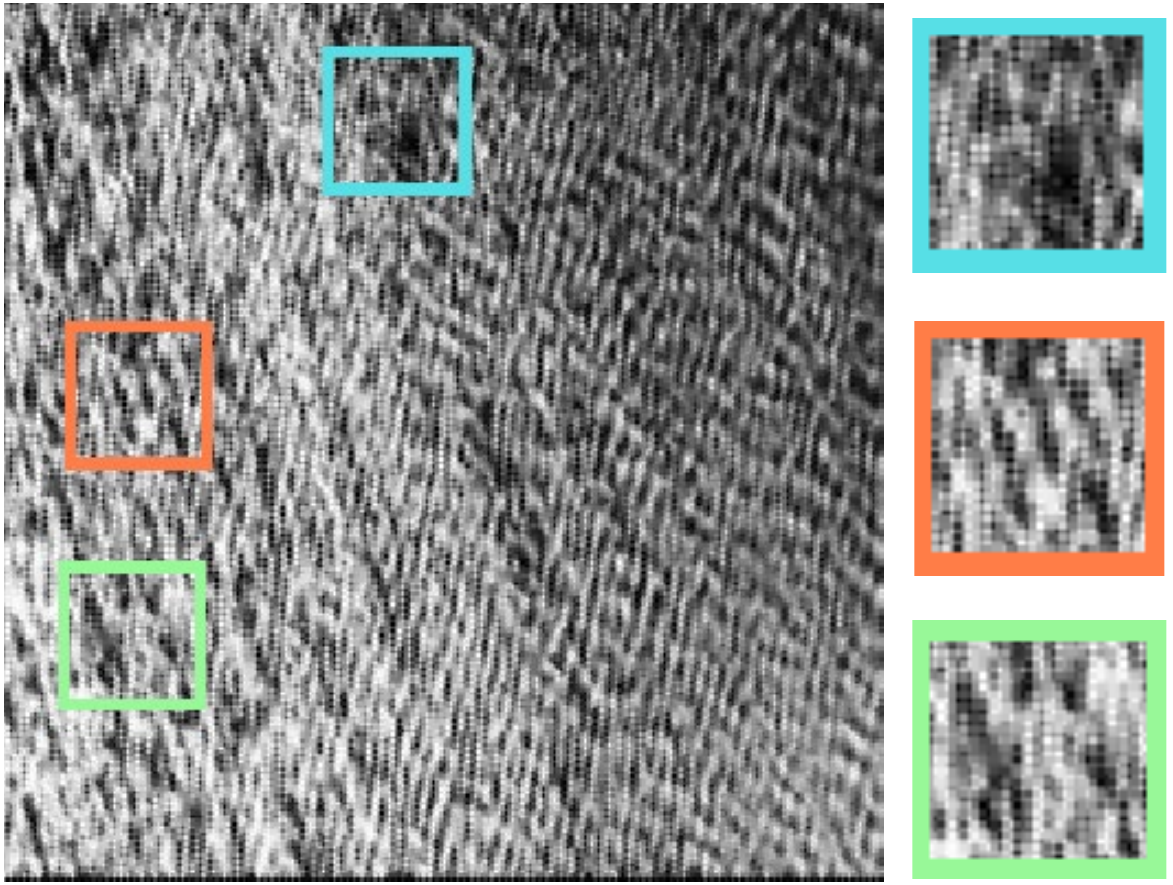}}
\\

\subfloat[Textures 1.4.03]{\includegraphics[width=0.33\columnwidth]{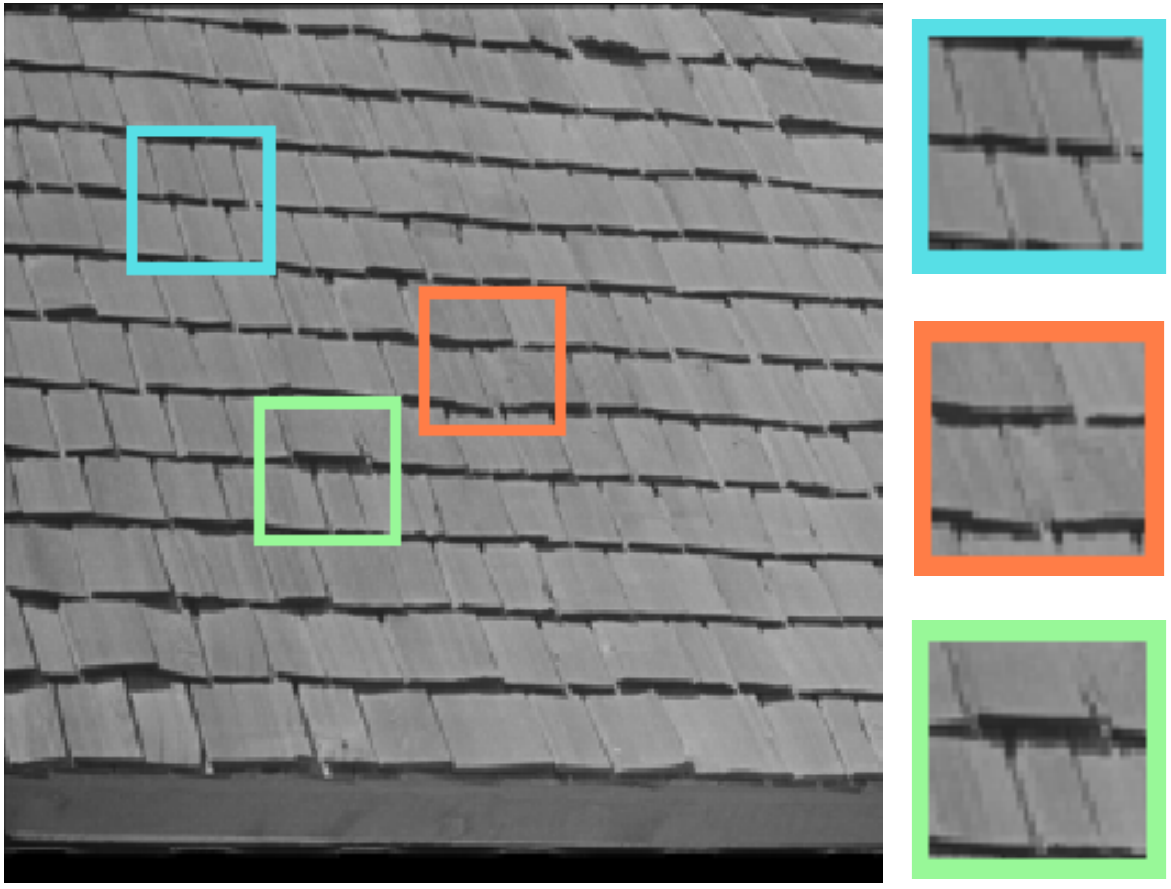}}
\hfil
\subfloat[RDN: 39.2941]{\includegraphics[width=0.33\columnwidth]{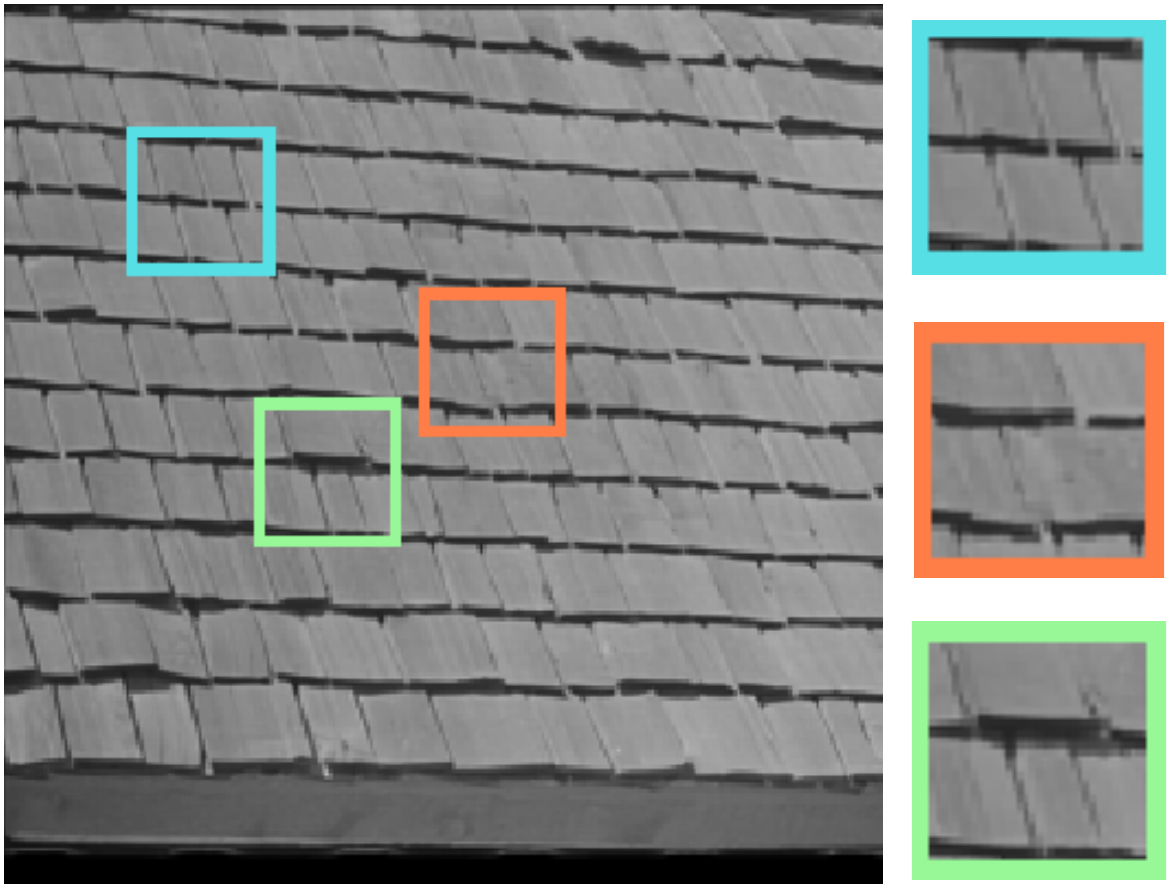}}
\hfil
\subfloat[LMI: 32.0380]{\includegraphics[width=0.33\columnwidth]{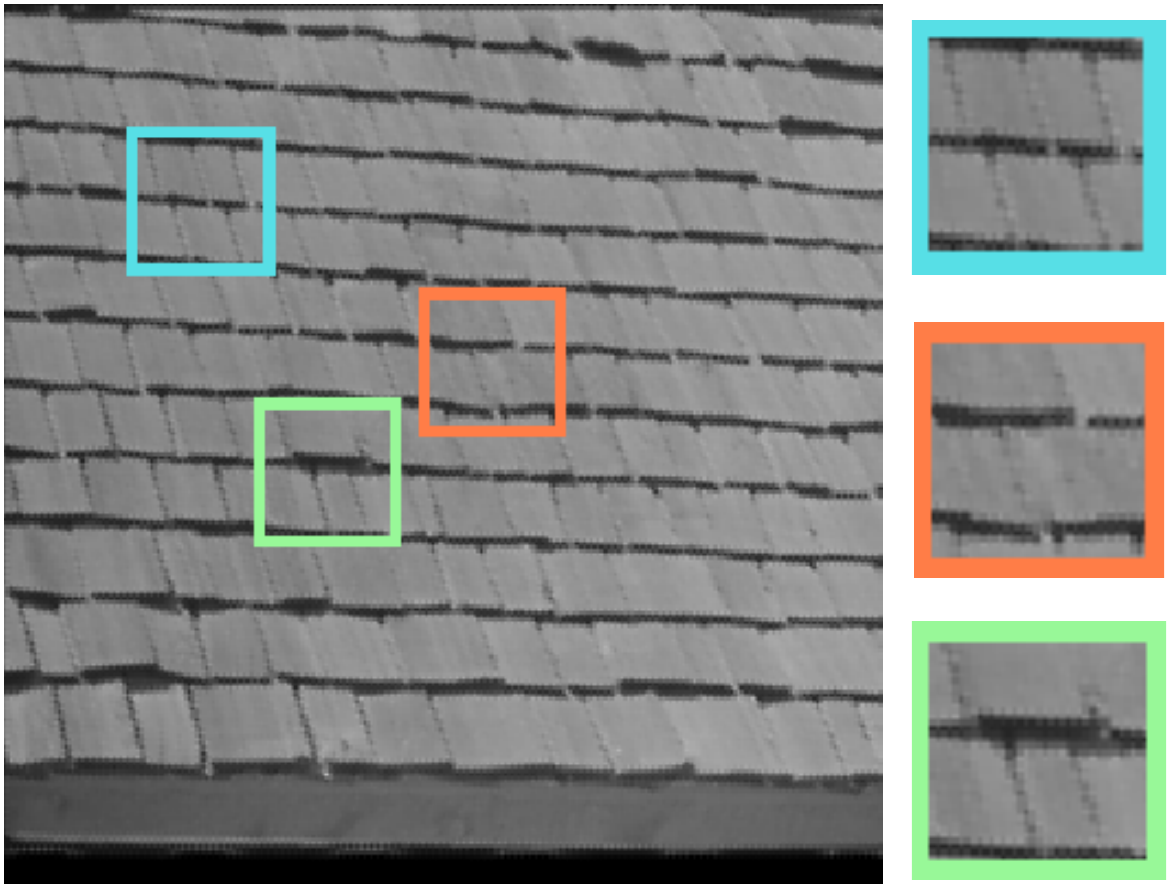}}

\caption{Visual comparisons for selected images from USC-SIPI `textures' volume. Numerical data expresses PSNR (in dB).}
\label{fig:vis_textures}
\vfil
\end{figure}

\begin{figure*}[t]
    \centering
    \includegraphics[width=0.65\textwidth]{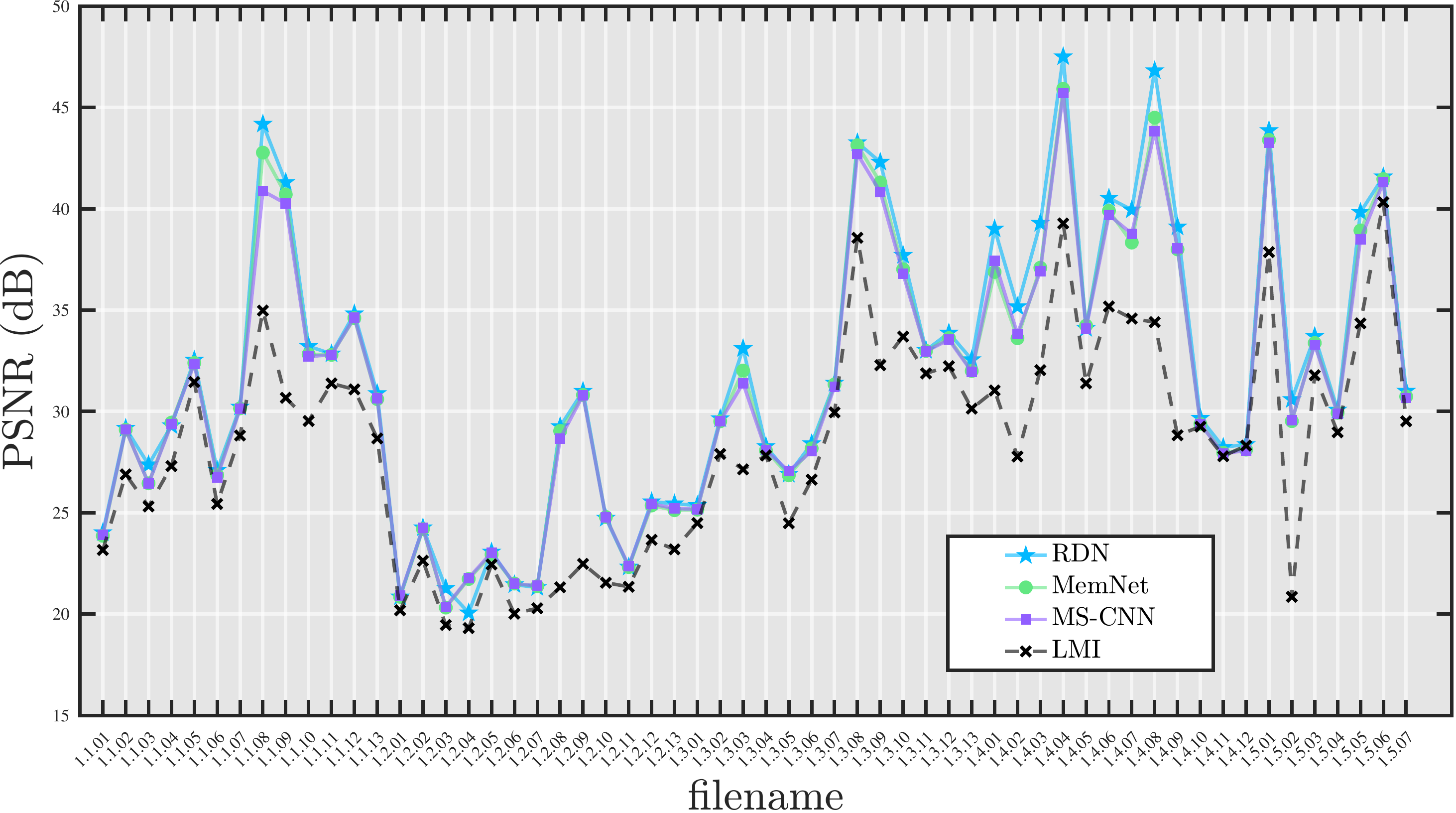}
    \caption{Visual quality assessment for first-layer prediction on Brodatz texture images.}
    \label{fig:texture}
\end{figure*}

\subsection{Comparison of Initialisation Strategies}
We compare the zero and local-mean initialisation strategies by the learning curves and the perceptual qualities. The learning curves are plotted in Figure~\ref{fig:init}. The curves represent the training loss over $100$ epochs for the input/target pairs ($\bm{x}_{\mathbb{B}}$, $\bm{x}$). It can be observed that while the convergence rate varies from model to model, zero initialisation reaches a slightly lower loss than the local-mean initialisation in the last epoch. A possible explanation is that although setting the pixel intensity to zero seems to be abrupt and oversimplified, this approach involves minimal human intervention in the machine learning pipeline. By contrast, local-mean initialisation can be viewed as introducing subjective prior knowledge (i.e. the smoothness prior) in the training process. The perceptual qualities of predicted images are measured in Figure~\ref{fig:init_vis}. Through observing visual quality measurements on different models, we conclude that there is virtually no difference between the impacts of two initialisation strategies upon predictive accuracy. While local-mean initialisation provides a rough estimation in advance, its contribution to predictive accuracy is negligible. Since zero initialisation has a virtue of low computational complexity, we adopt it for the remaining experiments.

\subsection{Comparison of Training Strategies}
Predictive accuracies of the universal, independent, and causal training strategies for dual-layer prediction are depicted in Figure~\ref{fig:training}. The bars show the average PSNR and SSIM scores between the ground-truth and predicted images w.r.t. different training strategies. Steganographic distortion introduced in the first layer would propagate and decrease the predictive accuracy of the second layer. As a result, the second-layer scores are generally lower than the first-layer scores due to the distributional shift. The causal training achieves a comparatively high accuracy for the second-layer prediction, whereas the performance gap between the other two strategies is not distinct. This suggests that the causal training using a training set with a distribution similar to that of the test set can indeed alleviate the distributional shift to some extent. The narrow performance gap between the universal and independent training can be attributed to the translation-invariance property of CNNs. It suggests that training a single model can be as good as training two separate models. Although causal training requires additional computations when constructing the second training set, it is proved to be the most effective among the three training strategies. Hence, we opt for the causal training strategy for the remaining experiments.

\subsection{Evaluation of Visual Quality}
Visual comparisons between a heurstic model and a neural network model are shown in Figures~\ref{fig:vis_USCSIPI} and~\ref{fig:vis_textures}. We evaluate the PSNR scores for images predicted from the LMI and RDN. The reported scores are related to the first-layer prediction. A close inspection of zoomed-in views reveals that the RDN is better able to retrieve textural areas with reasonable accuracy in comparison with the na{\"i}ve LMI, owing to the ability of neural networks to learn rich patterns. Figure~\ref{fig:texture} compares predictive accuracy of the RDN, MemNet, MS-CNN and LMI on texture images. The results reflect a significant improvement yielded by the RDN model over other predictive models, confirming the capability of the former to predict textural patterns.

\begin{figure*}[t!]
\centering
\subfloat[Aeroplane]{\includegraphics[width=0.58\columnwidth]{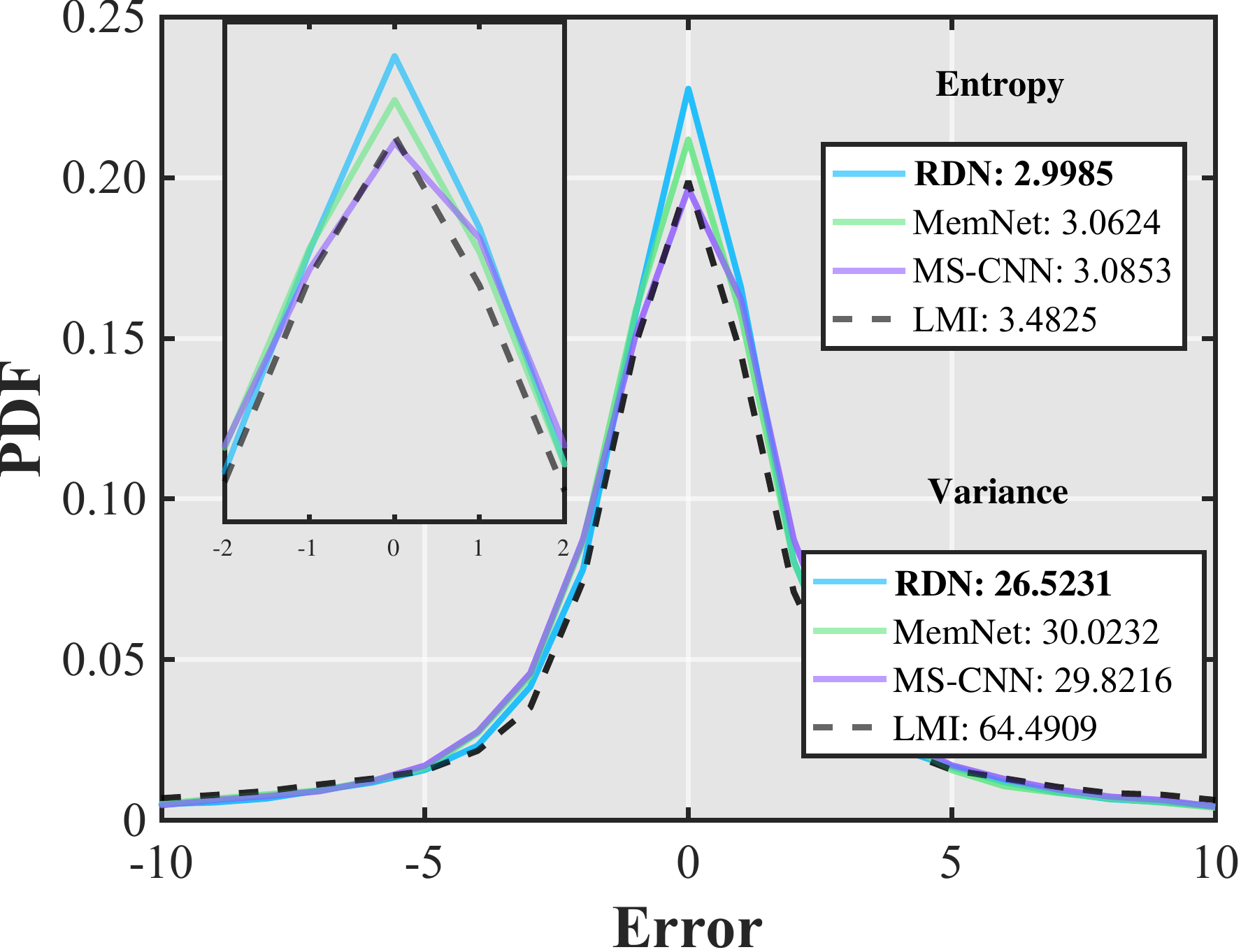}}
\hfil
\subfloat[Lena]{\includegraphics[width=0.58\columnwidth]{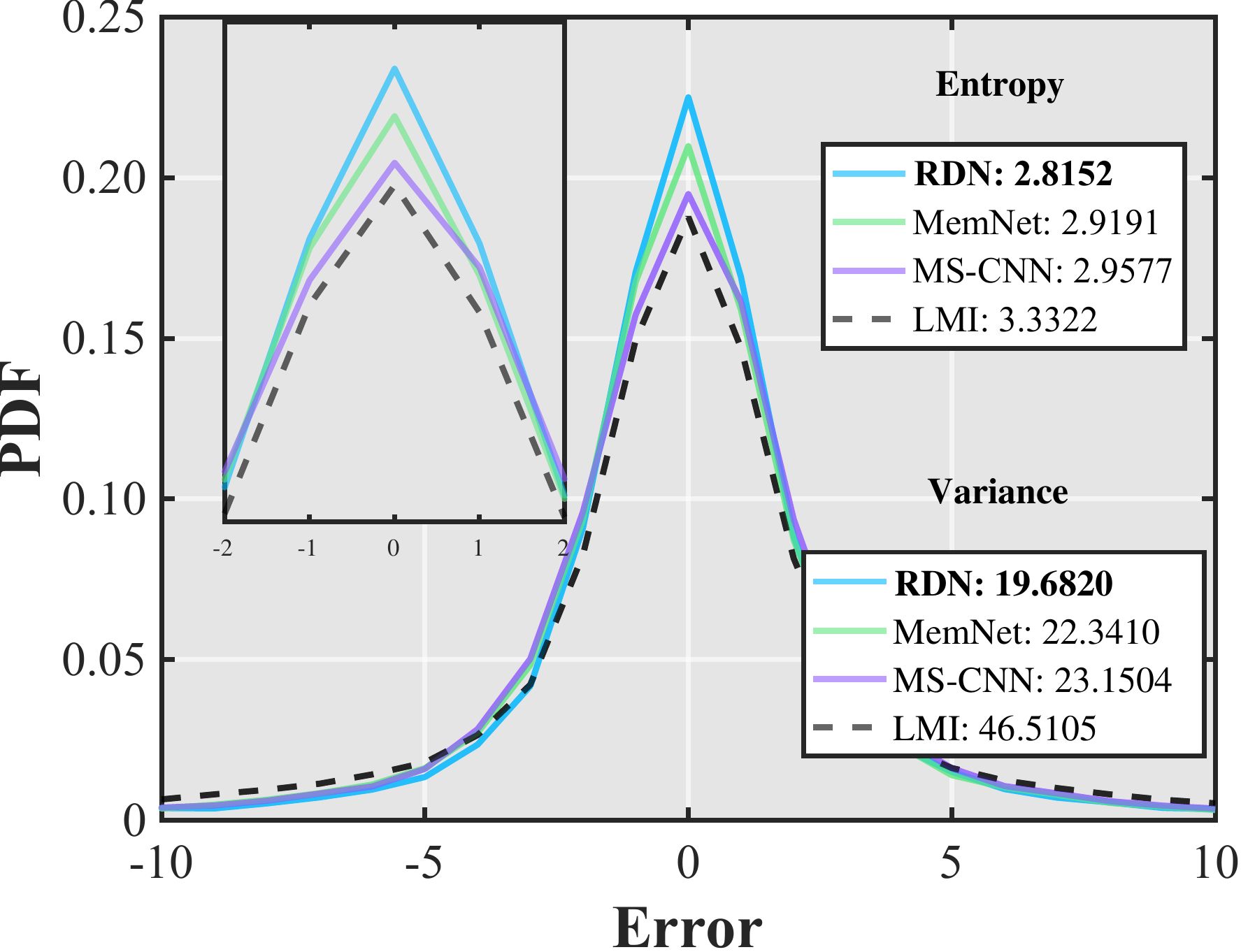}}
\hfil
\subfloat[Mandrill]{\includegraphics[width=0.58\columnwidth]{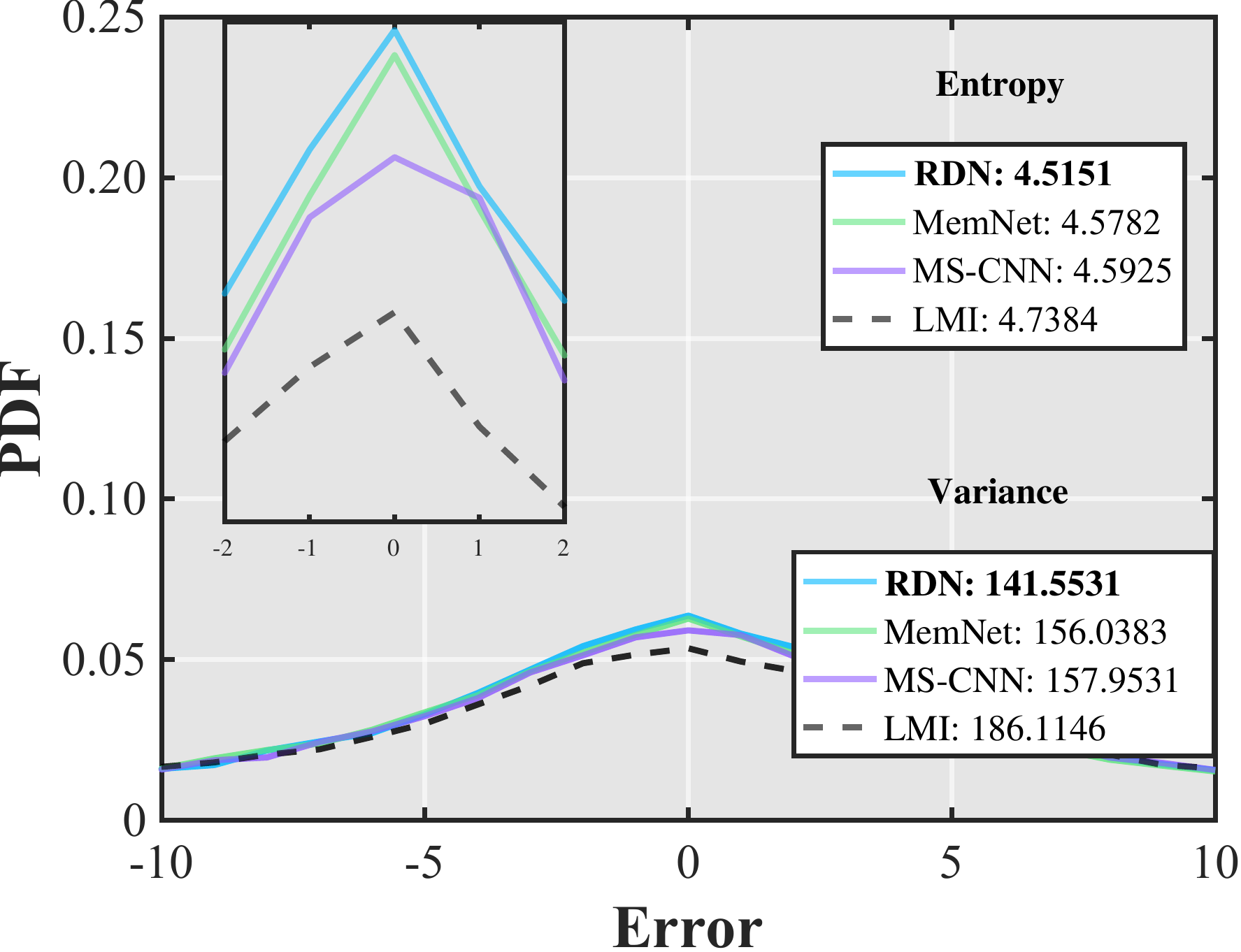}}
\caption{Probability distribution function of errors. Numerical data expresses entropy and variance.}
\label{fig:PDF}   
\subfloat[Aeroplane]{\includegraphics[width=0.58\columnwidth]{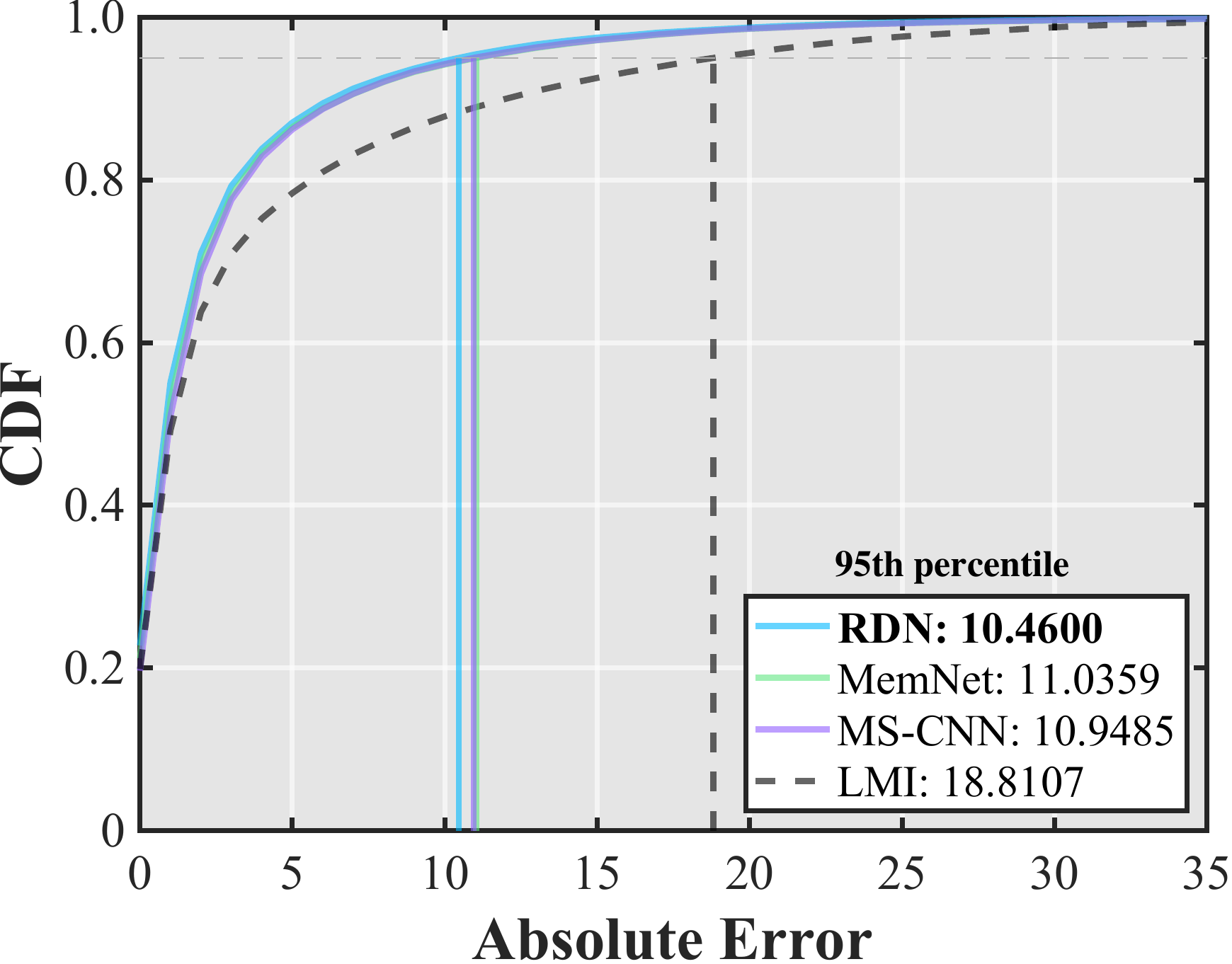}}
\hfil
\subfloat[Lena]{\includegraphics[width=0.58\columnwidth]{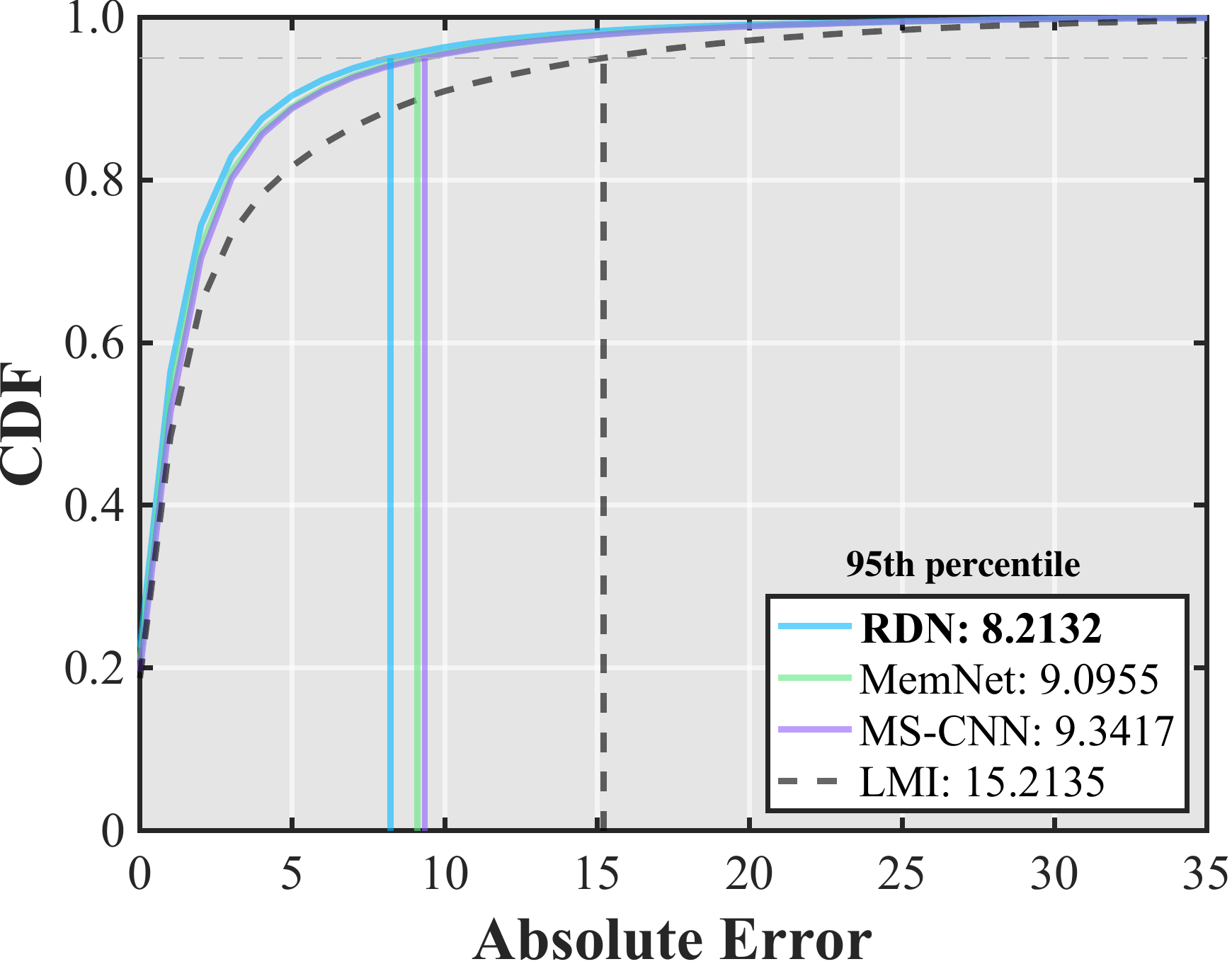}}
\hfil
\subfloat[Mandrill]{\includegraphics[width=0.58\columnwidth]{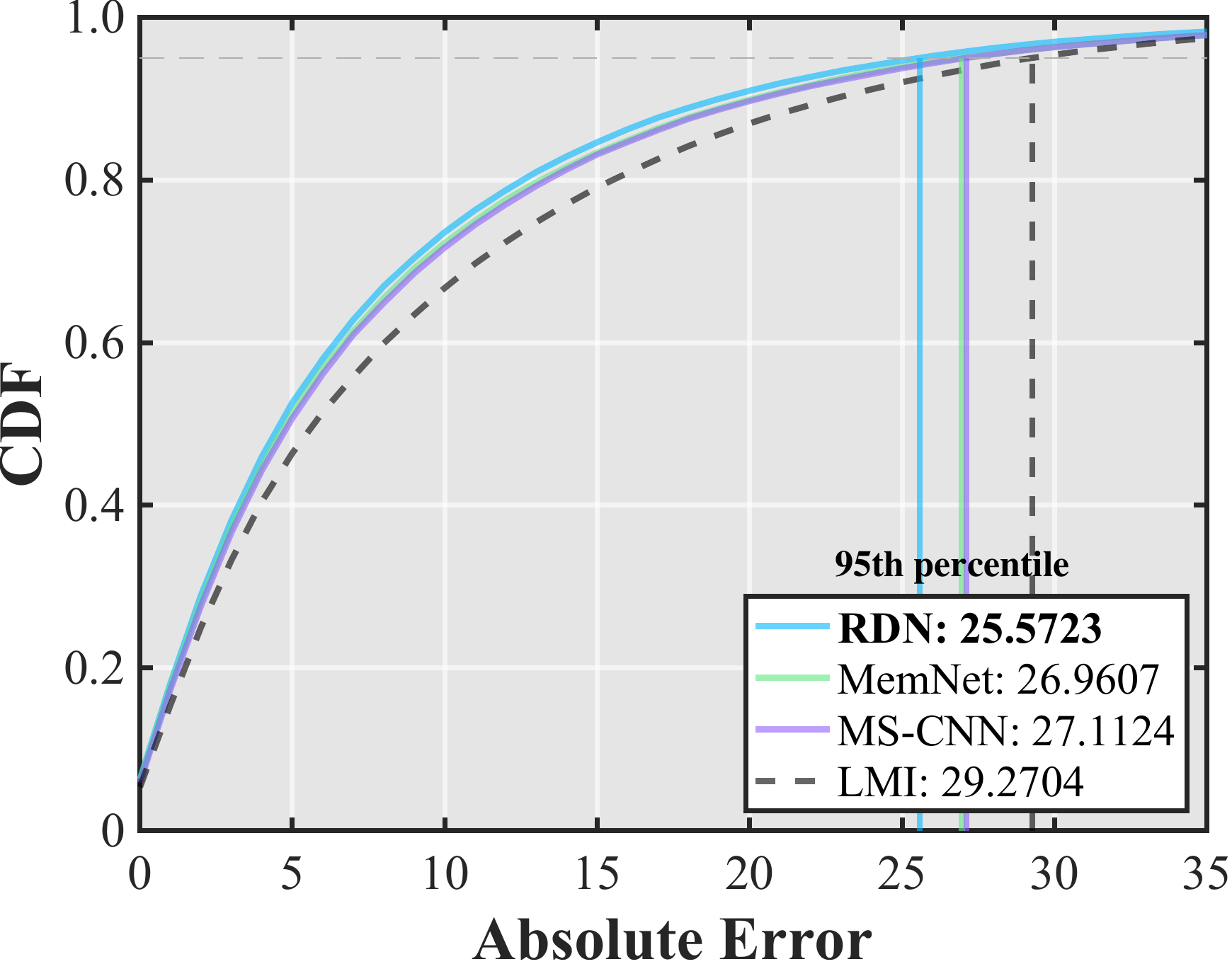}}
\caption{Cumulative distribution function of errors. Numerical data expresses 95\textsuperscript{th} percentile.}
\label{fig:CDF}
\subfloat[Aeroplane]{\includegraphics[width=0.6\columnwidth]{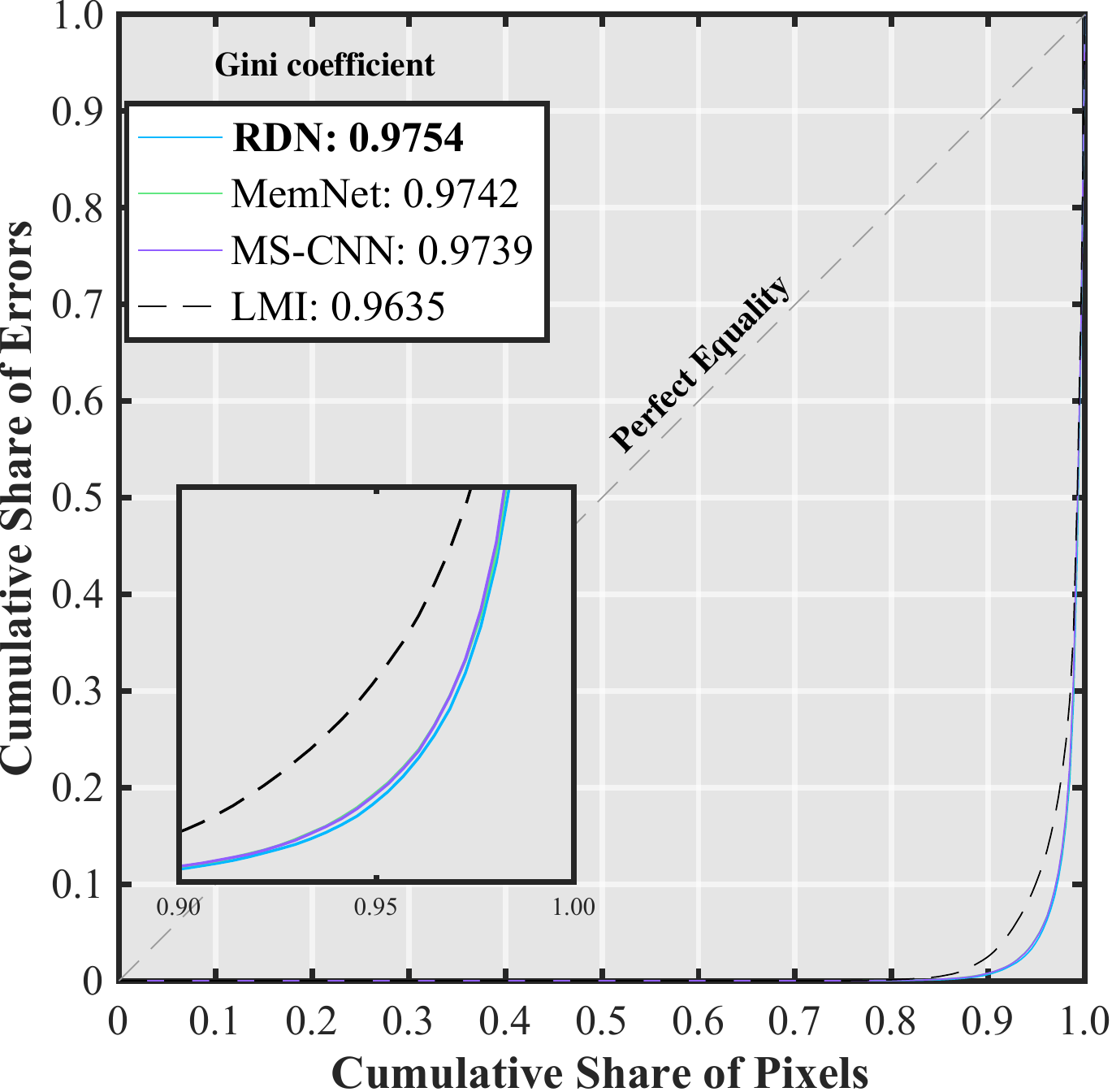}}
\hfil
\subfloat[Lena]{\includegraphics[width=0.6\columnwidth]{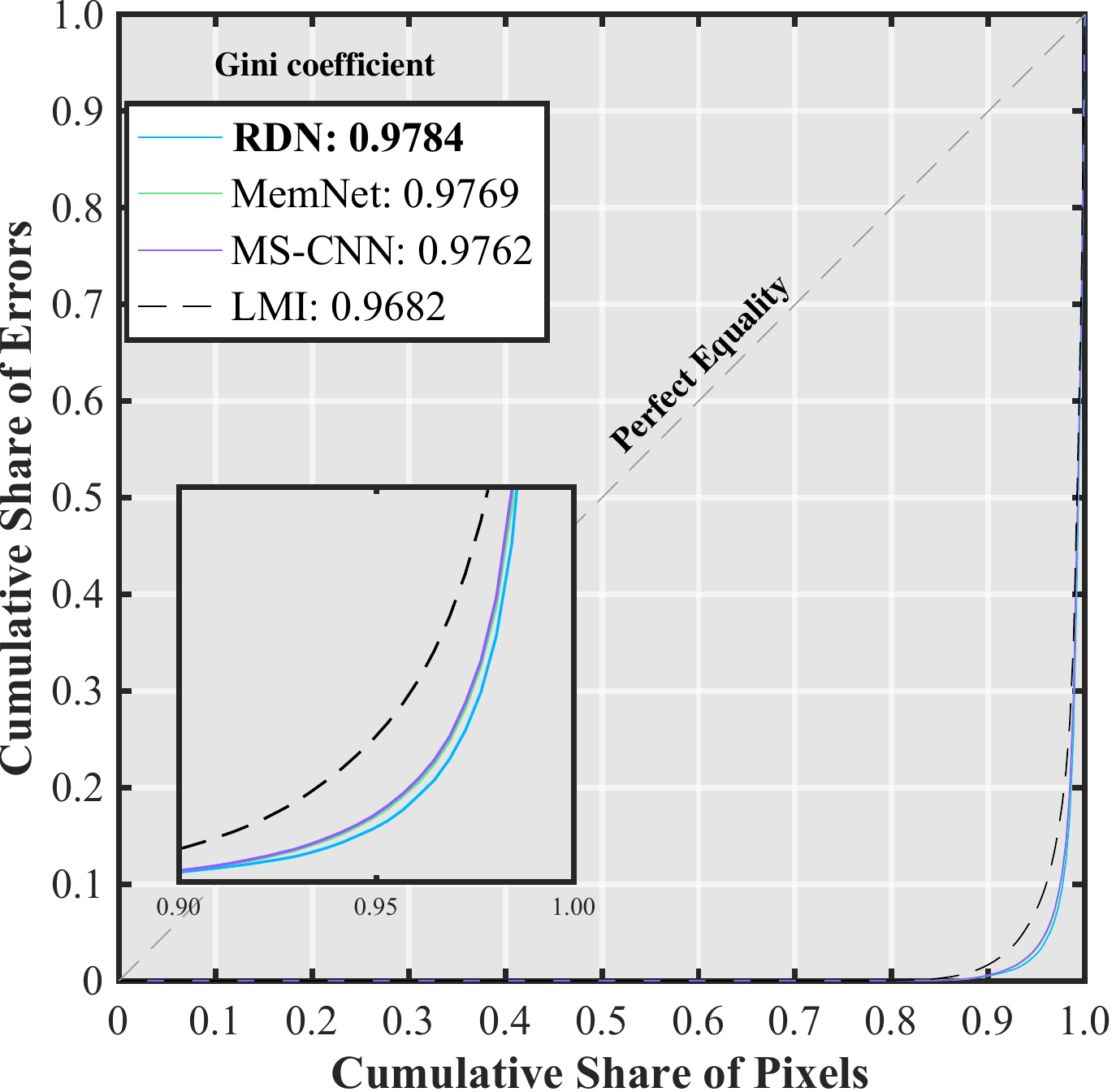}}
\hfil
\subfloat[Mandrill]{\includegraphics[width=0.6\columnwidth]{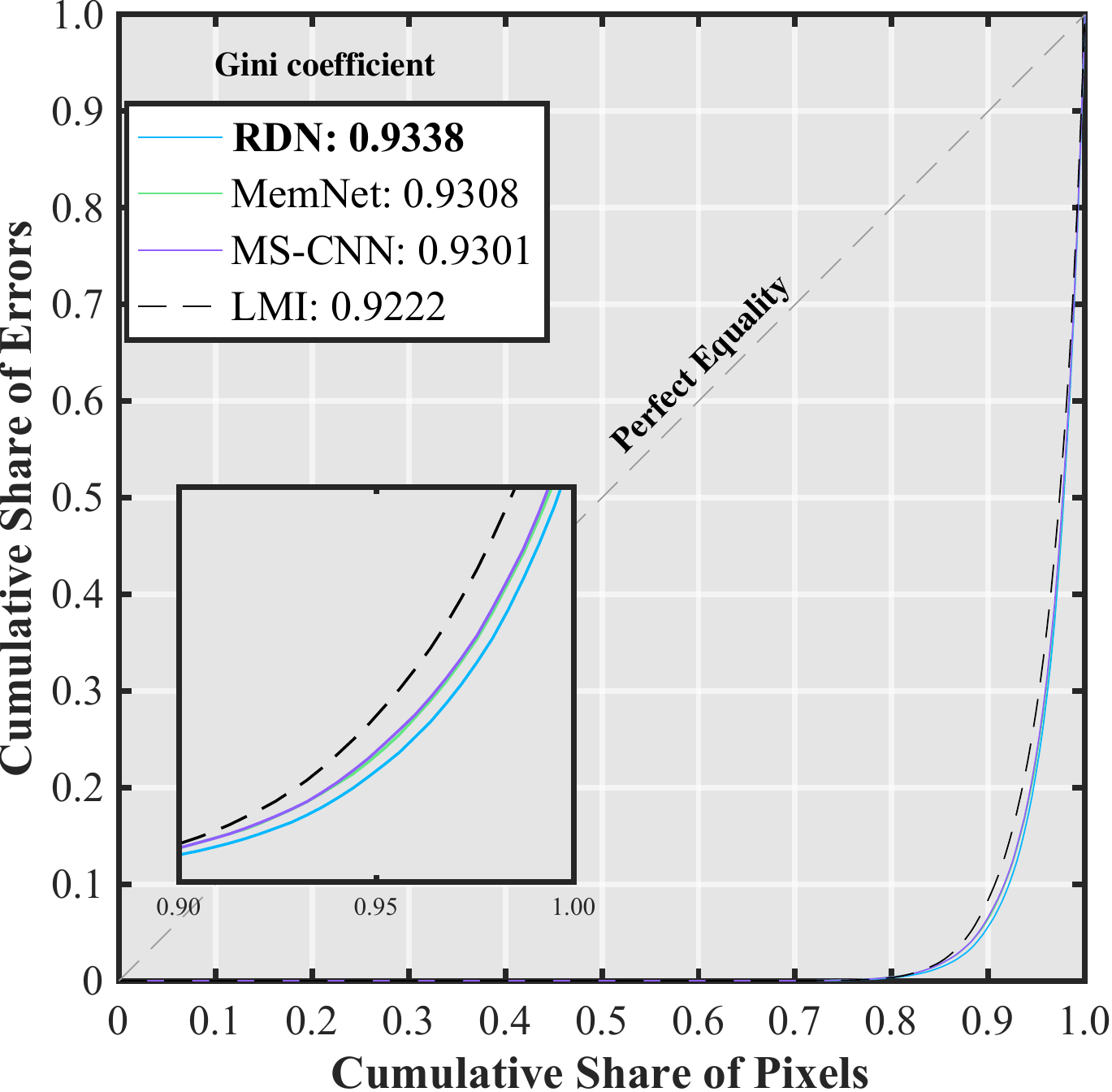}}
\caption{Lorenz curve of errors. Numerical data expresses Gini coefficient.}
\label{fig:Gini}
\end{figure*}

\begin{figure*}[t!] 
\centering
\subfloat[Aeroplane ($\vartheta = 1$)]{\includegraphics[width=0.58\columnwidth]{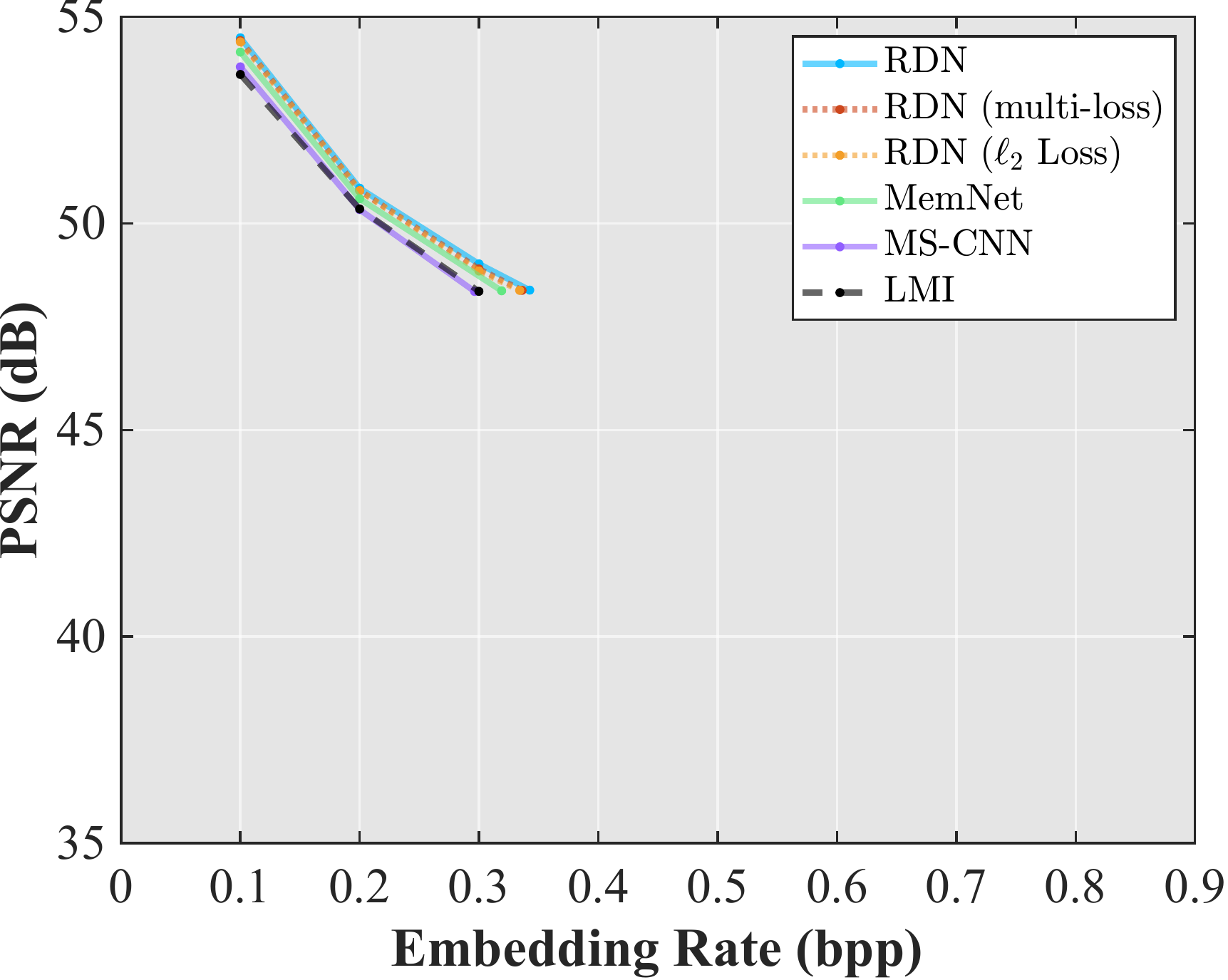}}
\hfil
\subfloat[Aeroplane ($\vartheta = 2$)]{\includegraphics[width=0.58\columnwidth]{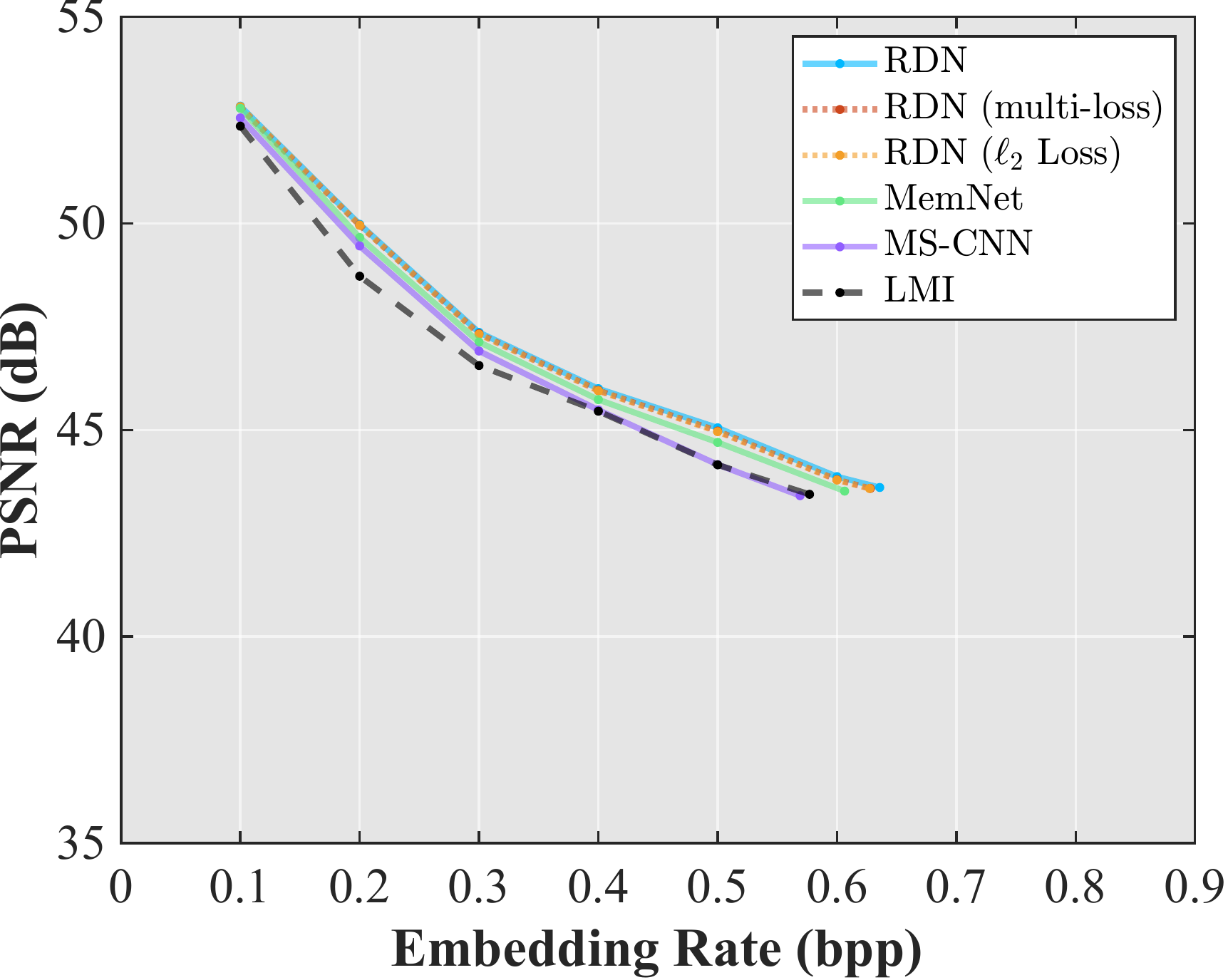}}
\hfil
\subfloat[Aeroplane ($\vartheta = 3$)]{\includegraphics[width=0.58\columnwidth]{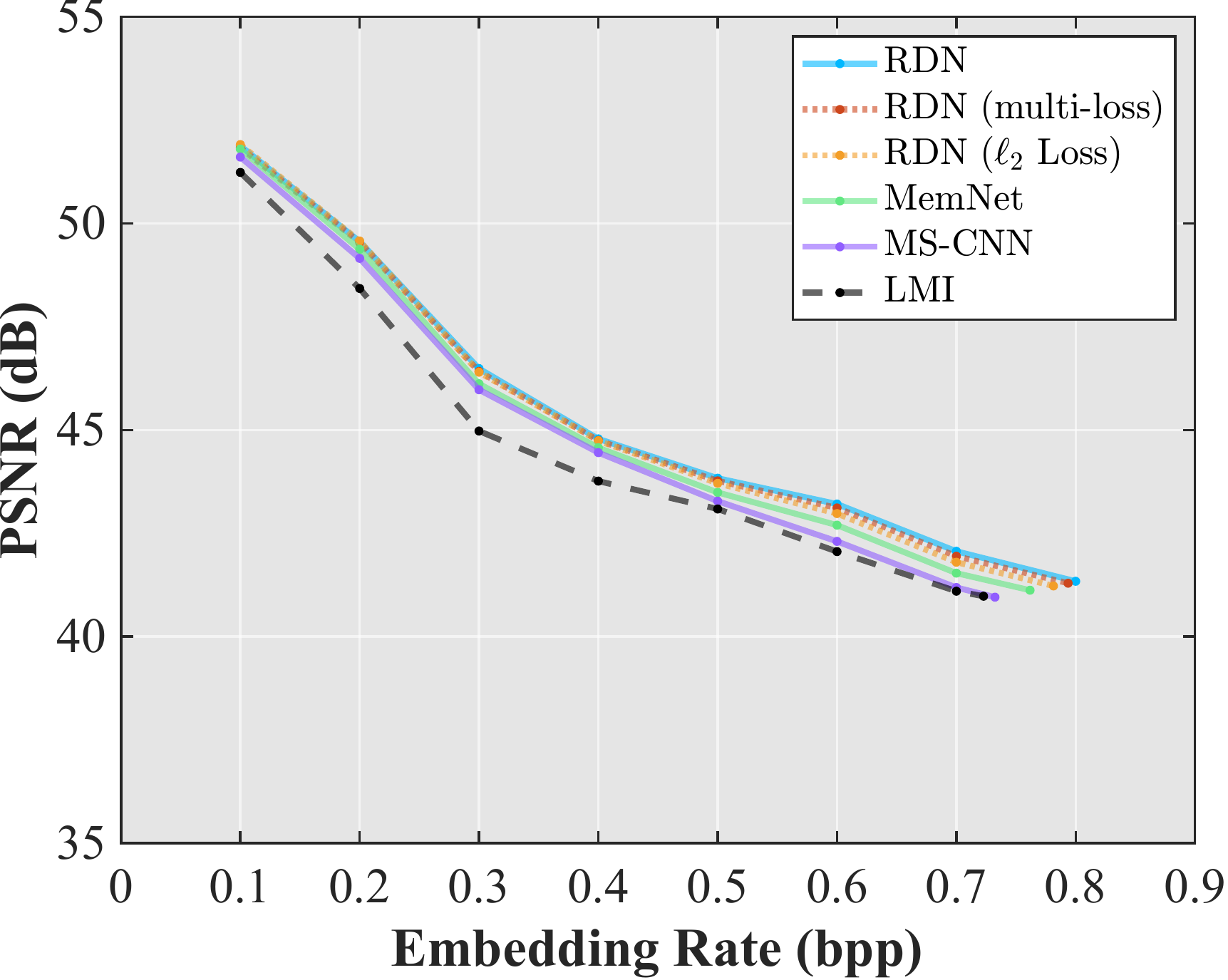}}
\\
\subfloat[Lena ($\vartheta = 1$)]{\includegraphics[width=0.58\columnwidth]{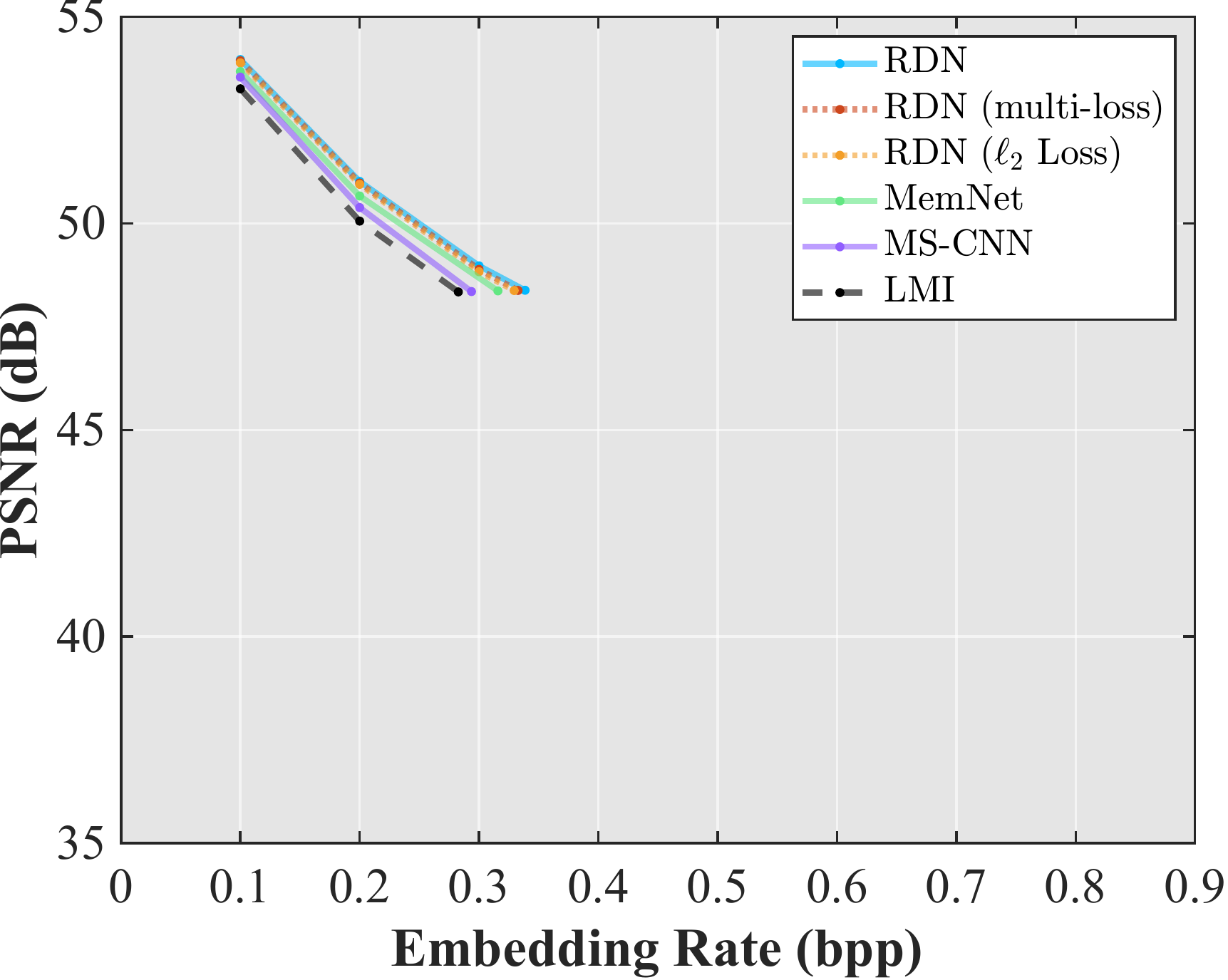}}
\hfil
\subfloat[Lena ($\vartheta = 2$)]{\includegraphics[width=0.58\columnwidth]{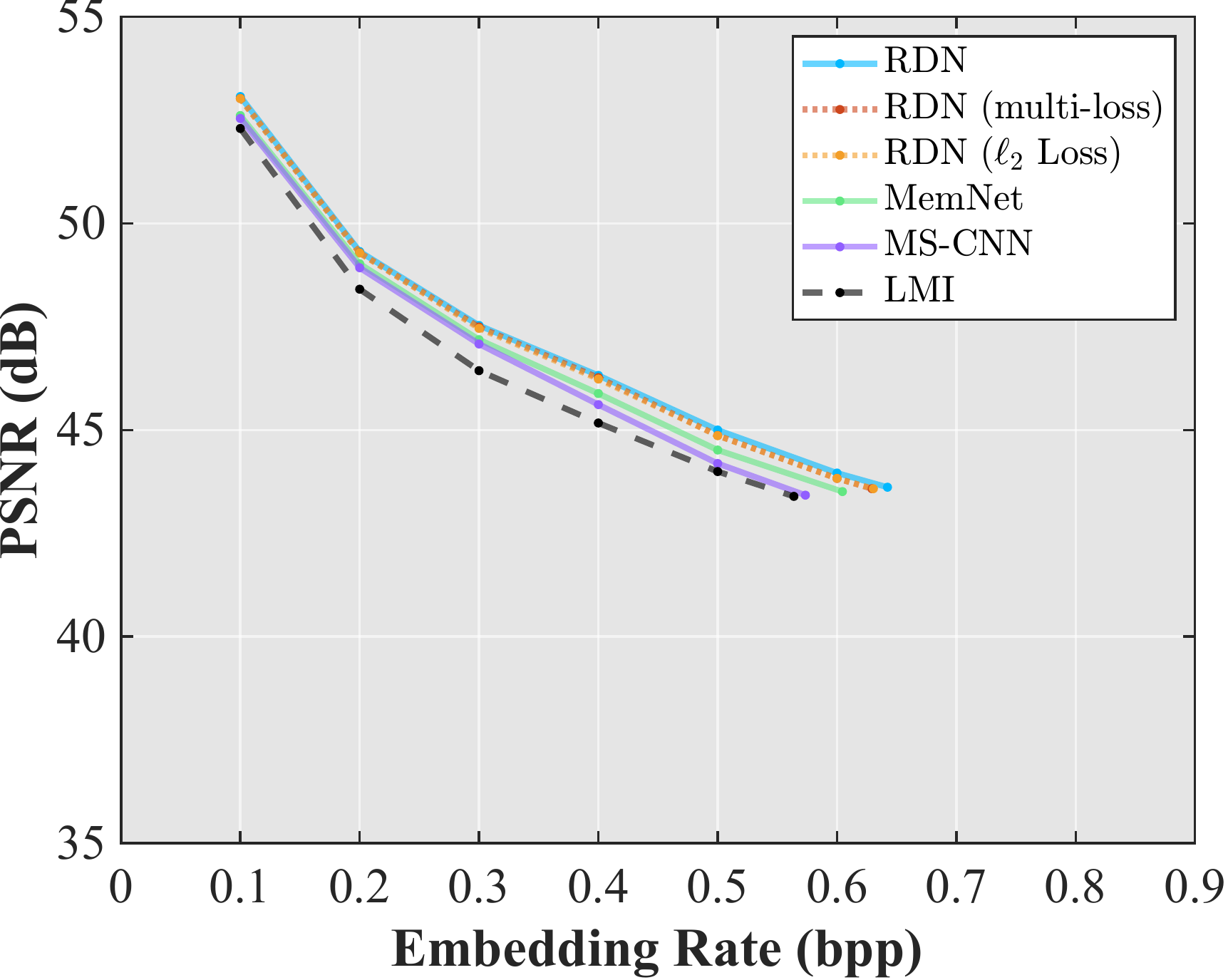}}
\hfil
\subfloat[Lena ($\vartheta = 3$)]{\includegraphics[width=0.58\columnwidth]{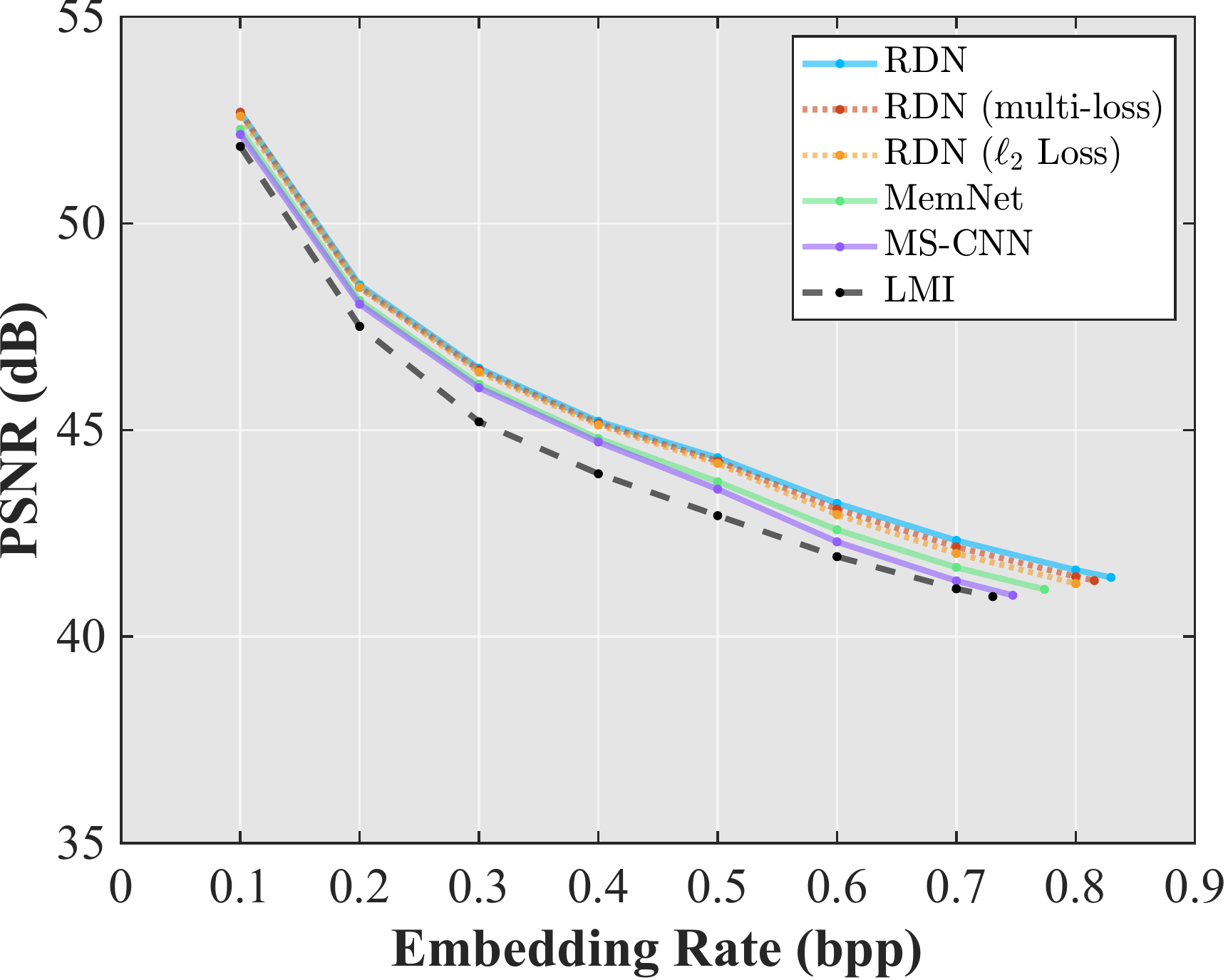}}
\\
\subfloat[Mandrill ($\vartheta = 1$)]{\includegraphics[width=0.58\columnwidth]{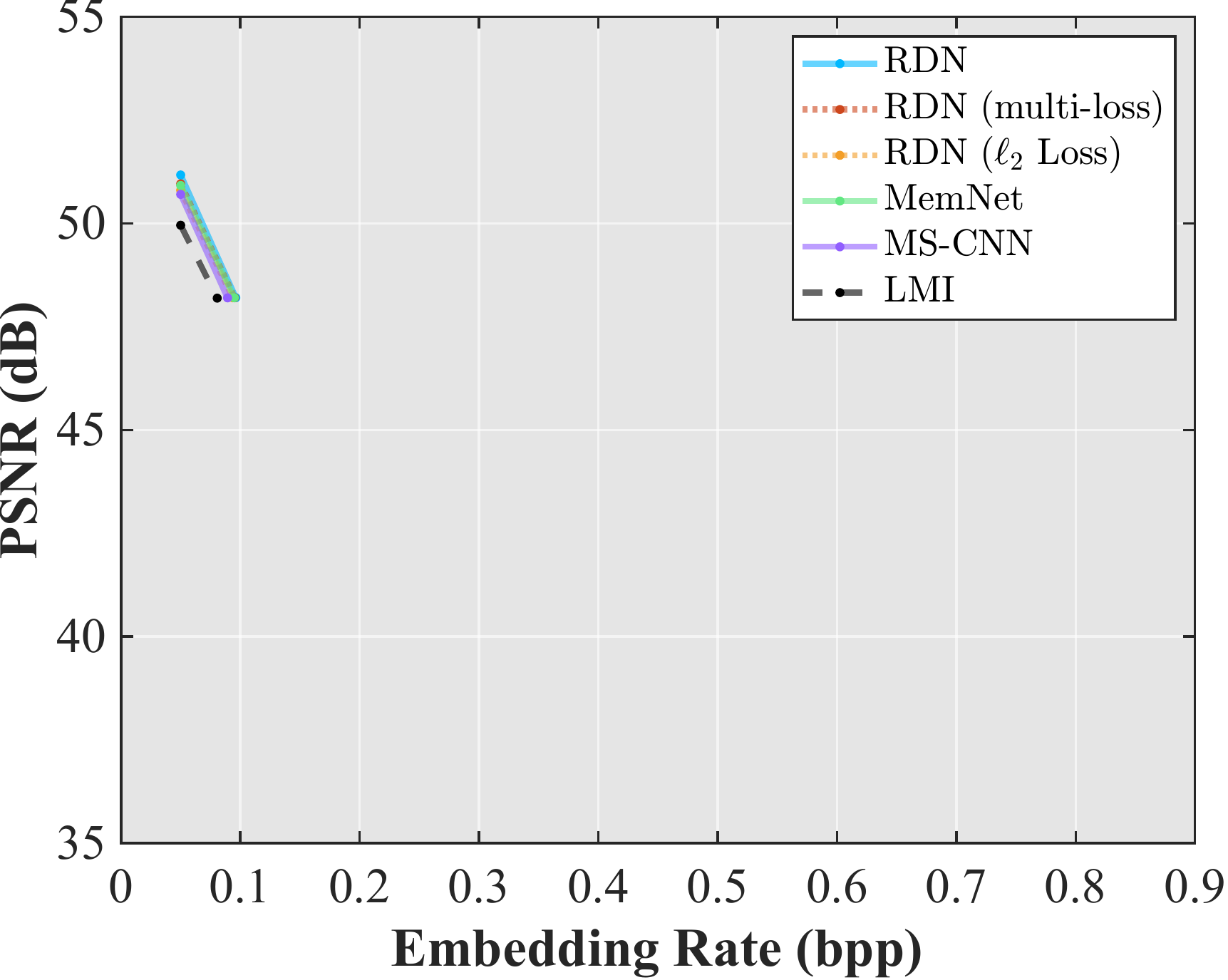}}
\hfil
\subfloat[Mandrill ($\vartheta = 2$)]{\includegraphics[width=0.58\columnwidth]{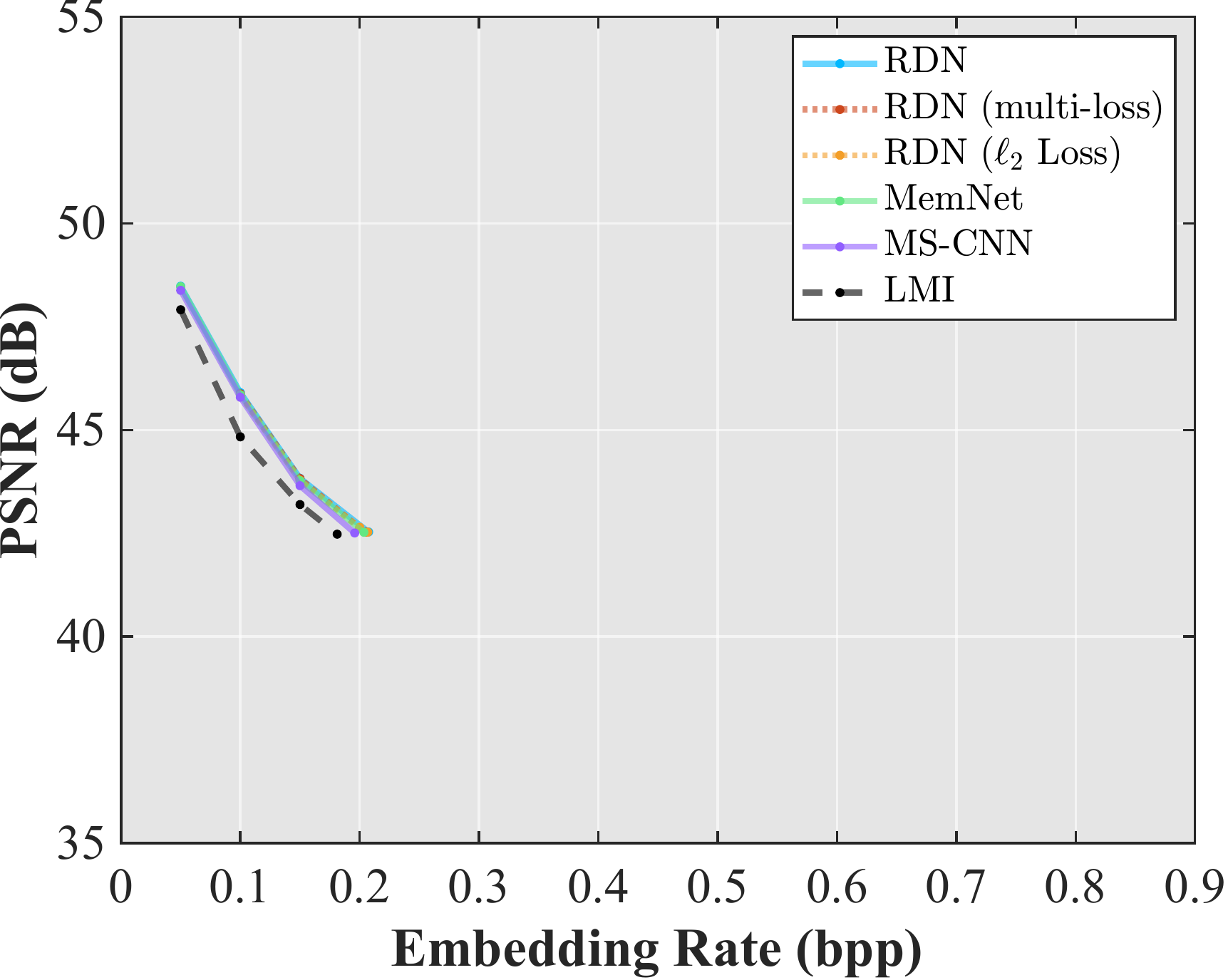}}
\hfil
\subfloat[Mandrill ($\vartheta = 3$)]{\includegraphics[width=0.58\columnwidth]{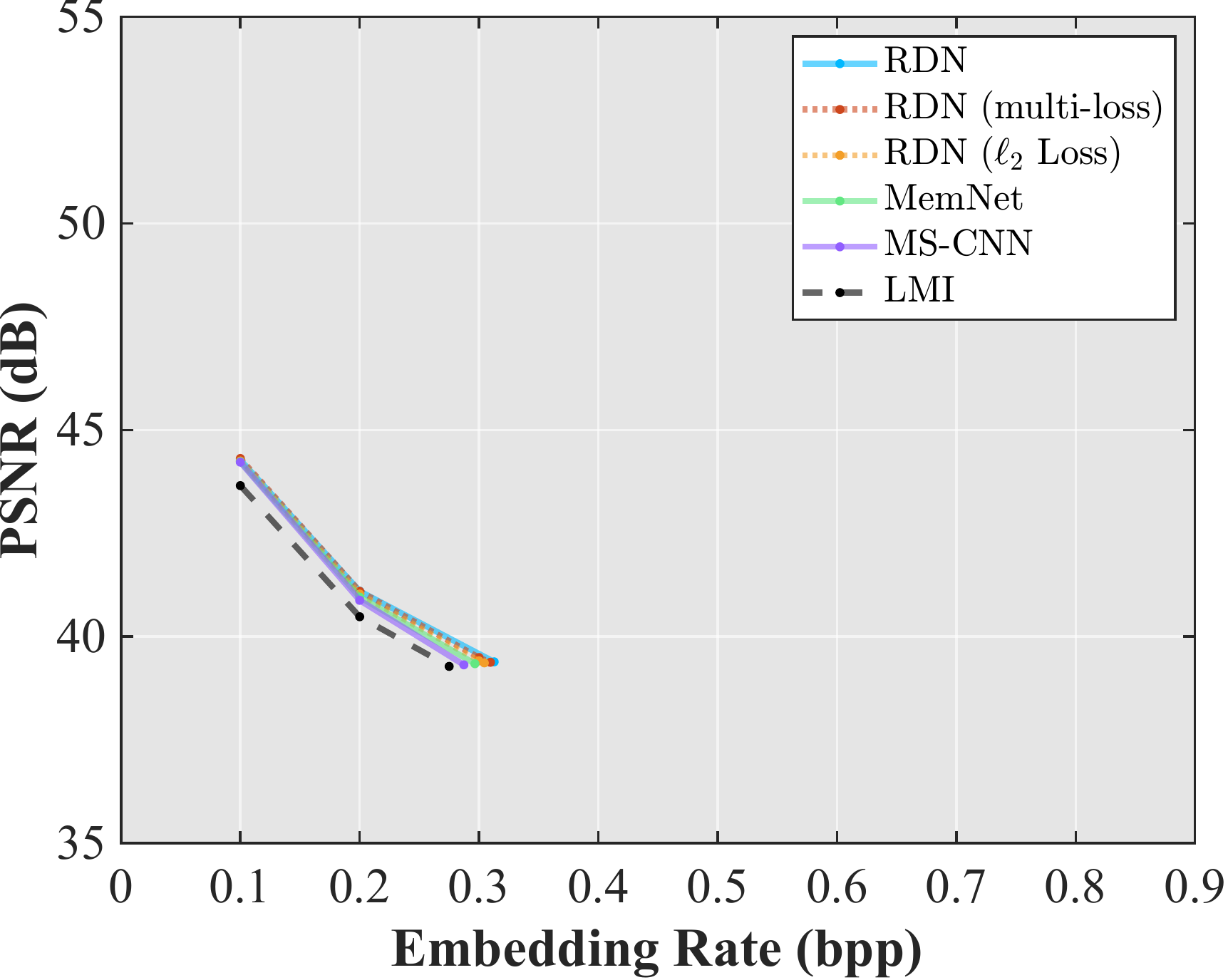}}

\caption{Rate-distortion curves for different settings of stego-channel parameter $\vartheta$.}
\label{fig:rate_dist}
\end{figure*}

\subsection{Evaluation of Distribution of Errors}
The distribution of prediction errors plays a pivotal role in steganographic performance. Normally, errors would follow a zero-centred Laplacian distribution and a more accurate model results in a more concentrated distribution. According to the coding algorithm, a more concentrated distribution can lead to a better steganographic rate-distortion trade-off. To analyse the error distribution, we examine the probability distribution function (PDF), cumulative distribution function (CDF), and Lorenz curve of errors, as shown in Figures~\ref{fig:PDF},~\ref{fig:CDF}, and~\ref{fig:Gini}, respectively. We use the entropy, variance, percentile statistics, and Gini coefficient to measure the degree of error concentration~\cite{Hart:1971aa}. Shannon's entropy is maximised for a uniform distribution and the converse is equally true: a smaller entropy means a more concentrated distribution~\cite{1948_6773024}. The variance is a measure of the spread of the data around the mean. The 95\textsuperscript{th} percentile indicates the maximum error magnitude below which $95\%$ of errors fall. The Gini coefficient is a measure of statistical dispersion intended to represent the inequality of the error magnitude within an image. A coefficient of $0$ expresses perfect equality (dispersion), whereas a coefficient of $1$ corresponds to maximal inequality (concentration). For the overall degree of concentration, the RDN ranked in the first tier, followed by the MemNet and the MS-CNN in the second tier, with the LMI ranked last.

\subsection{Evaluation of Rate-Distortion Performance}
The steganographic rate-distortion curves are plotted in Figure~\ref{fig:rate_dist}. The models with a higher predictive accuracy indeed achieve a better rate-distortion trade-off. The perceptual quality of stego images is inversely proportional to the embedding rate since distortions accumulate along the embedding process. The maximum capacity depends on the image content. Images with highly textured details would have lower capacity. A rise of $\vartheta$ increases the maximum embedding rate at the cost of compromising the rate-distortion trade-off. Amongst all models, the RDN stands out, achieving state-of-the-art steganographic performance. Apart from the $\ell_1$ loss, we investigate the applicability of other common loss functions in the field of image restoration. We implement the $\ell_2$ loss and the multi-loss on the RDN model. The latter is comprised of $\ell_1$ loss, feature-space loss and adversarial loss. While an extra computational cost is devoted to the implmentation of the multi-loss training, the results suggest that there is marginal, if any, gain in steganographic performance by applying such loss function. This reinforces the argument that the prediction-error modulation algorithm relies primarily on pixel-wise distance.

\section{Limitations of the Study}\label{sec:limit}
While deep learning has revolutionised the research field of reversible steganography and led to major technological breakthroughs in terms of capacity and imperceptibility, novel application scenarios may be offered if secrecy and robustness are taken into further consideration. The current use of neural networks in the modular framework is confined to predictive analytics. By relaxing the constraint on perfect reversibility, it might be possible to apply the end-to-end framework to automatically learn to create stego objects that are undetectable under steganalysis tools and robust against common multimedia processing operations. Apropos of predictive analytics, an important aspect, apart from predictive accuracy, is predictive uncertainty. If the uncertainty about predictions can be estimated, the rate-distortion performance may be further improved by selecting pixels which are predicted with high confidence.

\section{Conclusion}\label{sec:con}
In this work, we have discussed unexplored issues such as the impact of intensity initialisation on predictive accuracy and the influence of distributional shift in dual-layer prediction. Experimental results have revealed that setting pixel intensity to zero, albeit seemingly arbitrary, renders a steadily low loss over several epochs. In addition, it has been found that training models in a causal way can, to some extent, ameliorate the distributional shift in deployment since it minimises the discrepancy in the distributions of training and test sets. The state-of-the-art predictive accuracy and steganographic rate-distortion performance can be achieved by applying advanced pixel-level computer vision models. We envision a promising paradigm shift in reversible steganography ushered in by deep learning and hope that this paper can prove instructive for future research.



\bibliographystyle{Transactions-Bibliography/IEEEtran}
\bibliography{./Transactions-Bibliography/IEEEabrv,./Bib/myBib_abbrv}

\begin{thebibliography}{10}
\providecommand{\url}[1]{#1}
\csname url@samestyle\endcsname
\providecommand{\newblock}{\relax}
\providecommand{\bibinfo}[2]{#2}
\providecommand{\BIBentrySTDinterwordspacing}{\spaceskip=0pt\relax}
\providecommand{\BIBentryALTinterwordstretchfactor}{4}
\providecommand{\BIBentryALTinterwordspacing}{\spaceskip=\fontdimen2\font plus
\BIBentryALTinterwordstretchfactor\fontdimen3\font minus
  \fontdimen4\font\relax}
\providecommand{\BIBforeignlanguage}[2]{{%
\expandafter\ifx\csname l@#1\endcsname\relax
\typeout{** WARNING: IEEEtran.bst: No hyphenation pattern has been}%
\typeout{** loaded for the language `#1'. Using the pattern for}%
\typeout{** the default language instead.}%
\else
\language=\csname l@#1\endcsname
\fi
#2}}
\providecommand{\BIBdecl}{\relax}
\BIBdecl

\bibitem{668971}
R.~Anderson and F.~Petitcolas, ``On the limits of steganography,'' \emph{IEEE
  J. Sel. Areas Commun.}, vol.~16, no.~4, pp. 474--481, 1998.

\bibitem{2005_1511007}
J.~{Fridrich}, M.~{Goljan}, P.~{Lisonek}, and D.~{Soukal}, ``Writing on wet
  paper,'' \emph{IEEE Trans. Signal Process.}, vol.~53, no.~10, pp. 3923--3935,
  2005.

\bibitem{1997_650120}
I.~J. {Cox}, J.~{Kilian}, F.~T. {Leighton}, and T.~{Shamoon}, ``Secure spread
  spectrum watermarking for multimedia,'' \emph{IEEE Trans. Image Process.},
  vol.~6, no.~12, pp. 1673--1687, 1997.

\bibitem{1999_771070}
D.~{Kundur} and D.~{Hatzinakos}, ``Digital watermarking for telltale tamper
  proofing and authentication,'' \emph{Proceedings of the IEEE}, vol.~87,
  no.~7, pp. 1167--1180, 1999.

\bibitem{2001_Fridrich_Invertible}
J.~Fridrich, M.~Goljan, and R.~Du, ``Invertible authentication,'' in
  \emph{Proc. {SPIE} Conf. Secur. Watermarking Multimedia Contents (SWMC)}, San
  Jose, CA, USA, 2001, pp. 197--208.

\bibitem{2003_1196739}
C.~{De Vleeschouwer}, J.-F. {Delaigle}, and B.~{Macq}, ``Circular
  interpretation of bijective transformations in lossless watermarking for
  media asset management,'' \emph{IEEE Trans. Multimedia}, vol.~5, no.~1, pp.
  97--105, 2003.

\bibitem{2013_6329433}
G.~{Coatrieux}, W.~{Pan}, N.~{Cuppens-Boulahia}, F.~{Cuppens}, and C.~{Roux},
  ``Reversible watermarking based on invariant image classification and dynamic
  histogram shifting,'' \emph{IEEE Trans. Inf. Forensics Secur.}, vol.~8,
  no.~1, pp. 111--120, 2013.

\bibitem{2016_RDH_Survey}
Y.-Q. {Shi}, X.~{Li}, X.~{Zhang}, H.~{Wu}, and B.~{Ma}, ``Reversible data
  hiding: Advances in the past two decades,'' \emph{IEEE Access}, vol.~4, pp.
  3210--3237, 2016.

\bibitem{2015_Perturb_Goodfellow}
I.~Goodfellow, J.~Shlens, and C.~Szegedy, ``Explaining and harnessing
  adversarial examples,'' in \emph{Proc. Int. Conf. Learn. Represent. (ICLR)},
  San Diego, CA, USA, 2015, pp. 1--11.

\bibitem{2016_DeepFool}
S.-M. {Moosavi-Dezfooli}, A.~{Fawzi}, and P.~{Frossard}, ``{DeepFool}: {A}
  simple and accurate method to fool deep neural networks,'' in
  \emph{Proceedings of {IEEE} Conference on Computer Vision and Pattern
  Recognition (CVPR)}, Las Vegas, NV, USA, 2016, pp. 2574--2582.

\bibitem{Poisoning17}
L.~Mu\~{n}oz{-}Gonz\'{a}lez, B.~Biggio, A.~Demontis, A.~Paudice,
  V.~Wongrassamee, E.~C. Lupu, and F.~Roli, ``Towards poisoning of deep
  learning algorithms with back-gradient optimization,'' in \emph{Proc. ACM
  Workshop Artif. Intell. Secur. (AISec)}, Dallas, TX, USA, 2017, pp. 27--38.

\bibitem{9753668}
C.-C. Chang, ``Reversible linguistic steganography with {B}ayesian masked
  language modeling,'' \emph{IEEE Trans. Comput. Soc. Syst.}, pp. 1--10, 2022.

\bibitem{1948_6773024}
C.~E. Shannon, ``A mathematical theory of communication,'' \emph{Bell Syst.
  Tech. J.}, vol.~27, no.~3, pp. 379--423, 1948.

\bibitem{2011_5762603}
X.~{Li}, B.~{Yang}, and T.~{Zeng}, ``Efficient reversible watermarking based on
  adaptive prediction-error expansion and pixel selection,'' \emph{IEEE Trans.
  Image Process.}, vol.~20, no.~12, pp. 3524--3533, 2011.

\bibitem{2014_6746082}
I.~{Dragoi} and D.~{Coltuc}, ``Local prediction based difference expansion
  reversible watermarking,'' \emph{IEEE Trans. Image Process.}, vol.~23, no.~4,
  pp. 1779--1790, 2014.

\bibitem{Hwang:2016aa}
H.~J. Hwang, S.~Kim, and H.~J. Kim, ``Reversible data hiding using least square
  predictor via the {LASSO},'' \emph{{EURASIP} J. Image Video Process.}, vol.
  2016, no.~1, 2016, {{A}rt. no. 42}.

\bibitem{2018_8578360}
Y.~Zhang, Y.~Tian, Y.~Kong, B.~Zhong, and Y.~Fu, ``Residual dense network for
  image super-resolution,'' in \emph{Proc. {IEEE/CVF} Conf. Comput. Vis.
  Pattern Recognit. (CVPR)}, Salt Lake City, UT, USA, 2018, pp. 2472--2481.

\bibitem{8964437}
Y.~{Zhang}, Y.~Tian, Y.~Kong, B.~Zhong, and Y.~Fu, ``Residual dense network for
  image restoration,'' \emph{IEEE Trans. Pattern Anal. Mach. Intell.}, vol.~43,
  no.~7, pp. 2480--2495, 2021.

\bibitem{10.1093/mind/LIX.236.433}
A.~M. Turing, ``Computing machinery and intelligence,'' \emph{Mind}, vol. LIX,
  no. 236, pp. 433--460, 1950.

\bibitem{LeCun:2015aa}
Y.~LeCun, Y.~Bengio, and G.~E. Hinton, ``Deep learning,'' \emph{Nature}, vol.
  521, no. 7553, pp. 436--444, 2015.

\bibitem{Zhu:2018aa}
J.~Zhu, R.~Kaplan, J.~Johnson, and F.-F. Li, ``{HiDDeN}: {H}iding data with
  deep networks,'' in \emph{Proc. Eur. Conf. Comput. Vis. (ECCV)}, Munich,
  Germany, 2018, pp. 682--697.

\bibitem{2017_8017430}
W.~{Tang}, S.~{Tan}, B.~{Li}, and J.~{Huang}, ``Automatic steganographic
  distortion learning using a generative adversarial network,'' \emph{IEEE
  Signal Process. Lett.}, vol.~24, no.~10, pp. 1547--1551, 2017.

\bibitem{2019_8603808}
W.~{Tang}, B.~{Li}, S.~{Tan}, M.~{Barni}, and J.~{Huang}, ``{CNN}-based
  adversarial embedding for image steganography,'' \emph{IEEE Trans. Inf.
  Forensics Secur.}, vol.~14, no.~8, pp. 2074--2087, 2019.

\bibitem{2020_8735922}
J.~{Yang}, D.~{Ruan}, J.~{Huang}, X.~{Kang}, and Y.-Q. {Shi}, ``An embedding
  cost learning framework using {GAN},'' \emph{IEEE Trans. Inf. Forensics
  Secur.}, vol.~15, pp. 839--851, 2020.

\bibitem{Volkhonskiy:2019aa}
D.~Volkhonskiy, I.~Nazarov, and E.~Burnaev, ``Steganographic generative
  adversarial networks,'' in \emph{Proc. Int. Conf. Mach. Vis. (ICMV)}, vol.
  11433, Amsterdam, Netherlands, 2019, pp. 1--15.

\bibitem{NIPS2017_6791}
J.~Hayes and G.~Danezis, ``Generating steganographic images via adversarial
  training,'' in \emph{Proc. Adv. Neural Inf. Process. Syst. (NeurIPS)}, Long
  Beach, CA, USA, 2017, pp. 1951---1960.

\bibitem{2018_8403208}
D.~{Hu}, L.~{Wang}, W.~{Jiang}, S.~{Zheng}, and B.~{Li}, ``A novel image
  steganography method via deep convolutional generative adversarial
  networks,'' \emph{IEEE Access}, vol.~6, pp. 38\,303--38\,314, 2018.

\bibitem{Luo:2020aa}
X.~Luo, R.~Zhan, H.~Chang, F.~Yang, and P.~Milanfar, ``Distortion agnostic deep
  watermarking,'' in \emph{Proc. {IEEE/CVF} Conf. Comput. Vis. Pattern
  Recognit. (CVPR)}, Seattle, WA, USA, 2020, pp. 13\,545--13\,554.

\bibitem{Wengrowski_2019_CVPR}
E.~Wengrowski and K.~Dana, ``Light field messaging with deep photographic
  steganography,'' in \emph{Proc. {IEEE} Conf. Comput. Vis. Pattern Recognit.
  (CVPR)}, Long Beach, CA, USA, 2019, pp. 1515--1524.

\bibitem{Fang:2021aa}
H.~Fang, D.~Chen, F.~Wang, Z.~Ma, H.~Liu, W.~Zhou, W.~Zhang, and N.~H. Yu,
  ``{TERA}: {S}creen-to-camera image code with transparency, efficiency,
  robustness and adaptability,'' \emph{IEEE Trans. Multimedia}, pp. 1--13,
  2021.

\bibitem{Tancik:2020aa}
M.~Tancik, B.~Mildenhall, and R.~Ng, ``{StegaStamp}: {I}nvisible hyperlinks in
  physical photographs,'' in \emph{Proc. {IEEE/CVF} Conf. Comput. Vis. Pattern
  Recognit. (CVPR)}, Seattle, WA, USA, 2020, pp. 2114--2123.

\bibitem{Duan:2019aa}
X.~Duan, K.~Jia, B.~Li, D.~Guo, E.~Zhang, and C.~Qin, ``Reversible image
  steganography scheme based on a {U-Net} structure,'' \emph{IEEE Access},
  vol.~7, pp. 9314--9323, 2019.

\bibitem{Jung:2019aa}
Z.~Zhang, G.~Fu, F.~Di, C.~Li, and J.~Liu, ``Generative reversible data hiding
  by image-to-image translation via {GANs},'' \emph{Secur. Commun. Netw.}, vol.
  2019, 2019, {{A}rt. no. 4932782}.

\bibitem{Lu_2021_CVPR}
S.-P. Lu, R.~Wang, T.~Zhong, and P.~L. Rosin, ``Large-capacity image
  steganography based on invertible neural networks,'' in \emph{Proc.
  {IEEE/CVF} Conf. Comput. Vis. Pattern Recognit. (CVPR)}, virtual, 2021, pp.
  10\,816--10\,825.

\bibitem{Fridrich:2002aa}
J.~Fridrich, M.~Goljan, and R.~Du, ``Lossless data embedding\textemdash{N}ew
  paradigm in digital watermarking,'' \emph{{EURASIP} J. Adv. Signal Process.},
  vol. 986842, no. 2002, pp. 185--196, 2002.

\bibitem{2020_9245471}
C.-C. Chang, ``Adversarial learning for invertible steganography,'' \emph{IEEE
  Access}, vol.~8, pp. 198\,425--198\,435, 2020.

\bibitem{2017_pix2pix_8100115}
P.~{Isola}, J.-Y. {Zhu}, T.~{Zhou}, and A.~A. {Efros}, ``Image-to-image
  translation with conditional adversarial networks,'' in \emph{Proc. {IEEE}
  Conf. Comput. Vis. Pattern Recognit. (CVPR)}, Honolulu, HI, USA, 2017, pp.
  5967--5976.

\bibitem{Hu:2021aa}
R.~Hu and S.~Xiang, ``{CNN} prediction based reversible data hiding,''
  \emph{IEEE Signal Process. Lett.}, vol.~28, pp. 464--468, 2021.

\bibitem{Chang:2021aa}
C.-C. Chang, ``Neural reversible steganography with long short-term memory,''
  \emph{Secur. Commun. Netw.}, vol. 2021, 2021, {{A}rt. no. 5580272}.

\bibitem{8237748}
Y.~{Tai}, J.~{Yang}, X.~{Liu}, and C.~{Xu}, ``{MemNet}: {A} persistent memory
  network for image restoration,'' in \emph{Proc. {IEEE} Int. Conf. Comput.
  Vis. (ICCV)}, Venice, Italy, 2017, pp. 4549--4557.

\bibitem{8723565}
W.~Yang, X.~Zhang, Y.~Tian, W.~Wang, J.-H. Xue, and Q.~Liao, ``Deep learning
  for single image super-resolution: {A} brief review,'' \emph{IEEE Trans.
  Multimedia}, vol.~21, no.~12, pp. 3106--3121, 2019.

\bibitem{Wilson:1923aa}
E.~B. Wilson, ``First and second laws of error,'' \emph{J. Amer. Statist.
  Assoc.}, vol.~18, no. 143, pp. 841--851, 1923.

\bibitem{https://doi.org/10.48550/arxiv.2202.13133}
C.-C. Chang, ``Nonlinear discrete optimisation of reversible steganographic
  coding,'' arXiv, 2022.

\bibitem{2009_4811982}
V.~{Sachnev}, H.~J. {Kim}, J.~{Nam}, S.~{Suresh}, and Y.-Q. {Shi}, ``Reversible
  watermarking algorithm using sorting and prediction,'' \emph{IEEE Trans.
  Circuits Syst. Video Technol.}, vol.~19, no.~7, pp. 989--999, 2009.

\bibitem{4775883}
Z.~Wang and A.~C. Bovik, ``Mean squared error: {L}ove it or leave it? {A} new
  look at signal fidelity measures,'' \emph{IEEE Signal Process. Mag.},
  vol.~26, no.~1, pp. 98--117, 2009.

\bibitem{7797130}
H.~Zhao, O.~Gallo, I.~Frosio, and J.~Kautz, ``Loss functions for image
  restoration with neural networks,'' \emph{IEEE Trans Comput. Imag.}, vol.~3,
  no.~1, pp. 47--57, 2017.

\bibitem{2017_8099502}
C.~{Ledig}, L.~{Theis}, F.~{Husz{\'a}r}, J.~{Caballero}, A.~{Cunningham},
  A.~{Acosta}, A.~{Aitken}, A.~{Tejani}, J.~{Totz}, Z.~{Wang}, and W.~{Shi},
  ``Photo-realistic single image super-resolution using a generative
  adversarial network,'' in \emph{Proc. {IEEE} Conf. Comput. Vis. Pattern
  Recognit. (CVPR)}, Honolulu, HI, USA, 2017, pp. 105--114.

\bibitem{Simonyan15}
K.~Simonyan and A.~Zisserman, ``Very deep convolutional networks for
  large-scale image recognition,'' in \emph{Proc. Int. Conf. Learn.
  Representations (ICLR)}, San Diego, CA, USA, 2015, pp. 1--14.

\bibitem{ILSVRC15}
O.~Russakovsky, J.~Deng, H.~Su, J.~Krause, S.~Satheesh, S.~Ma, Z.~Huang,
  A.~Karpathy, A.~Khosla, M.~Bernstein, A.~C. Berg, and F.-F. Li, ``{ImageNet}
  large scale visual recognition challenge,'' \emph{Int. J. Comput. Vis.}, vol.
  115, no.~3, pp. 211--252, 2015.

\bibitem{NIPS2014_5423}
I.~Goodfellow, J.~Pouget-Abadie, M.~Mirza, B.~Xu, D.~Warde-Farley, S.~Ozair,
  A.~Courville, and Y.~Bengio, ``Generative adversarial nets,'' in
  \emph{Proceedings of Advances in Neural Information Processing Systems
  (NeurIPS)}, Montreal, QC, Canada, 2014, pp. 2672--2680.

\bibitem{2016_7780459}
K.~{He}, X.~{Zhang}, S.~{Ren}, and J.~{Sun}, ``Deep residual learning for image
  recognition,'' in \emph{Proc. {IEEE} Conf. Comput. Vis. Pattern Recognit.
  (CVPR)}, 2016, pp. 770--778.

\bibitem{2017_7839189}
K.~{Zhang}, W.~{Zuo}, Y.~{Chen}, D.~{Meng}, and L.~{Zhang}, ``Beyond a
  {G}aussian denoiser: {R}esidual learning of deep {CNN} for image denoising,''
  \emph{IEEE Transactions on Image Processing}, vol.~26, no.~7, pp. 3142--3155,
  2017.

\bibitem{2003_1227616}
J.~Tian, ``Reversible data embedding using a difference expansion,'' \emph{IEEE
  Trans. Circuits Syst. Video Technol.}, vol.~13, no.~8, pp. 890--896, 2003.

\bibitem{2007_4099409}
D.~M. {Thodi} and J.~J. {Rodriguez}, ``Expansion embedding techniques for
  reversible watermarking,'' \emph{IEEE Trans. Image Process.}, vol.~16, no.~3,
  pp. 721--730, 2007.

\bibitem{2008Fallahpour}
M.~Fallahpour, ``Reversible image data hiding based on gradient adjusted
  prediction,'' \emph{{IEICE} Electron. Exp.}, vol.~5, no.~20, pp. 870--876,
  2008.

\bibitem{5313862}
L.~Luo, Z.~Chen, M.~Chen, X.~Zeng, and Z.~Xiong, ``Reversible image
  watermarking using interpolation technique,'' \emph{IEEE Trans. Inf.
  Forensics Secur.}, vol.~5, no.~1, pp. 187--193, 2010.

\bibitem{2011_BOSSbase}
P.~Bas, T.~Filler, and T.~Pevn{\'y}, ``{B}reak our steganographic system: {T}he
  ins and outs of organizing {BOSS},'' in \emph{Proc. Int. Workshop Inf. Hiding
  (IH)}, Prague, Czech Republic, 2011, pp. 59--70.

\bibitem{1979_Lanczos}
C.~E. Duchon, ``Lanczos filtering in one and two dimensions,'' \emph{J. Appl.
  Meteorol.}, vol.~18, no.~8, pp. 1016--1022, 1979.

\bibitem{2006_USC_SIPI}
A.~G. Weber, ``The {USC-SIPI} image database: {V}ersion 5,'' USC Viterbi School
  Eng., Signal Image Process. Inst., Los Angeles, CA, USA, Tech. Rep. 315,
  2006.

\bibitem{BrodatzTextures}
P.~Brodatz, \emph{Textures: {A} Photographic Album for Artists and
  Designers}.\hskip 1em plus 0.5em minus 0.4em\relax New York, NY, USA: Dover
  Publications, 1966.

\bibitem{iet:/content/journals/10.1049/el_20080522}
Q.~Huynh-Thu and M.~Ghanbari, ``Scope of validity of {PSNR} in image/video
  quality assessment,'' \emph{Electron. Lett.}, vol.~44, no.~13, pp. 800--801,
  2008.

\bibitem{1284395}
Z.~Wang, A.~Bovik, H.~Sheikh, and E.~Simoncelli, ``Image quality assessment:
  {F}rom error visibility to structural similarity,'' \emph{IEEE Trans. Image
  Process.}, vol.~13, no.~4, pp. 600--612, 2004.

\bibitem{Hart:1971aa}
P.~E. Hart, ``Entropy and other measures of concentration,'' \emph{J. Roy.
  Statist. Soc. Ser. A}, vol. 134, no.~1, pp. 73--85, 1971.

\end{thebibliography}

\vspace{-20 mm}
\begin{IEEEbiography}[{\includegraphics[width=1in,height=1.25in,clip,keepaspectratio]{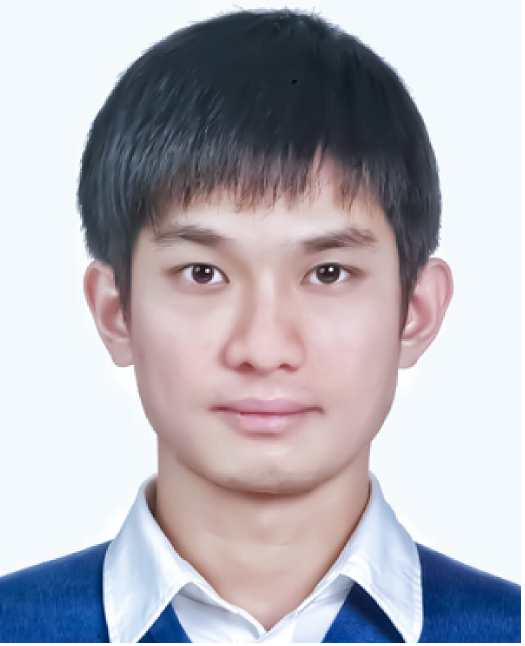}}]
{Ching-Chun Chang}
received his PhD in Computer Science from the University of Warwick, UK, in 2019. He participated in a short-term scientific mission supported by European Cooperation in Science and Technology Actions at the Faculty of Computer Science, Otto-von-Guericke-Universit\"{a}t Magdeburg, Germany, in 2016. He was granted the Marie-Curie fellowship and participated in a research and innovation staff exchange scheme supported by Marie Sk\l{}odowska-Curie Actions at the Department of Electrical and Computer Engineering, New Jersey Institute of Technology, USA, in 2017. He was a Visiting Scholar with the School of Computer and Mathematics, Charles Sturt University, Australia, in 2018, and with the School of Information Technology, Deakin University, Australia, in 2019. He was a Research Fellow with the Department of Electronic Engineering, Tsinghua University, China, in 2020. He is currently a Postdoctoral Fellow with the National Institute of Informatics, Japan. His research interests include steganography, forensics, biometrics, cybersecurity, computer vision, computational linguistics and artificial intelligence.
\end{IEEEbiography}

\begin{IEEEbiography}[{\includegraphics[width=1in,height=1.25in,clip,keepaspectratio]{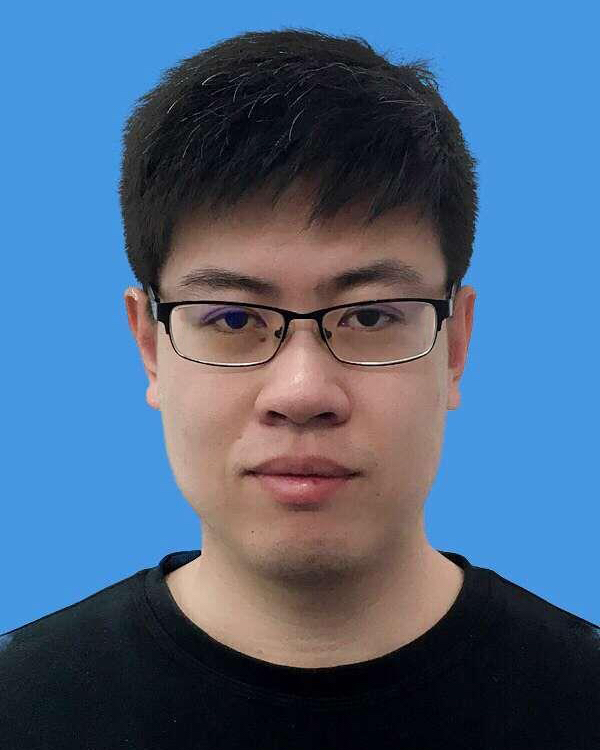}}]
{Xu Wang}
received his Ph.D. degree from Feng Chia University in 2022. He is currently a Lecturer with the School of Information Science and Engineering, University of Jinan, Jinan, China. His research interests include image processing, multimedia security, information hiding, and deep learning.
\end{IEEEbiography}

\begin{IEEEbiography}[{\includegraphics[width=1in,height=1.25in,clip,keepaspectratio]{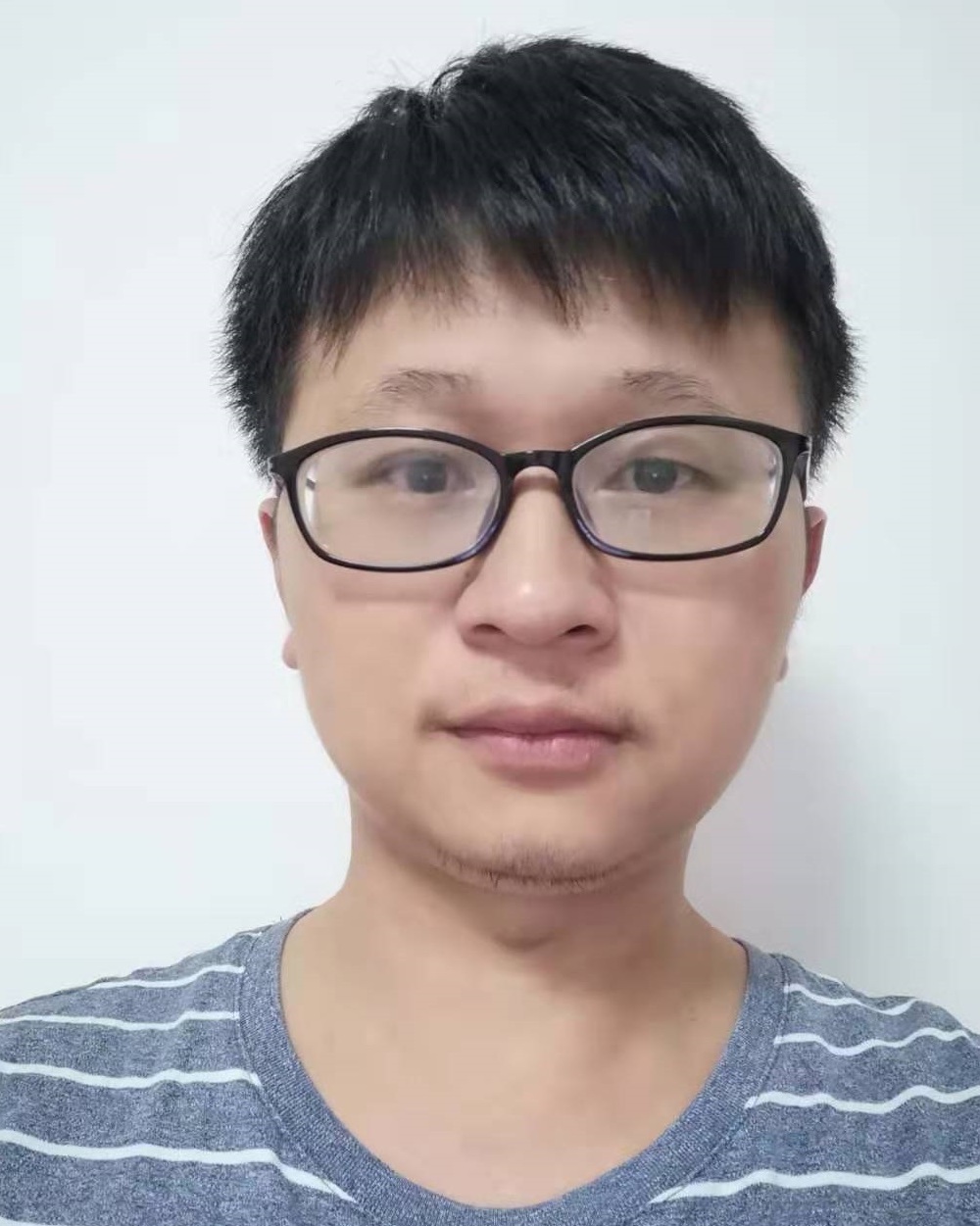}}]
{Sisheng Chen}
received the B.S. degree from Shan Dong University, in 2005, and the M.S. degree in applied mathematics from Fujian Normal University, in 2008. He is currently pursuing the Ph.D. degree in information engineering and computer science, Feng Chia University, Taichung, Taiwan. He is also a Lecturer with the Fujian Polytechnic Normal University. His research interests include multimedia security, image processing, and deep learning.
\end{IEEEbiography}

\begin{IEEEbiography}[{\includegraphics[width=1in,height=1.25in,clip,keepaspectratio]{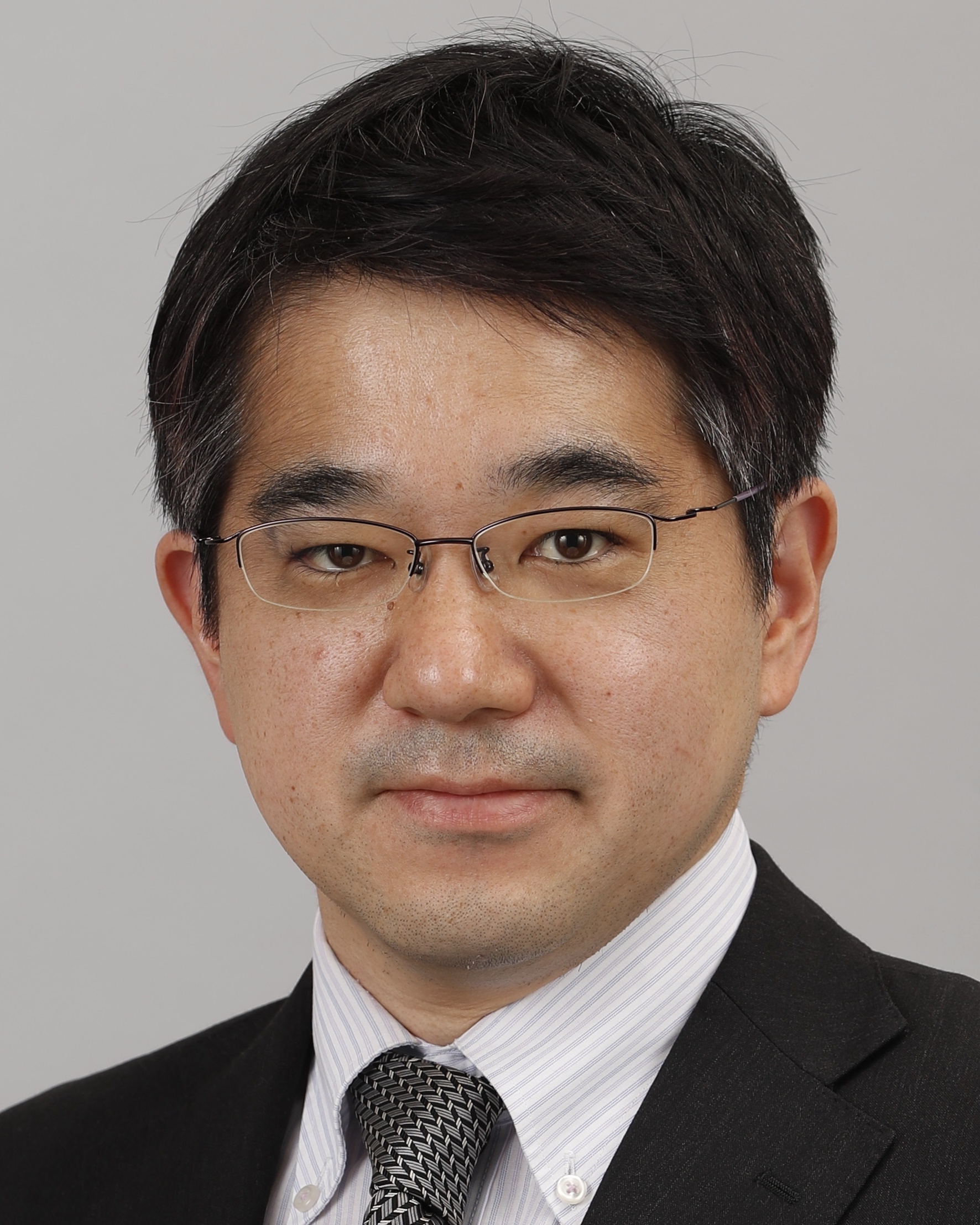}}]
{Isao Echizen}
received B.S., M.S., and D.E. degrees from the Tokyo Institute of Technology, Japan, in 1995, 1997 and 2003, respectively. He joined Hitachi, Ltd. in 1997 and until 2007 was a research engineer in the company's systems development laboratory. He is currently a Director and a Professor of the Information and Society Research Division, the National Institute of Informatics (NII), a Director of the Global Research Center for Synthetic Media, the NII, a Professor in the Department of Information and Communication Engineering, Graduate School of Information Science and Technology, the University of Tokyo, a Deputy Dean in the School of Multidisciplinary Sciences, the Graduate University For Advanced Studies (SOKENDAI), and a Professor in the  Department of Informatics, School of Multidisciplinary Sciences, SOKENDAI, Japan.  He was a Visiting Professor at the Tsuda University, Japan, at the University of Freiburg, Germany, and at the University of Halle-Wittenberg, Germany. He is currently engaged in research on multimedia security and multimedia forensics. He currently serves as a Research Director in CREST FakeMedia project, Japan Science and Technology Agency (JST). He received the Best Paper Award from the IPSJ in 2005 and 2014, the Fujio Frontier Award and the Image Electronics Technology Award in 2010, the One of the Best Papers Award from the Information Security and Privacy Conference in 2011, the IPSJ Nagao Special Researcher Award in 2011, the DOCOMO Mobile Science Award in 2014, the Information Security Cultural Award in 2016, and the IEEE Workshop on Information Forensics and Security Best Paper Award in 2017. He was a Member of the Information Forensics and Security Technical Committee of the IEEE Signal Processing Society. He is the IEEE Senior Member, and the Japanese representative on IFIP TC11 (Security and Privacy Protection in Information Processing Systems), a Member-at-Large of Board-of-Governors of APSIPA, and an Editorial Board Member of the IEEE Transactions on Dependable and Secure Computing, the EURASIP Journal on Image and Video processing, and the Journal of Information Security and Applications, Elsevier.
\end{IEEEbiography}

\begin{IEEEbiography}[{\includegraphics[width=1in,height=1.25in,clip,keepaspectratio]{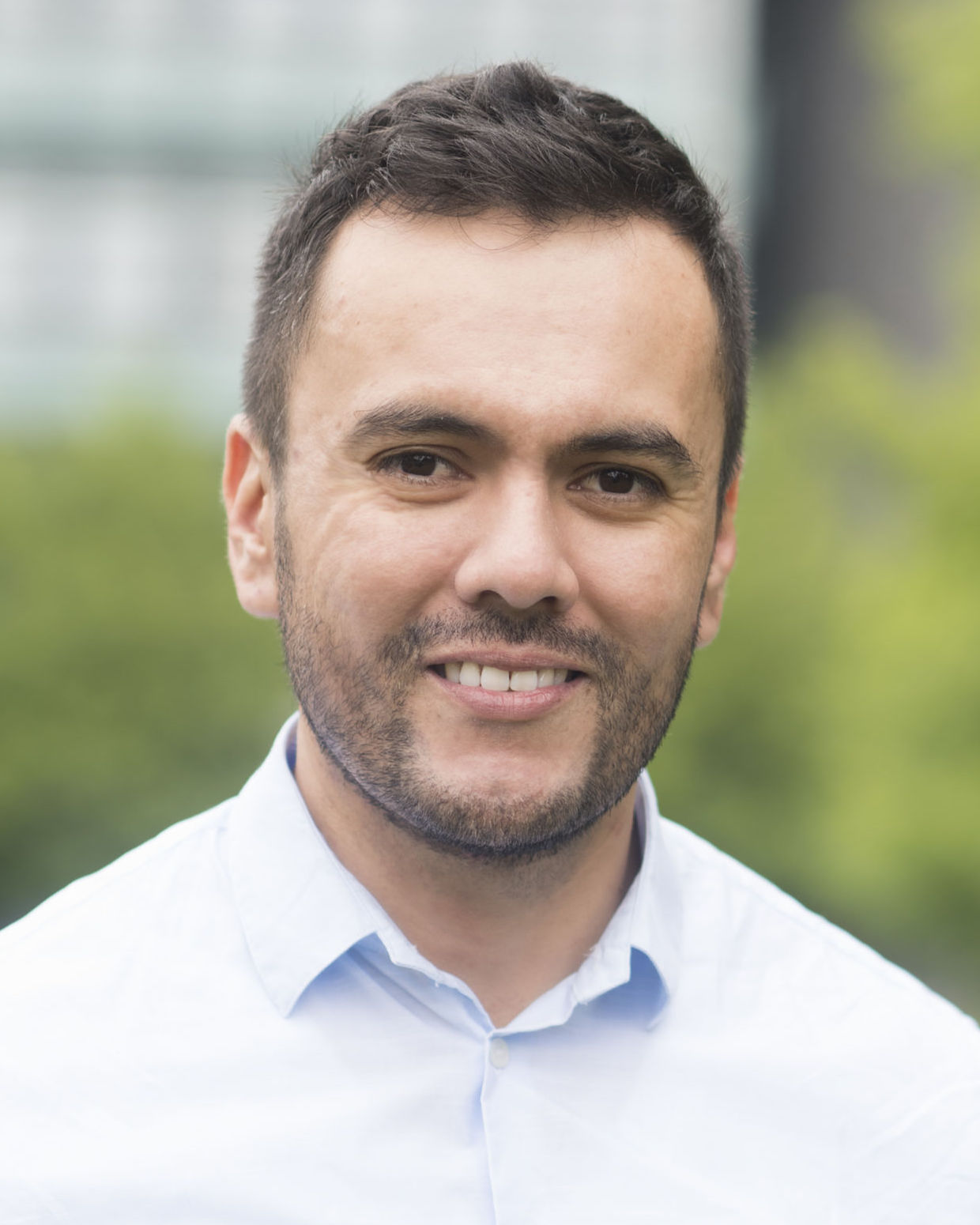}}]
{Victor Sanchez}
received his M.Sc. degree from the University of Alberta, Canada, in 2003, and his Ph.D. degree from the University of British Columbia, Canada, in 2010. From 2011 to 2012, he was with the Video and Image Processing Laboratory, University of California at Berkeley, as a Postdoctoral Researcher. In 2012, he was a Visiting Lecturer with the Group on Interactive Coding of Images, Universitat Aut\`{o}noma de Barcelona. From 2018 to 2019, he was a Visiting Scholar with the School of Electrical and Information Engineering, University of Sydney, Australia. He is currently an Associate Professor with the Department of Computer Science, University of Warwick, UK. His main research interests include signal and information processing with applications to multimedia analysis, image and video coding, security and communications. He has authored several technical articles and book chapters in these areas. His research has been funded by the Consejo Nacional de Ciencia y Tecnolog\'{i}a, Mexico; the Natural Sciences and Engineering Research Council of Canada; the Canadian Institutes of Health Research; the FP7 and the H2020 Programs of the European Union; the Engineering and Physical Sciences Research Council, UK; and the Defence and Security Accelerator, UK.
\end{IEEEbiography}

\begin{IEEEbiography}[{\includegraphics[width=1in,height=1.25in,clip,keepaspectratio]{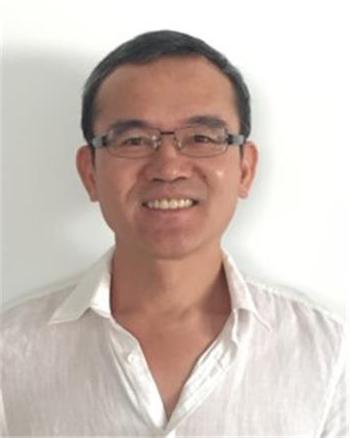}}]
{Chang-Tsun Li}
received the BSc degree in Electrical Engineering from the National Defence University, Taiwan, the MSc degree in Computer Science from the US Naval Postgraduate School, USA, and the PhD degree in Computer Science from the University of Warwick, UK. He is currently a Professor of Cyber Security at Deakin University, Australia. He has had over 20 years research experience in multimedia forensics and security, biometrics, machine learning, data analytics, computer vision, image processing, pattern recognition, bioinformatics and content-based image retrieval. The outcomes of his research have been translated into award-winning commercial products protected by a series of international patents and have been used by a number of law enforcement agencies, national security institutions and companies around the world, including INTERPOL (Lyon, France), UK Home Office, Metropolitan Police Service (UK), Sussex Police Service (UK), Guildford Crown Court (UK), and US Department of Homeland Security. The research team of the CSCRC-funded Development of Australian Cyber Criteria Assessment (DACCA) project under his leadership is the winner of the 2021 AISA Cyber Security Researcher of the Year Award. He is currently Chair of Computational Forensics Technical Committee of the International Association of Pattern Recognition (IAPR), Member of IEEE Information Forensics and Security Technical Committee, Associate Editor of IEEE Transactions on Circuit and System for Video Technology, IET Biometrics and the EURASIP Journal of Image and Video Processing.
\end{IEEEbiography}








\EOD
\end{document}